\documentclass[useAMS,usenatbib,twocolumn]{mn2e}
\pdfoutput=1
\usepackage{amssymb}
\usepackage{amsmath}
\usepackage{natbib}
\usepackage[pdftex]{graphicx}
\usepackage{subfigure}
\usepackage{booktabs}
\usepackage{tabularx}
\usepackage{float}
\usepackage{afterpage}
\begin{document}
\label{firstpage}
\title[Nonlinear tides]
{Nonlinear tides in a homogeneous rotating planet or star: global simulations of the elliptical instability}
          \author[A. J. Barker]{Adrian J. Barker\thanks{Email address: ajb268@cam.ac.uk} \\
Department of Applied Mathematics and Theoretical Physics, University of Cambridge, Centre for Mathematical Sciences, \\ Wilberforce Road, Cambridge CB3 0WA, UK}
\date{}
\pagerange{\pageref{firstpage}--\pageref{lastpage}} \pubyear{2016}
\maketitle

\begin{abstract}
I present results from the first global hydrodynamical simulations of the elliptical instability in a tidally deformed gaseous planet (or star) with a free surface. The elliptical instability is potentially important for tidal evolution of the shortest-period hot Jupiters. I model the planet as a spin-orbit aligned or anti-aligned, and non-synchronously rotating, tidally deformed, homogeneous fluid body. A companion paper presented an analysis of the global modes and instabilities of such a planet. Here I focus on the nonlinear evolution of the elliptical instability. This is observed to produce bursts of turbulence that drive the planet towards synchronism with its orbit in an erratic manner. If the planetary spin is initially anti-aligned, the elliptical instability also drives spin-orbit alignment on a similar timescale as the spin synchronisation. The instability generates differential rotation inside the planet in the form of zonal flows, which play an important role in the saturation of the instability, and in producing the observed burstiness. These results are broadly consistent with the picture obtained using a local Cartesian model (where columnar vortices played the role of zonal flows). I also simulate the instability in a container that is rigid (but stress-free) rather than free, finding broad quantitative agreement. The dissipation resulting from the elliptical instability could explain why the shortest-period hot Jupiters tend to have circular orbits inside about 2--3 days, and predicts spin-synchronisation (and spin-orbit alignment) out to about 10--15 days. However, other mechanisms must be invoked to explain tidal circularisation for longer orbital periods.
\end{abstract}
	
	
\section{Introduction}

Gravitational tidal interactions between short-period planets and their host stars can play an important role in the evolution of the orbit and internal rotations of both bodies. As probably the clearest example, dissipation of planetary tidal flows is thought to explain why the shortest-period hot Jupiters have preferentially circular orbits, unlike the population of Jovian planets with orbital periods longer than ten days, which have a wide range of eccentricities (e.g.~\citealt{Rasio1996,WF2015}). Over the last decade, much work has been devoted to understand the mechanisms of tidal dissipation in fluid bodies, but many uncertainties remain \citep{Gio2004,Wu2005b,IvanovPap2007,GoodmanLackner2009,PapIv2010,FBBO2014,Ogilvie2014}.

One of the major uncertainties in the theory of tides is the importance of nonlinear fluid effects. These may be particularly important for the tides in the shortest-period hot Jupiters because of their large amplitudes (e.g. WASP-19 b has a dimensionless tidal amplitude $A\sim 0.05$ using Eq.~\ref{deftidalamp} below, which can no longer be treated as a small parameter), so that linear theory (e.g.~\citealt{Wu2005b,Gio2004,IvanovPap2007,PapIv2010,Ogilvie2013}) may no longer accurately describe the tidal response. Nonlinear fluid effects are likely to play a crucial role whenever tidal forcing excites small-scale waves (typically restored by buoyancy and/or rotation), since nonlinearities become important for much smaller amplitudes for these waves than for large-scale tidal flows, resulting in wave breaking \citep{Barker2010,Barker2011} or subtler parametric instabilities \citep{BO2011,Weinberg2012}, as well as localised angular momentum deposition (e.g.~\citealt{FBBO2014}). In addition, nonlinear tidal effects can drive instabilities of the large-scale non-wavelike tidal flows, which would not be predicted by a linear tidal theory.
In this work, and in the companion paper, we study one such instability: the elliptical instability, which occurs in fluids with elliptical streamlines \citep{Kerswell2002}, such as in tidally deformed planets or stars.

Previous work using a local computational model \citep{BL2013,BL2014} has demonstrated that this instability could be important for tidal dissipation inside planets with the shortest orbital periods (which have the largest dimensionless tidal amplitudes) -- in particular, it may explain why hot Jupiters with orbital periods shorter than about 2 days have preferentially circular orbits. The elliptical instability has also been studied in impressive laboratory experiments \citep{Lacaze2004,LeBars2007,LeBars2010}, as well as in global numerical simulations in a rigid ellipsoidal container \citep{Cebron2010,Cebron2013}. However, simulations of the instability in a global model with a realistic free surface that are appropriate for this problem have not yet been undertaken (though see \citealt{Ou2004,Ou2007} for a different application). This is the primary aim of this paper. Studying global effects is required in order to determine how they could modify the outcome of the instability. If global effects could modify the dissipative properties of the flow, this might elevate the importance of this instability for astrophysics, by allowing it to be important for longer orbital periods.

I adopt the simplest global model in which to self-consistently study nonlinear tidal effects in planets (or stars): a rotating and tidally deformed homogeneous ellipsoidal fluid body. More realistic models should be considered in future investigations, but this model has enormous theoretical advantages over more complicated ones due to its tractability, and there is an existing body of work that we can apply to aid our understanding of its properties \citep{L1989a,L1989b,ST1992,LL1996,LL1996a}. In addition, it is the cleanest configuration in which to study the elliptical instability in isolation. This is because the lowest order (quadrupolar) tidal potential does not directly excite global inertial modes in a homogeneous incompressible body (at least for aligned spin and orbit; \citealt{GoodmanLackner2009,Ogilvie2009,PapIv2010}), so enhanced tidal dissipation (over viscous damping of the global tidal flow) can only occur via this instability. In a companion paper \citep{Barker2015a}, we studied the global modes and instabilities of such a planet. In this work I present the results of global numerical simulations, using a spectral element method, to study the nonlinear outcome of the elliptical instability. My aim is to understand its nonlinear evolution and to determine its astrophysical relevance for tidal dissipation. 

I briefly describe the model in \S 2 -- though see \cite{Barker2015a} for further details -- before describing the code used and various code tests in \S 3 and 4. The main results are presented in \S 5 and 6, and a discussion (where these results are applied to the tidal evolution of extrasolar planets and close binary stars) and conclusion is presented in \S 7 and 8.

\section{Simplified model}

In \cite{Barker2015a}, we constructed a simple model in which we can study nonlinear tides in a gaseous planet or star, which I will briefly outline here. I focus on the spin--orbit synchronisation (and spin--orbit alignment) problem for an aligned (or purely anti-aligned) circular orbit for simplicity. This is because in this case there exists a natural frame in which the equilibrium shape of the ellipsoid is fixed (in the absence of instabilities), which is advantageous computationally. I use Cartesian coordinates $(x,y,z)$ centred on the planet of mass $m_p$ and unperturbed radius $R_p$ such that the $z$-axis is aligned with its spin axis, and the angular velocity of the fluid is $\Omega\geq 0$. The star has mass $m_\star$, about which the planet orbits with angular velocity $\boldsymbol{n}=n\boldsymbol{e}_z$. I allow the sign of $n$ to be positive or negative so that both purely aligned (prograde) and anti-aligned (retrograde) orbits can be studied. In the frame that rotates at the rate $\boldsymbol{n}$ (the ``bulge frame"), the governing equations are 
\begin{eqnarray}
\label{eqs1}
&&\left(\partial_{t} + \boldsymbol{u}\cdot \nabla\right)\boldsymbol{u} + 2\boldsymbol{n}\times \boldsymbol{u} = -\nabla \Pi + \nu \nabla^{2}\boldsymbol{u}, \\
&& \nabla \cdot \boldsymbol{u} = 0, \\
&& \Pi = p+\Phi-\frac{1}{2}|\boldsymbol{n}\times\boldsymbol{x}|^{2}+\Psi.
\label{eqs3}
\end{eqnarray}
where $p$ is a pressure, $\nu$ is the kinematic viscosity, $\Phi$ is a fixed gravitational potential and $\Psi$ is an imposed tidal potential. The fluid has uniform density ($\rho\equiv1$) and is incompressible. Viscosity can be thought to crudely represent the effects of turbulent convection on dissipating large-scale tidal flows (e.g.~\citealt{Zahn1966,Goldreich1977,Penev2007,Penev2009,OgilvieLesur2012}), though I take $\nu$ to be uniform and independent of frequency in this work, both for simplicity, and because the four-decade old controversy over the frequency dependence of $\nu$ has not been resolved. 

Equilibrium is maintained by fluid pressure, central gravity and centrifugal and tidal forces. I adopt a fixed gravitational potential for the planet
\begin{eqnarray}
\label{fixedpot}
\Phi(\boldsymbol{x})=\frac{1}{2}\omega_{d}^{2}r^2,
\end{eqnarray}
where $r^2=x^2+y^2+z^2$, and the dynamical frequency is $\omega_d= \sqrt{\frac{Gm_p}{R_p^3}}$. Using a fixed potential might be thought to represent the gravity of a centrally condensed body, though in this case the body is strictly uniform. I neglect the self-gravity of the fluid for computational convenience, but its inclusion would be unlikely to significantly change the results (the elliptical instability excites inertial waves, which only weakly perturb the gravitational potential).

The tidal potential (to lowest order) is
\begin{eqnarray}
\Psi = \frac{A\omega_d^2}{2} \left( r^2 - 3 (\hat{\boldsymbol{a}_\star}\cdot \boldsymbol{x})^2\right),
\end{eqnarray}
where $\hat{\boldsymbol{a}_\star}=(1,0,0)$ defines the direction to the star, which is stationary in the bulge frame (because the orbit is circular). I define 
\begin{eqnarray}
\label{deftidalamp}
A&=& \frac{m_{\star}}{m_{p}}\left(\frac{R_p}{a_\star}\right)^3,
\end{eqnarray}
which is a measure of the dimensionless tidal amplitude, with $a_\star$ being the distance to the star, and adopt units such that the tidally unperturbed non-rotating planet would have radius $R_p=1$ and set $\omega_d=1$. 

The planet has volume $V$ and its surface is the triaxial ellipsoid
\begin{eqnarray}
\label{1}
\frac{x^2}{a^2}+\frac{y^2}{b^2}+\frac{z^2}{c^2}=1,
\end{eqnarray}
where $a,b$ and $c$ are the semi-axes of the ellipsoid, which are stationary in the bulge ($\boldsymbol{n}$) frame (in the absence of viscosity and instabilities). The basic laminar tidal flow is
\begin{eqnarray}
\label{basicflow}
\boldsymbol{U}_{0}(\boldsymbol{x})=\gamma\left(-\frac{a}{b}y,\frac{b}{a}x,0\right), 
\end{eqnarray}
where $\gamma=\Omega-n$, which is an exact inviscid solution that is steady (in the absence of instabilities) and results from the non-synchronous rotation of the fluid ($\Omega\ne n$) in its ellipsoidal volume. When $A\ne 0$ (and $\gamma\ne 0$), the planet is similar to a Roche-Riemann ellipsoid \citep{C1987}. The stability of a similar configuration (a Riemann S-type ellipsoid) has been studied in the absence of a tidal deformation by \cite{LL1996,LL1996a}, and most recently we have studied the stability of our Roche-Riemann-like ellipsoid (including the tidal deformation) in the companion paper \citep{Barker2015a}.

The shape of the planet can be determined from the three input parameters ($A,\Omega,n$) by \citep{Barker2015a}
\begin{eqnarray}
\label{shape}
\epsilon &=& \frac{3A}{2(1-\gamma^2-n^2)-A}, \\
c^2 &=& \frac{2\left[ (2A+\gamma^2+n^2-1)(A-\gamma^2-n^2+1)+f\right]}{(A+1)(A+2(\gamma^2+n^2-1))},
\end{eqnarray}
with
\begin{eqnarray}
f=2\gamma n \sqrt{(1-2A-\gamma^2-n^2)(1+A-\gamma^2-n^2)},
\end{eqnarray}
and where $\epsilon$ is a measure of the tidal deformation, defined by $a=\sqrt{1+\epsilon}$ and $b=\sqrt{1-\epsilon}$. Note that $\epsilon\approx \frac{3A}{2}$ for small $\gamma$, $n$ and $A$.

Note that $\boldsymbol{U}_{0}$ is not an exact solution in the presence of viscosity, because it does not satisfy the stress-free boundary condition at the free surface. Viscosity leads to a weak tidal torque, even in the absence of instability, which slowly synchronises the spin of the body with its orbit, and drives additional weak internal flows.
The mean viscous dissipation rate for the flow given by Eq.~\ref{basicflow} is
\begin{eqnarray}
D_{\mathrm{lam}}= \frac{2\nu}{V}\int_{V} e_{ij}e_{ij}\; \mathrm{d}V=\nu \gamma^2 \left(\frac{a}{b}-\frac{b}{a}\right)^2,
\label{disspred}
\end{eqnarray}
where $e_{ij}=\frac{1}{2}\left(\partial_i u_{j} + \partial_j u_{i}\right)$ (taking $u_i=U_{0,i}$). In the presence of viscosity, $D_{\mathrm{lam}}$ does not vanish unless $\gamma=0$ (or $a=b$). In the absence of perturbations and viscosity, our planet will remain in equilibrium for all time, because $\boldsymbol{U}_{0}$ is an exact inviscid solution. However, this flow has elliptical streamlines, so an infinitesimal perturbation will drive the elliptical instability for certain choices of $A,\Omega$ and $n$. This instability will also be excited in the presence of viscosity if its growth rate is sufficiently large (in practice, this requires us to consider tidal amplitudes that are slightly larger than those for most of the observed hot Jupiters in our simulations).

The planetary orbit is fixed in this model, which is a reasonable approximation when studying the synchronisation (and spin-orbit alignment) of the planet. This is because short-period planets typically have much less angular momentum contained in their spin compared with their orbits. I focus on the synchronisation problem in this work because it is the simplest computationally, as there exists a frame in which the geometry of the tidal flow is stationary, in the absence of instabilities. While I do not directly study the circularisation problem, the small-scale linear instability has similar properties in this case \citep{KerswellMalkus1998}, and I believe its nonlinear evolution is likely to share many properties with the synchronisation problem that is explicitly studied in this paper.

\subsection{Elliptical instability}

The elliptical instability is a fluid instability of elliptical streamlines, which can be driven if $\gamma\ne 0$ and $\epsilon\ne 0$ and draws upon the kinetic energy of the tidal flow. We have studied the global properties of this instability in detail in \cite{Barker2015a}, but to aid the reader who may not be familiar with its properties, I briefly introduce those which are relevant to understand the main results of \S~\ref{Nonlinear}. The instability results from the interaction of the elliptical deformation (which has frequency $2\gamma$ in the fluid frame, which rotates at the rate $\Omega$ about the $z$-axis) with a pair of inertial waves (which have frequencies in the fluid frame $|\omega_{i}|\leq 2|\Omega|$ for $i=1,2$). Instability is possible if the wave frequencies add up so that they are in resonance with the deformation. For instability to occur, we also require the two waves to have harmonic orders $\ell_1=\ell_2$ and azimuthal wavenumbers $m_1\pm m_2=2$, since the tidal deformation has $m=2$ \citep{Kerswell1993,LL1996}. The growth rate is typically 
\begin{eqnarray}
\sigma \sim \epsilon \gamma,
\end{eqnarray}
but exhibits an additional dependence on $n$ \citep{Craik1989}. The fastest growing modes typically have $|\omega_1|\approx |\omega_2|\approx \gamma$, and a necessary condition for instability (in the case of uniform rotation and $\epsilon\ll 1$) is that
\begin{eqnarray}
-\Omega \leq n \leq 3 \Omega,
\end{eqnarray}
since inertial waves cannot be excited outside of this frequency range \citep{Kerswell2002}. However, when $\epsilon$ is no longer tiny, instability is possible for modes that are not exactly resonant because each resonance has a finite width ($O(\epsilon\gamma)$). Thus, a given pair of waves can be unstable if $\omega_1\pm\omega_2 \approx 2\gamma \left(1+O(\epsilon)\right)$ -- in addition, for large enough $\epsilon$, instabilities involving three or more waves are possible in principle. More detailed results relating to the instability are presented in \cite{Barker2015a}.

\section{Numerical method and setup}

I solve Eqs.~\ref{eqs1}--\ref{eqs3} in their weak variational form with the efficiently parallelised spectral element code Nek5000 \citep{nek5000}. Spectral element methods combine the geometric flexibility of finite element methods with the accuracy of spectral methods, which makes them particularly suitable for studying tidal flows in realistic ellipsoidal geometries. I have previously used this code to study tidally forced inertial waves in spherical shells \citep{FBBO2014}. 

The computational domain is decomposed into $E$ non-overlapping hexahedral elements, and within each element, the velocity and pressure are represented as tensor-product Lagrange polynomials of orders $N$ and $N-2$, where the points are the Gauss-Lobatto-Legendre and Gauss-Legendre points, respectively (e.g.~\citealt{DevilleFischerMund2002}). The convergence is algebraic with increasing number of elements $E$ and exponential with increasing polynomial order $N$. The total number of grid points for each velocity component is $E N^3$. Temporal discretisation in Nek5000 is accomplished by a third order method based on a semi-implicit formulation, in which the nonlinear and Coriolis terms are treated explicitly, and the remaining linear terms are treated implicitly. Solutions are de-aliased following the $3/2$ rule i.e.~$3N/2$ grid points in each dimension are used for the non-linear terms, whereas only $N$ are used for the linear terms. I use an adaptive time-step based on the CFL condition with an appropriate safety factor, and I have ensured that the mesh and fluid motion are accurately computed by checking that the results do not depend on the time-step size for several cases.

\begin{figure}
  \begin{center}
      \subfigure{\includegraphics[trim=0cm 0cm 0cm 0cm, clip=true,width=0.4\textwidth]{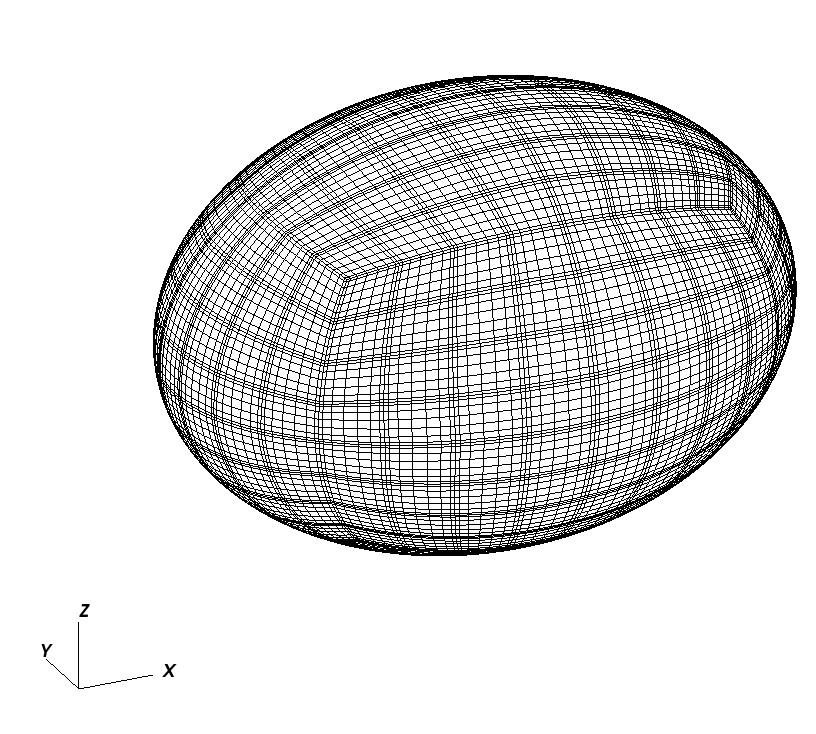} }
      \subfigure{\includegraphics[trim=0cm 0cm 0cm 1cm, clip=true,width=0.4\textwidth]{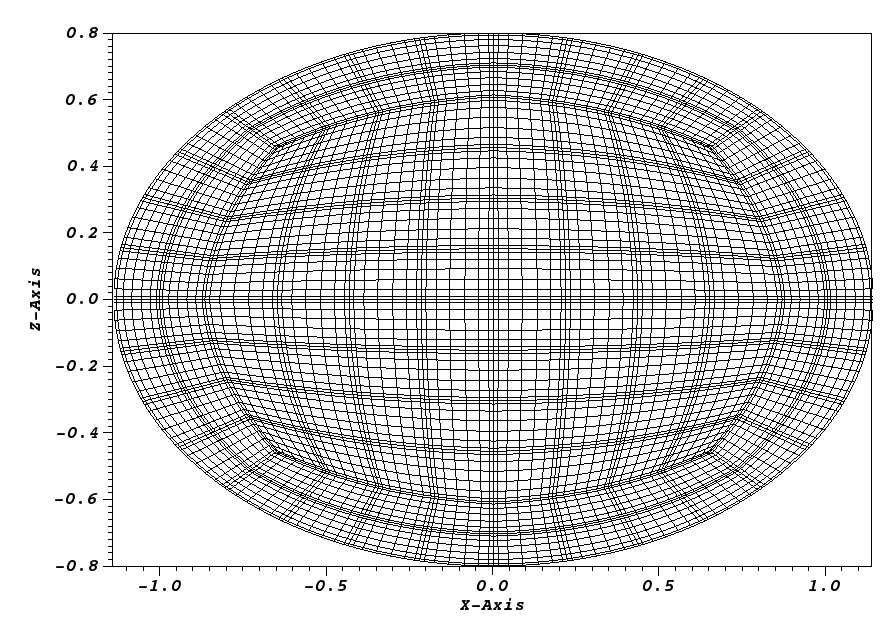} }
       \end{center}
  \caption{An example spectral element mesh for a highly deformed ellipsoid with $\epsilon=0.3$ and $c=0.8$ for illustration. This has $E=1280$ elements and $N=8$ grid points inside each element, which is the lowest resolution considered in this work. Top: external view. Bottom: slice through the mesh at $y=0$.}
  \label{0}
\end{figure}

The computational mesh has $E=1280$ elements, which results from the merger of a Cartesian mesh close to the origin and a pair of concentric spherical shells close to the external boundary. The resulting mesh describes a full sphere centred on the origin, but we can simply deform the mesh into one that describes a full ellipsoid by applying the transformation $(x,y,z)\rightarrow (ax,by,cz)$ for each grid point in the original mesh, where $a,b,c$ are chosen so that the initial state is an inviscid equilibrium for a given set of parameters ($A,\Omega,n$). This method works well for all $a,b,c$ considered in this work. I typically adopt $N=8$ to $14$ for the simulations presented in this work (corresponding to approximately $\sim 90^3$ to $150^3$ grid points). An example deformed mesh is shown in Fig.~\ref{0} for $N=8$.

Nek5000 solves for the mesh motion using an Arbitrary Lagrangian Eulerian (ALE) method, which allows us to treat the boundary condition at the surface of the planet as a free surface. Unfortunately, the extra computational work per time-step, together with the restriction on the time-step to ensure that the mesh motion is accurately captured, makes the code significantly more computationally demanding than a fixed grid version. This is the first time that such a boundary condition has been studied in nonlinear simulations of tidal flows in planets or stars. Later in this work, I will present the results of comparison simulations that have a rigid boundary (with the same initial container shape) on which an impenetrable but stress-free condition is applied to the fluid velocity. These simulations have also been performed using Nek5000 with otherwise the same setup as the free surface simulations.

\subsection{Testing surface gravity modes in Nek5000}
\label{tests}

In previous work, I have thoroughly tested Nek5000 on several problems with a fixed boundary (e.g.~\citealt{FBBO2014,BDL2014}). Here I outline several tests of its free surface capabilities, which were undertaken so that I could be comfortable in its application to the main problem of this paper. The simplest test is to compare the frequencies of surface gravity modes in a non-rotating spherical planet in a fixed gravitational potential (Eq.~\ref{fixedpot}), which are\footnote{Note that the $\ell=1$ mode would be a trivial mode with zero frequency if I had solved Poisson's equation for $\Phi$ instead of adopting a fixed potential. These frequencies match onto those of a self-gravitating body in the limit of large $\ell$.} (e.g.~\citealt{Barker2015a})
\begin{eqnarray}
\omega_\ell = \pm \sqrt{\ell} \omega_d.
\end{eqnarray}
To test these, I initialise a simulation with no flow at $t=0$, but with a mesh displacement $\boldsymbol{x} \rightarrow \boldsymbol{x} + \boldsymbol{\xi},$
where 
\begin{eqnarray}
\boldsymbol{\xi} = A_{\xi} f(r) r^{\ell-1}\mathrm{Re}\left[\tilde{Y}_{\ell}^{\ell}(\theta,\phi)\right] \boldsymbol{x}
\end{eqnarray}
with $f(r)=r^{12}$, and $\tilde{Y}_{\ell}^{\ell}$ is a spherical harmonic but with constants set to unity. (Note that this strictly has non-vanishing divergence -- alternatively, the exact eigenfunction for a surface gravity mode could have been used.) This is a radial displacement that is proportional to a sectoral harmonic (with $\ell=m$), and $f(r)$ guarantees that the perturbation is strongest at the surface and vanishes near $r=0$. I choose $A_{\xi}=0.005$ (so that we start in the linear regime) and run simulations for $\ell\in[1,6]$, using a resolution of $N=8$ and fixed time-step $\mathrm{dt}=0.003$ ($\nu=10^{-8}$ i.e.~viscosity is negligible). 

The mean kinetic energy for these simulations,
\begin{eqnarray}
K=\frac{1}{V}\int_V\frac{1}{2}|\boldsymbol{u}|^2\mathrm{d}\, V,
\end{eqnarray}
is plotted in the top panel of Fig.~\ref{1} for each $\ell\in[1,6]$, where $V$ is the fluid volume. This shows that oscillations with larger $\ell$ have larger frequencies, and that the waves remain negligibly damped for several periods (as expected for the chosen viscosity). In the bottom panel of Fig.~\ref{1}, I show the results of computing the Lomb-Scargle periodogram \citep{Press1992} of the RMS velocity $\sqrt{2K}$. The predicted frequencies are represented by the dashed vertical lines, and the excellent agreement with the dominant frequency in the simulations indicates that the code accurately captures these modes. This provides a significant test of the free surface capabilities of Nek5000.

\begin{figure}
  \begin{center}
      \subfigure{\includegraphics[trim=4.5cm 0cm 8.3cm 0cm, clip=true,width=0.35\textwidth]{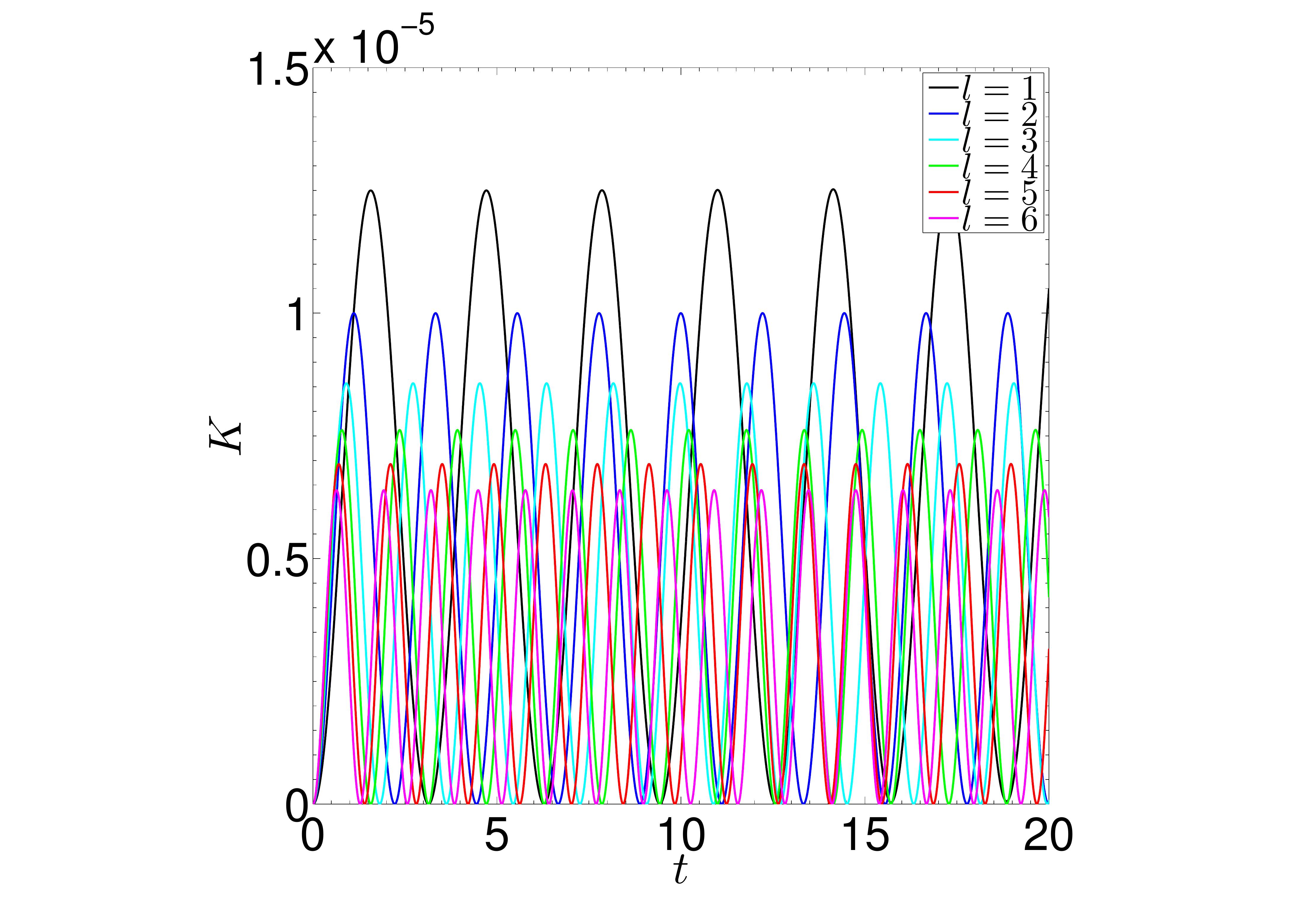} }
      \subfigure{\includegraphics[trim=4.5cm 0cm 8.3cm 0cm, clip=true,width=0.35\textwidth]{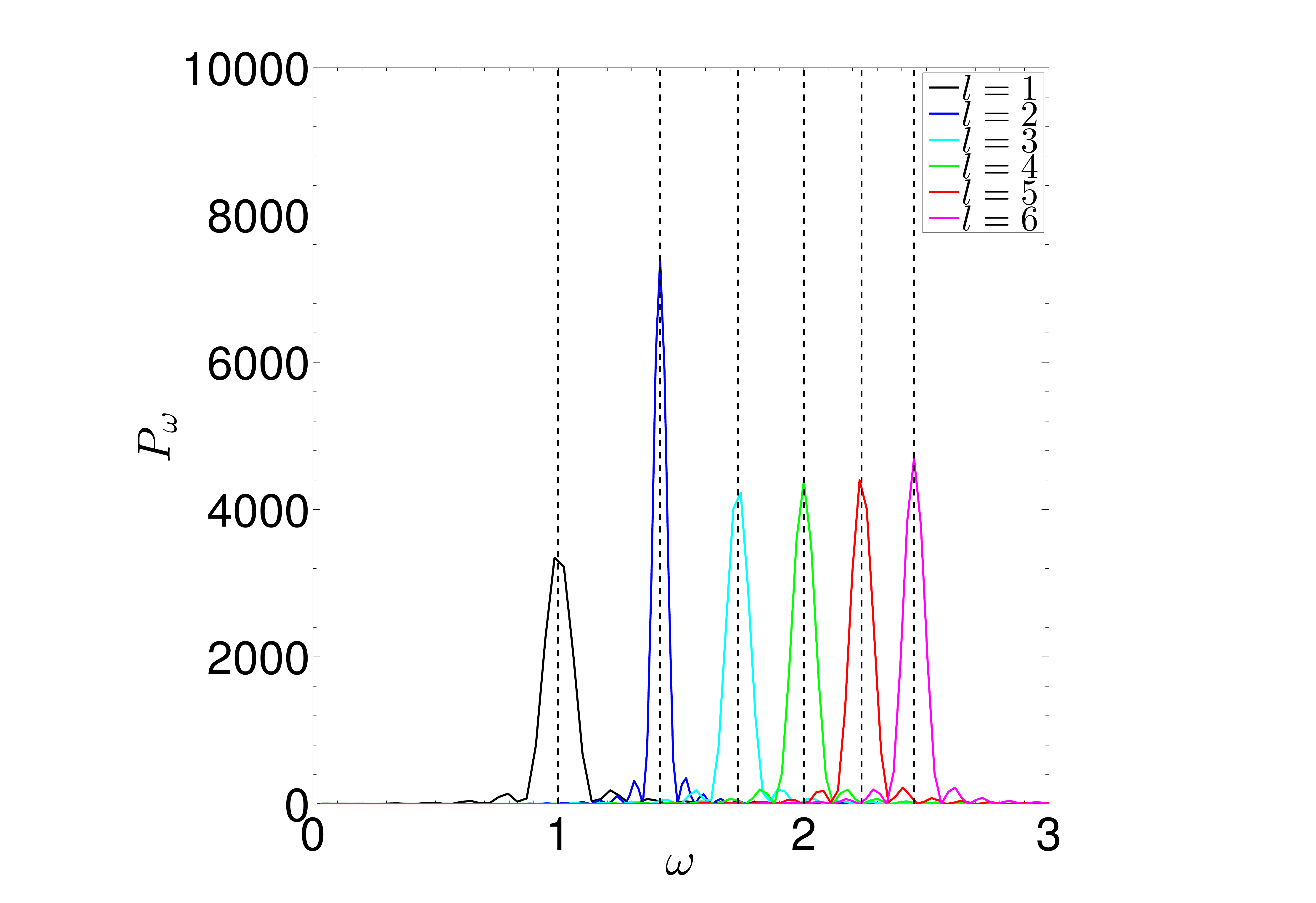} }
       \end{center}
  \caption{Test of the free surface capabilities of Nek5000 for initial surface deformations with $\ell\in[1,6]$ of a non-rotating sphere designed to excite surface gravity modes. Top: mean kinetic energy vs time. Bottom: Lomb-Scargle periodogram of $\sqrt{2K}$, showing the theoretical prediction $\omega_\ell=\sqrt{\ell}\omega_d$ as dashed vertical lines. This shows excellent agreement with the theoretical predictions. (The difference in significance of the peaks in the right panel is due to the different run-times considered).}
  \label{1}
\end{figure}

Several additional tests of the code were also carried out, including comparing the frequencies of surface gravity modes of a rotating ``Maclaurin-like spheroid" against theoretical predictions \citep{Braviner2014,Barker2015a}, and the evolution of the shape and simplest internal flows (linear in Cartesian coordinates) of an ellipsoid for various initial conditions (which have been tested against numerical integration of the ODEs derived within the formalism of \citealt{ST1992} and listed in Appendix B of \citealt{Barker2015a}). The agreement was excellent in all cases, so I omit these additional tests for brevity. Several further tests of the code are outlined in \S \ref{Nonlinear} in my discussion of simulation results. I can therefore be confident in applying Nek5000 to study the nonlinear evolution of the elliptical instability.

\subsection{Nonlinear simulations}

I initialise each simulation with the flow given by Eq.~\ref{basicflow}, to which I add small amplitude $\sim 10^{-5}$ random noise to each component of the velocity field at each grid point to allow instability to develop. All simulations are listed in Table \ref{table2} in Appendix~\ref{AppendixTable}, for reference. The initial shape and internal flow is an inviscid equilibrium, but it evolves weakly due to viscosity, and also when instability develops. The fact that the initial flow remains an equilibrium (prior to instability) in these simulations, apart from weak viscous evolution, indicates that the code correctly captures the basic equilibrium configuration. In addition, the initial mean viscous dissipation rate in each simulation accurately matches the prediction given by Eq.~\ref{disspred}, which I list in Table \ref{table2} and will explain in more detail in \S~\ref{Nonlinear}.

I study the instability as a function of ($A,\Omega,n$) as well as the kinematic viscosity $\nu$, which are treated as independent parameters. 
I will first present a comparison of the growth rate of the elliptical instability with theoretical predictions \citep{Barker2015a} in \S \ref{comparison}, where I also confirm the presence of a violent elliptical instability for retrograde spins outside the range in which it is usually thought to operate. I then illustrate the main results from the nonlinear simulations with several aligned (prograde) examples in \S\ref{Nonlinear}, and with several examples in which the planet has an initially anti-aligned (retrograde) spin in \S \ref{retrogradecases}. A comparison of free surface and rigid boundary simulations is presented in Appendix \ref{rigidsims}. 

\section{Comparison with theoretical predictions}
\label{comparison}

I begin by comparing the growth rates of the elliptical instability in the early stages of the global simulations with the predictions of the stability analysis of \cite{Barker2015a}. This set of simulations has $\Omega=0.2$ and $\nu=10^{-4}$ for various $n$, with either $A=0.1$ or $A=0.15$. The RMS vertical velocity is
\begin{eqnarray}
\langle u_z\rangle=\sqrt{\frac{1}{V}\int_V u_z^2 \, \mathrm{d}V},
\end{eqnarray}
which primarily quantifies the (vertical) energy in the inertial waves driven by the elliptical instability. In the linear growth phase, I fit a straight line to $\ln \langle u_z\rangle $ as a function of $t$, of the form $\ln \langle u_z\rangle \propto \sigma t$, to determine $\sigma$ (given that the instability excites waves with nonzero frequencies, a time interval that covers several wave periods must be chosen). 

\begin{figure}
  \begin{center}
      \subfigure{\includegraphics[trim=7cm 0cm 8cm 0cm, clip=true,width=0.35\textwidth]{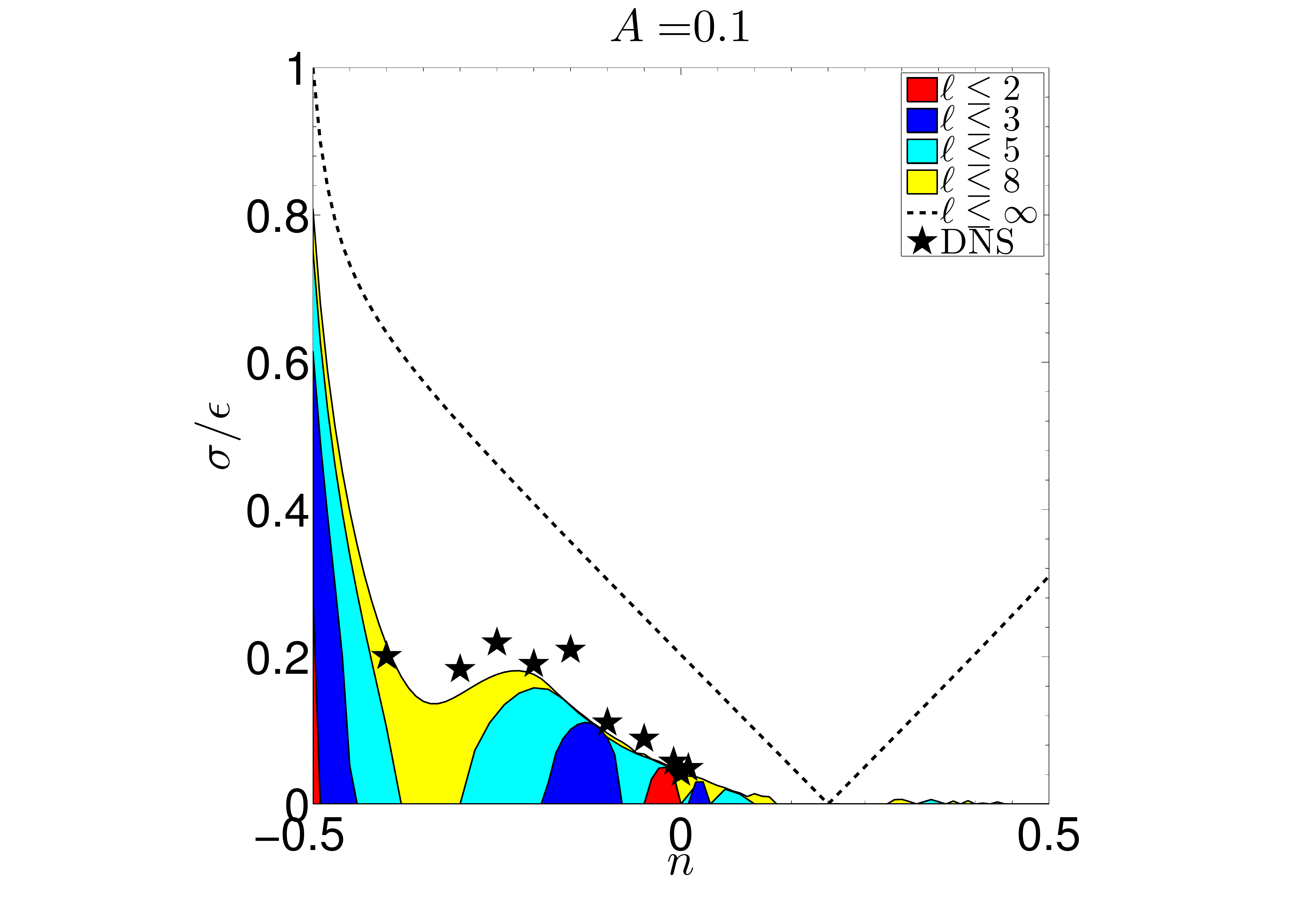} }
      \subfigure{\includegraphics[trim=7cm 0cm 8cm 0cm, clip=true,width=0.35\textwidth]{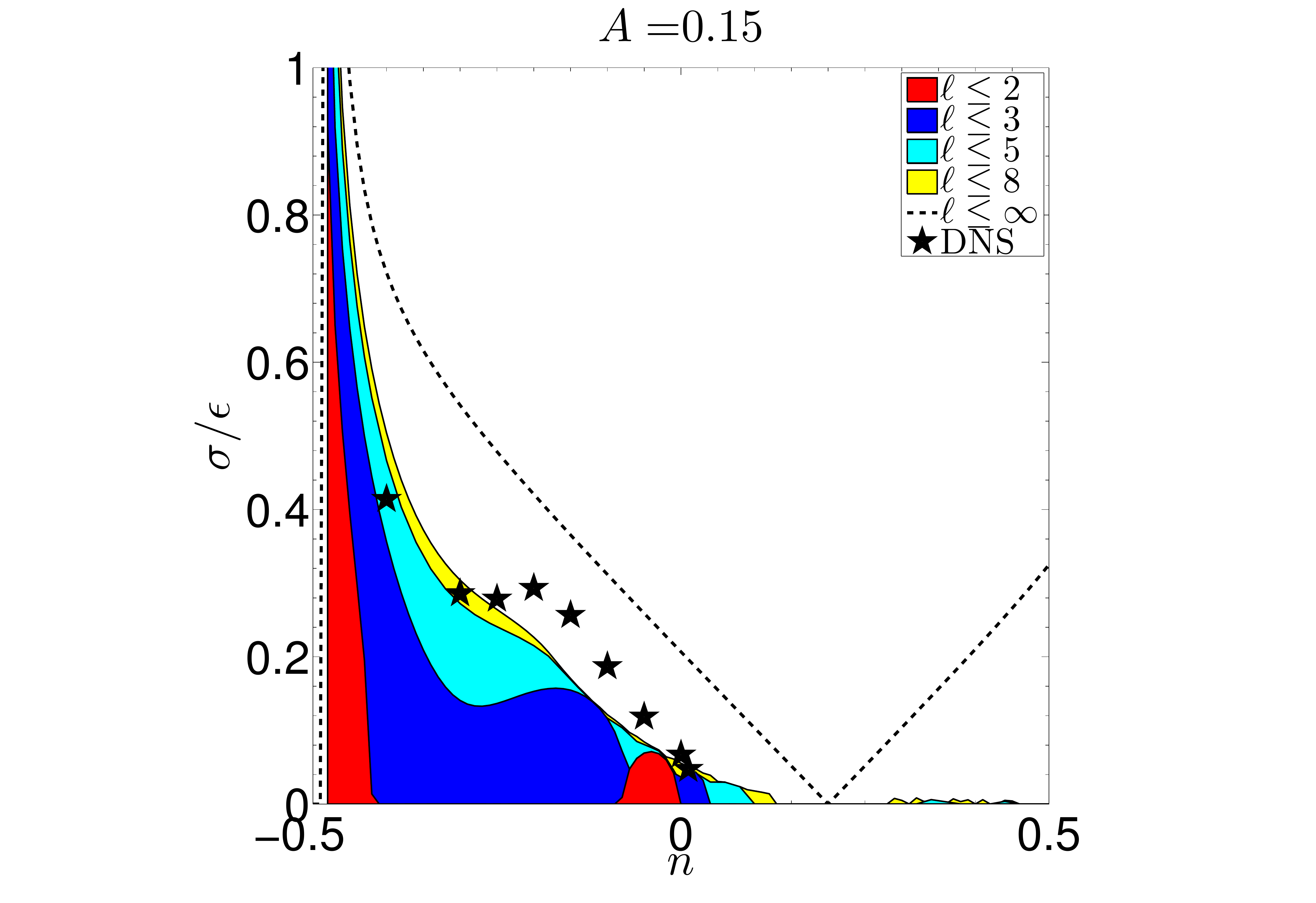} }
       \end{center}
  \caption{Growth rates normalised by $\epsilon$ as a function of $n$ for the elliptical instability for a planet rotating at the rate $\Omega=0.2$ for  $A=0.1$ (top) and $A=0.15$ (bottom) computed using global numerical simulations (DNS; black stars). I also plot predictions for the maximum growth rate from the inviscid global stability analysis \citep{Barker2015a} for modes with harmonic degrees up to a given $\ell$ (coloured shaded regions), as well as the upper bound from energetic considerations (\citealt{LL1996a}; region under the black dashed lines). All simulations were initialised with random noise and included viscosity with $\nu=10^{-4}$. The simulations agree reasonably well with the theoretical predictions. In addition, I confirm the presence of a violent elliptical instability for retrograde spins when $n\lesssim -\Omega=-0.2$ for these values of $A$.}
  \label{8}
\end{figure}

The numerical results for the growth rate (normalised by $\epsilon$) as a function of $n$ are plotted on Fig.~\ref{8} for both $A=0.1$ (top panel) and $A=0.15$ (bottom panel) as black stars. I plot the maximum growth rate based on the energetic upper bound \citep{LL1996a} as the black dashed lines. I also plot the maximum growth rate for all global modes up to a given harmonic degree $\ell$ from the inviscid stability analysis of \cite{Barker2015a} (based on the formalism of \citealt{LL1996}), as the shaded coloured regions\footnote{The predictions for the case with $\ell_{\mathrm{max}}=8$ are computed by assuming a rigid boundary, since the linear properties of the elliptical instability are found to be very similar in this case (and the $U$-basis functions were difficult to compute accurately for $\ell>5$; \citealt{Barker2015a}).}. The simulation results agree reasonably well with the inviscid theoretical predictions, and exhibit the same trends. This agreement is promising, particularly given that the simulations were initialised with random noise (rather than a ``clean" initialisation using the eigenfunction of the fastest growing mode), and they also include non-negligible viscosity. Similar agreement between simulations and theoretical predictions has been found for the related problem of elliptical instability driven by latitudinal libration \citep{Vant2015}.

The growth rate when the spin-over mode is excited ($\ell\leq2$) is represented by the red shaded region for $n<0$, and the regions that represent the excitation of other global inertial modes are illustrated by shading using different colours. The ellipsoidal shape is no longer defined if $n\lesssim -0.5$. This shows that more of the parameter space is unstable to modes with smaller spatial scales (larger $\ell)$. The fastest growth rates occur for retrograde (anti-aligned) spins, with the growth rate for prograde (aligned) cases being much smaller in general. This result is also found in a local plane-wave analysis of the elliptical instability \citep{Craik1989,Barker2015a}. Note that the growth rate for global modes with $\ell\leq 8$ is always smaller (by an $O(1)$ factor) than the energetic upper bound. In Fig.~\ref{8}, I also confirm the prediction\footnote{These simulations were in fact performed before the stability analysis, so this is more of an explanation than a prediction.} of a violent elliptical instability for retrograde spins even when $n\lesssim -\Omega$, when the usual elliptical instability of inertial modes is not normally thought to operate. This occurs only if the tidal amplitude is large enough to allow instability even when a pair of modes is not exactly in resonance. The nonlinear evolution of a simulation in this regime is presented in \S~\ref{violent}.

Now that I have confirmed the presence of elliptical instability in global simulations with growth rates that are consistent with theoretical predictions, I turn to discuss the nonlinear evolution of the elliptical instability. I present several illustrative example simulations in \S~\ref{Nonlinear} and \ref{retrogradecases}.

\section{Nonlinear simulations with a prograde spin}
\label{Nonlinear}

\subsection{An illustrative example: zonal flows as a saturation mechanism}
\label{example}

\begin{figure}
  \begin{center}
    \subfigure{\includegraphics[trim=6cm 0cm 8cm 0cm, clip=true,width=0.23\textwidth]{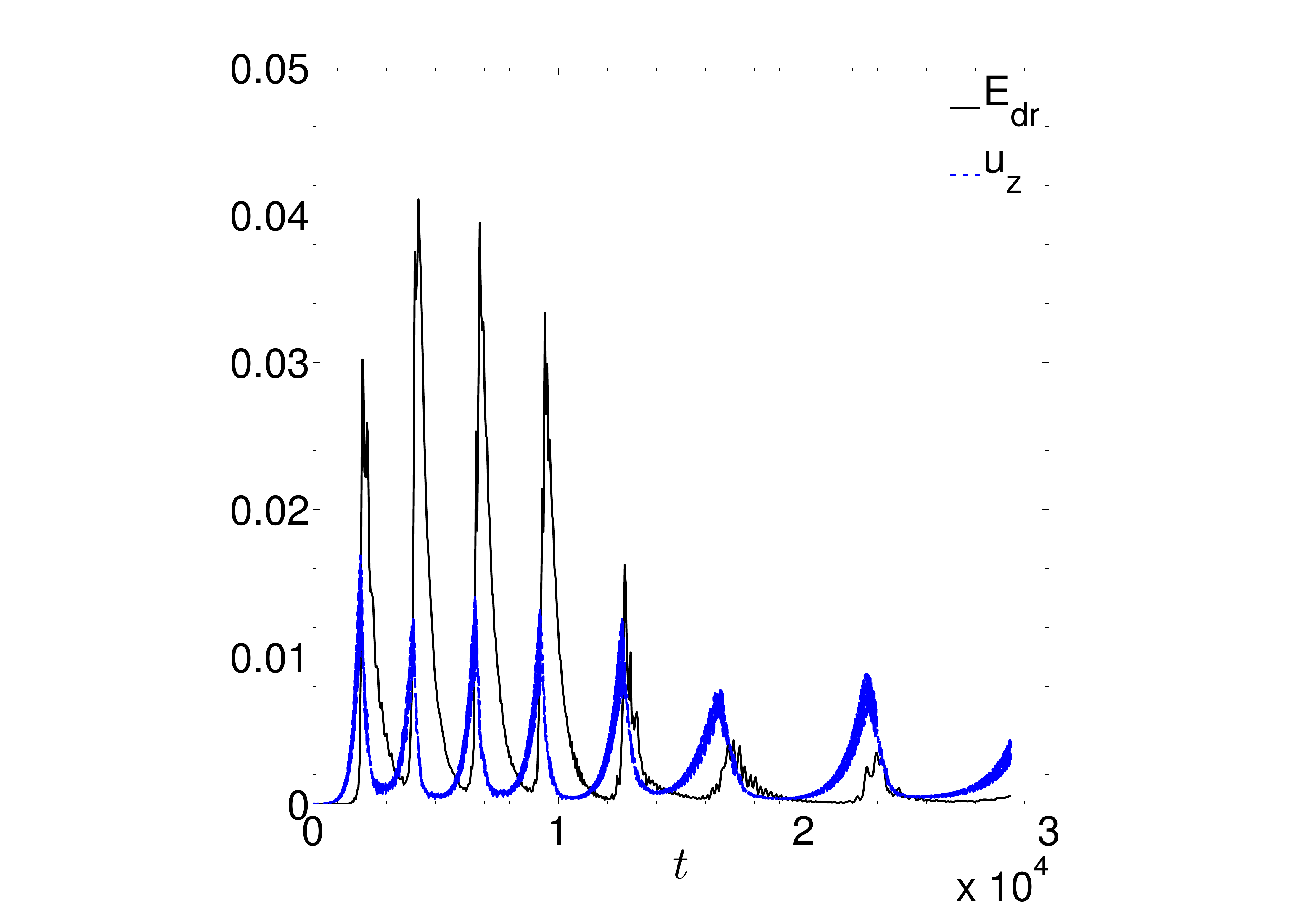} } 
    \subfigure{\includegraphics[trim=6cm 0cm 8cm 0cm, clip=true,width=0.23\textwidth]{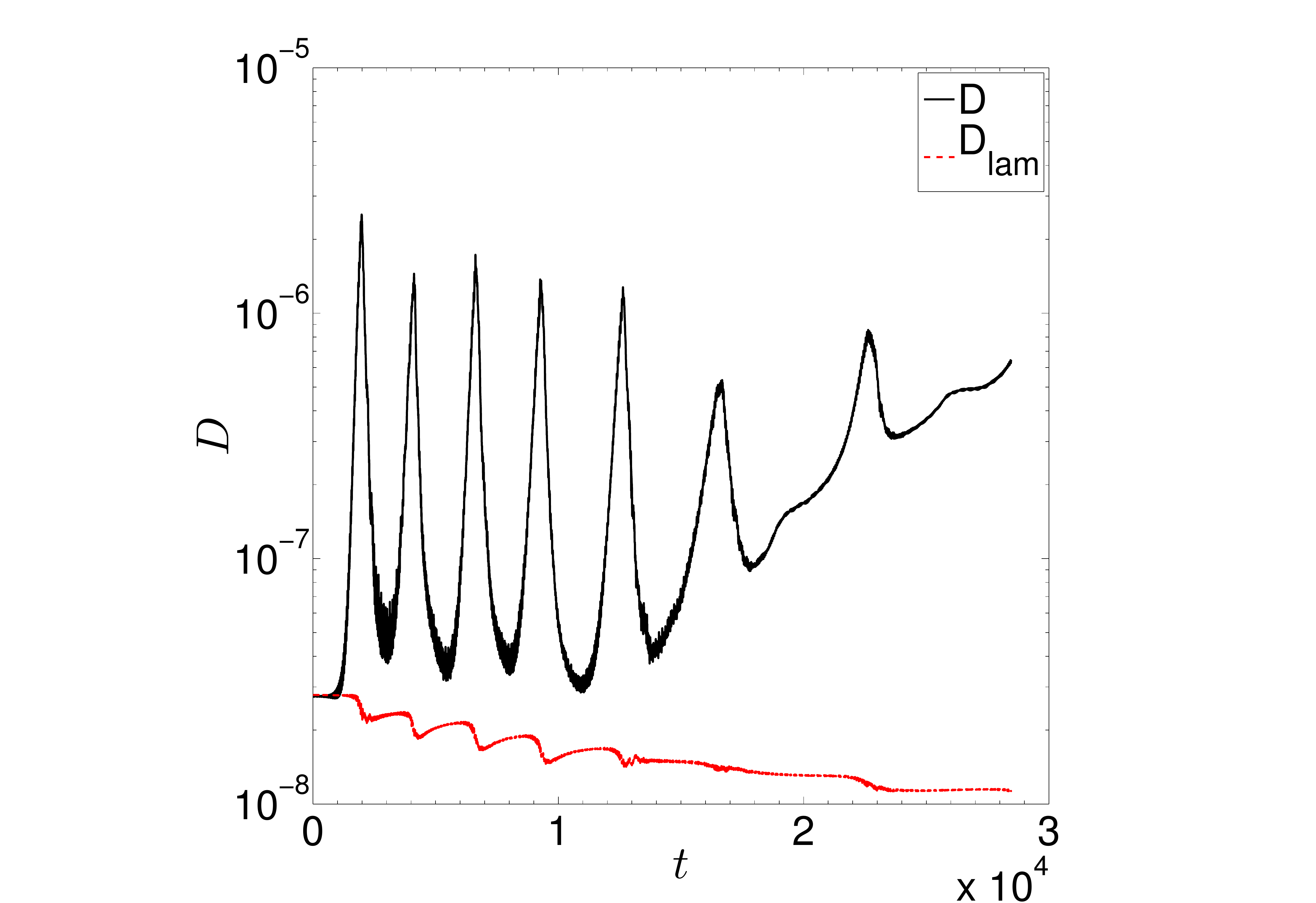} } 
    \subfigure{\includegraphics[trim=6cm 0cm 8cm 0cm, clip=true,width=0.23\textwidth]{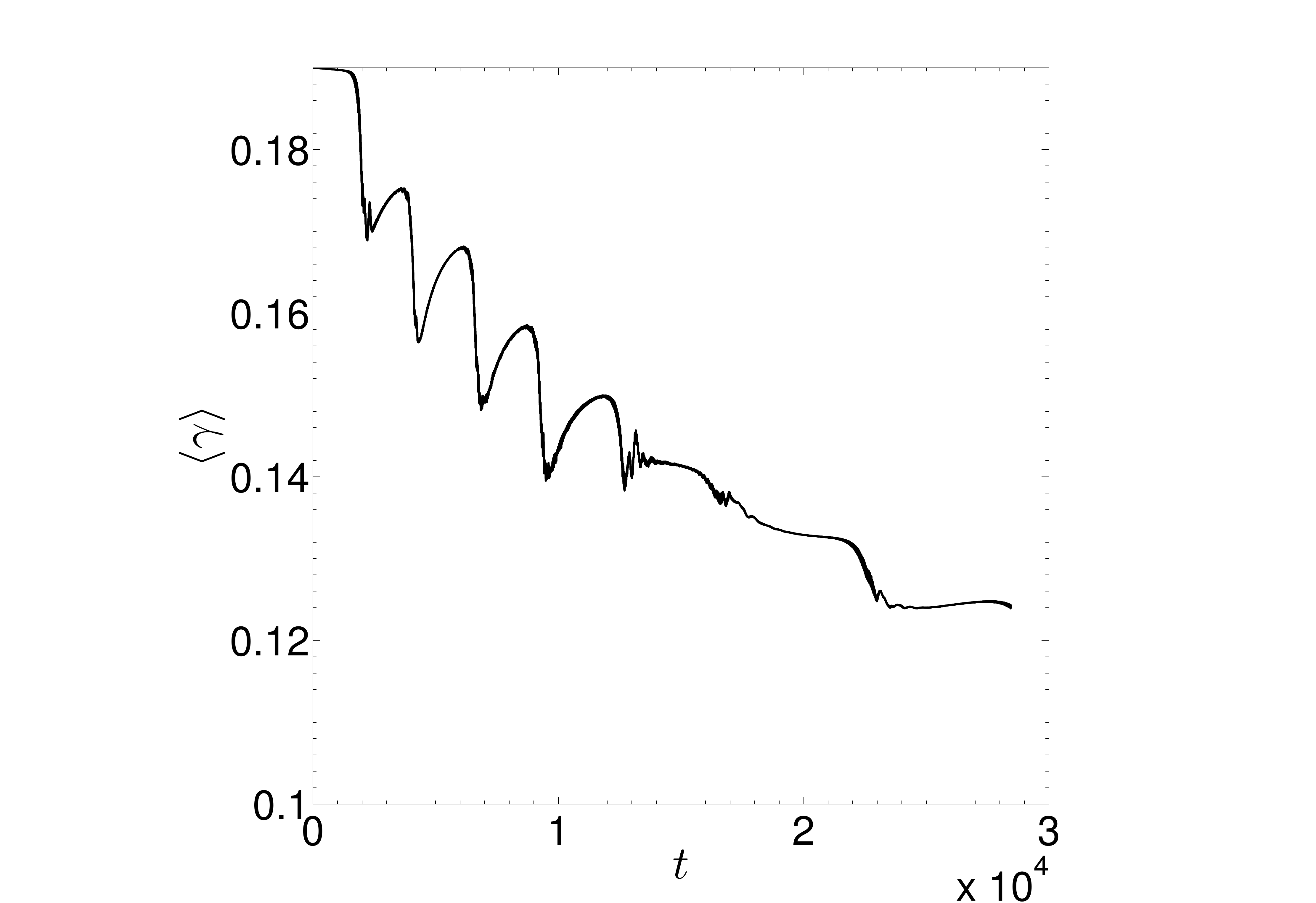} } 
    \subfigure{\includegraphics[trim=5.5cm 0cm 8cm 0cm, clip=true,width=0.23\textwidth]{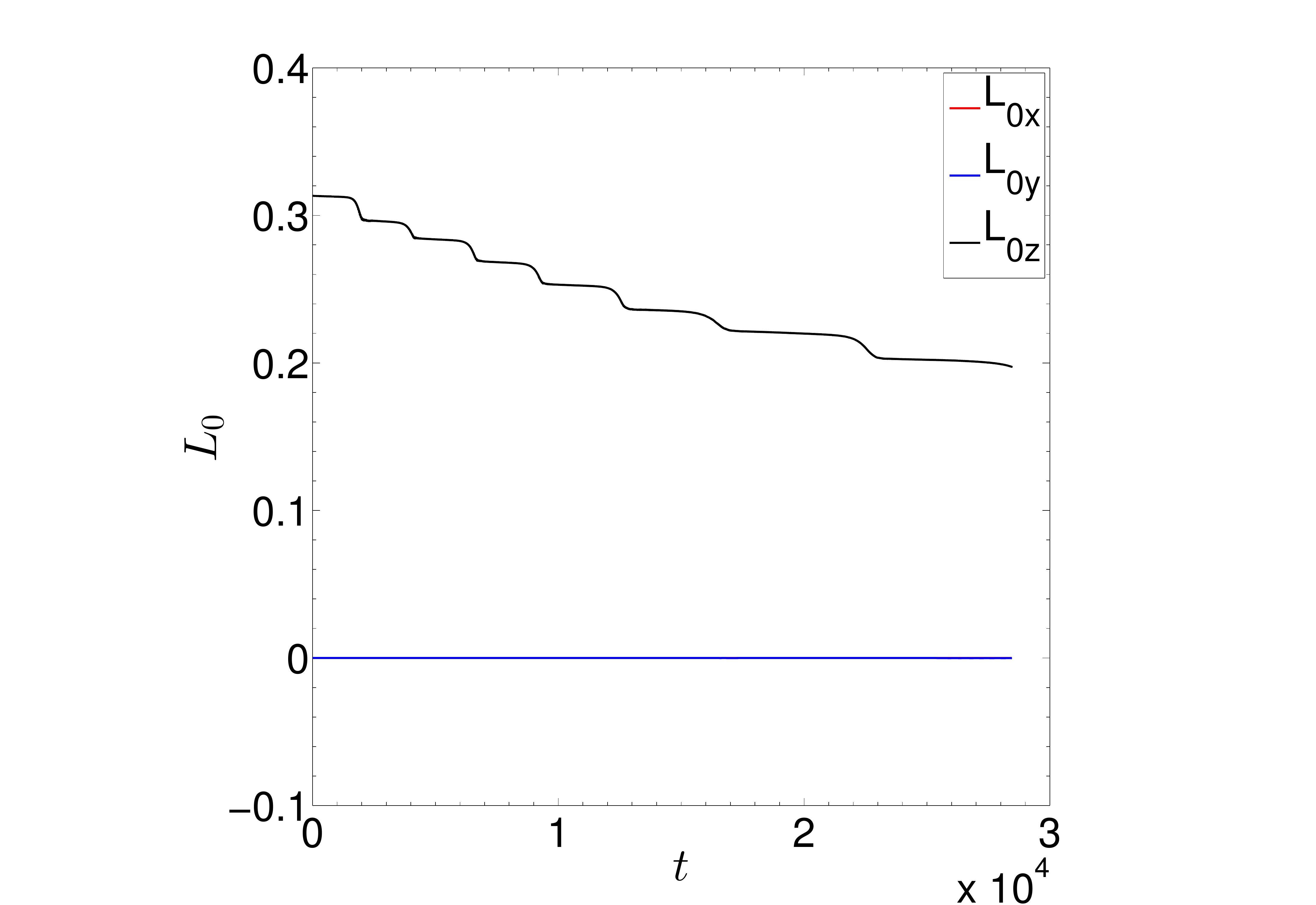} } 
       \end{center}
  \caption{Evolution of various flow quantities with time for a simulation with $\Omega=0.2,n=0.01,A=0.05$ and $\nu=3\times 10^{-5}$. Top left: comparison of RMS $u_z$ with the energy in the differential rotation, $E_\mathrm{dr}$. Top right: viscous dissipation rate (black line) and laminar viscous dissipation rate prediction (red line). Bottom left: mean asynchronism of the flow $\langle\gamma\rangle$. Bottom right: Cartesian components of the angular momentum of the fluid in the inertial frame. This figure illustrates that the elliptical instability can lead to enhanced tidal dissipation. It also shows the importance of differential rotation (zonal flows) as a saturation mechanism for the elliptical instability.}
  \label{2}
\end{figure}

I now describe the results of an example simulation, in which the planet initially rotates in a prograde sense with $\Omega=0.2$ and orbital angular frequency $n=0.01$, with $A=0.05$ and $\nu=3 \times 10^{-5}$ (computed using a resolution of $N=10$). This configuration has an initial shape with $\epsilon=0.08$, $a=1.040$, $b=0.959$, $c=0.941$. Since the initial tidal flow (Eq.~\ref{basicflow}) does not satisfy the stress-free condition at the surface, there is a weak viscously driven circulation in the fluid interior (leading to RMS vertical velocities $\langle u_z\rangle \sim 10^{-5}$), and viscous dissipation that leads to a gradual synchronisation of the spin and orbit. The predicted mean viscous dissipation rate at $t=0$ (Eq.~\ref{disspred}) is well matched by the simulation result ($D=2.2\times 10^{-6}$) shown in the top right panel of Fig.~\ref{2} by the agreement of the black (simulated) and red (predicted) lines during this stage.

The time evolution of various mean quantities is plotted in Fig.~\ref{2}. The initial elliptical instability grows to large amplitudes by $t\sim 1500$, by which time it enhances the dissipation, producing rapid partial synchronisation of the spin and orbit from $\gamma\approx0.19$ to $\gamma\approx 0.17$. After this initial burst, the turbulence temporarily dies away only to recur in a cyclic manner. Each burst of instability corresponds with a period of enhanced dissipation, which is shown by the second panel of Fig.~\ref{2}, where I plot
\begin{eqnarray}
D=\frac{2\nu}{V}\int_V e_{ij} e_{ij}  \, \mathrm{d}V.
\end{eqnarray}
In the third panel, I plot 
\begin{eqnarray}
\langle \gamma(t) \rangle=\frac{1}{V}\int_V \frac{u_\phi}{R} \mathrm{d}\, V,
\end{eqnarray}
the evolving mean asynchronism of the flow. During the burst phases, the spin rapidly undergoes a partial synchronisation with the orbit, which is repeated during subsequent bursts. The tidal synchronisation does not occur smoothly, instead occurring in an erratic manner, dominated by these short-lived bursts. Outside of these turbulent bursts, dissipation appears to be due to viscous dissipation of the differential rotation in the flow, and not purely laminar viscous dissipation of the global tidal flow i.e.~Eq.~\ref{disspred} (shown as the red dashed line, where I have replaced $\gamma\rightarrow \langle \gamma(t) \rangle$).

\begin{figure}
  \begin{center}
    \subfigure{\includegraphics[trim=6cm 0cm 5cm 0cm, clip=true,width=0.23\textwidth]{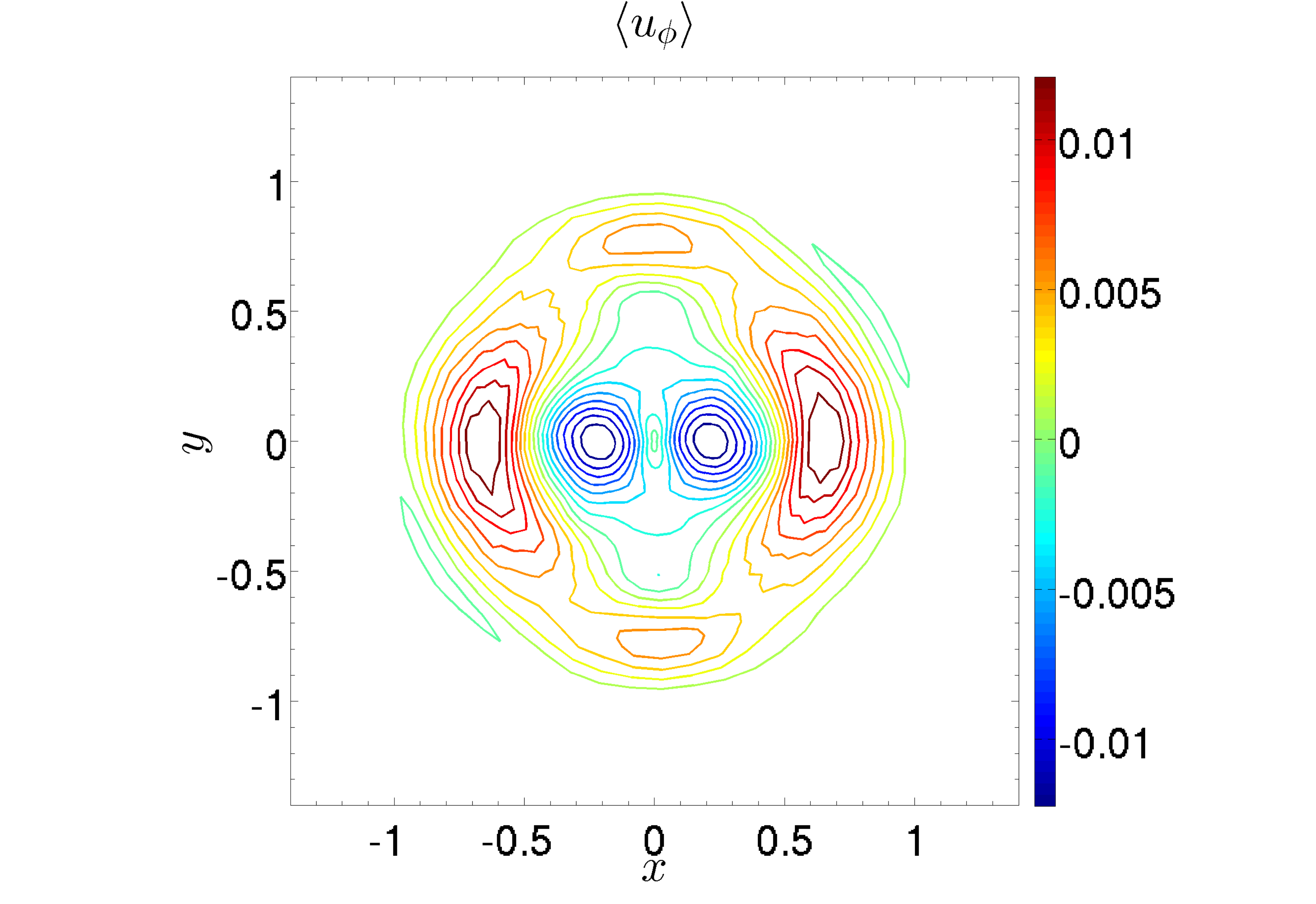} } 
    \subfigure{\includegraphics[trim=6cm 0cm 5cm 0cm, clip=true,width=0.23\textwidth]{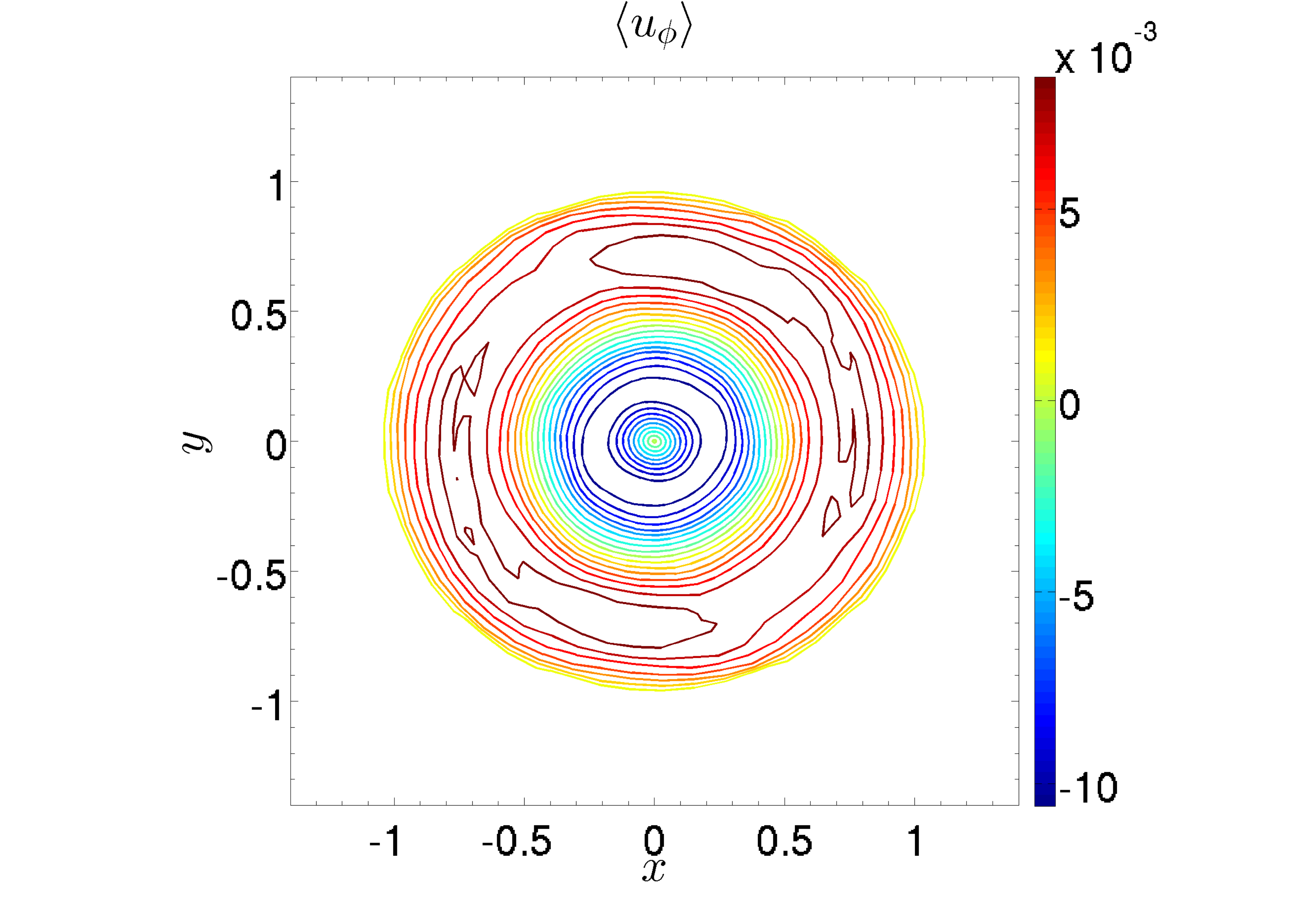} } 
    \subfigure{\includegraphics[trim=4cm 0cm 7cm 0cm, clip=true,width=0.23\textwidth]{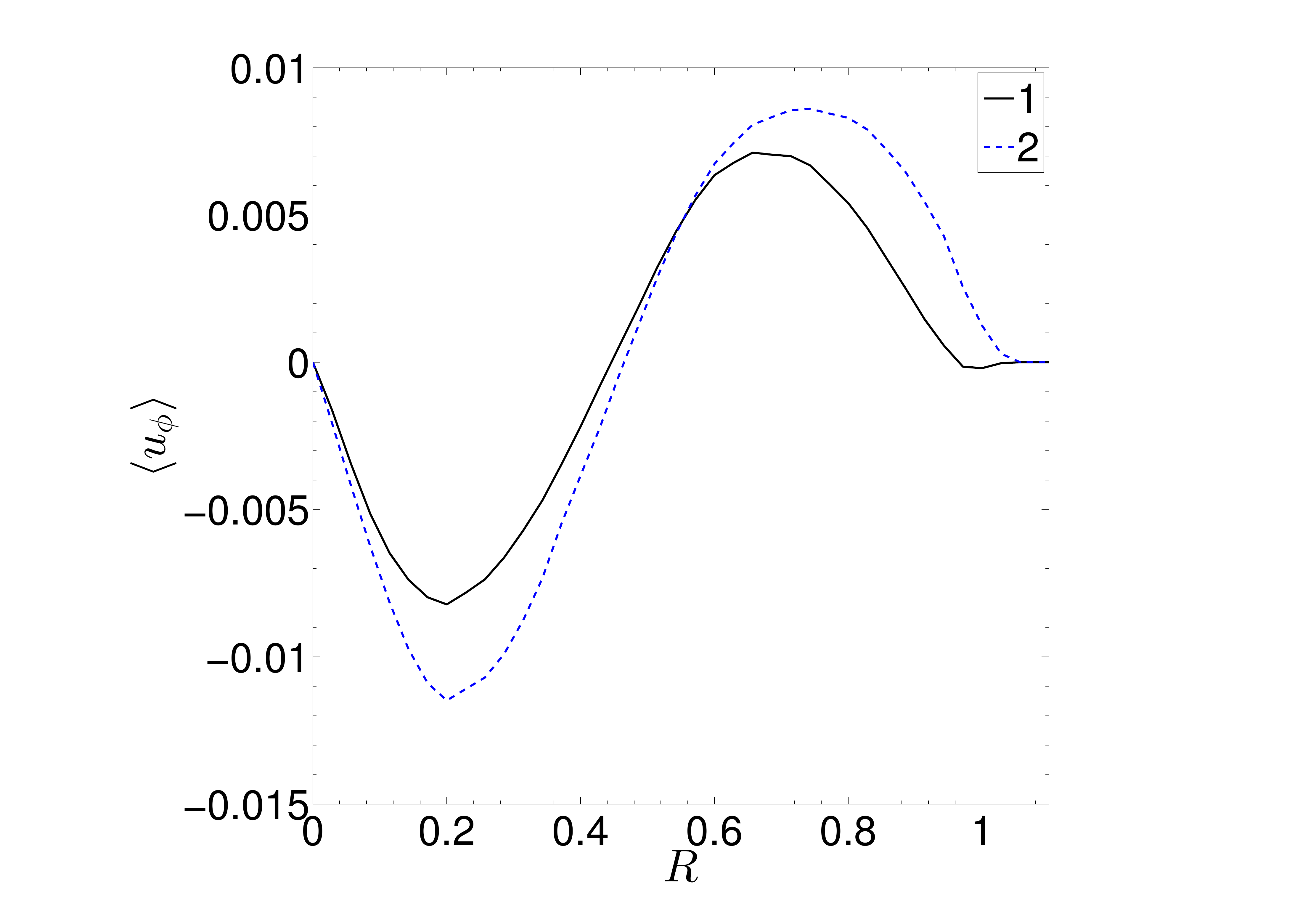} } 
    \subfigure{\includegraphics[trim=6cm 0cm 7cm 0cm, clip=true,width=0.23\textwidth]{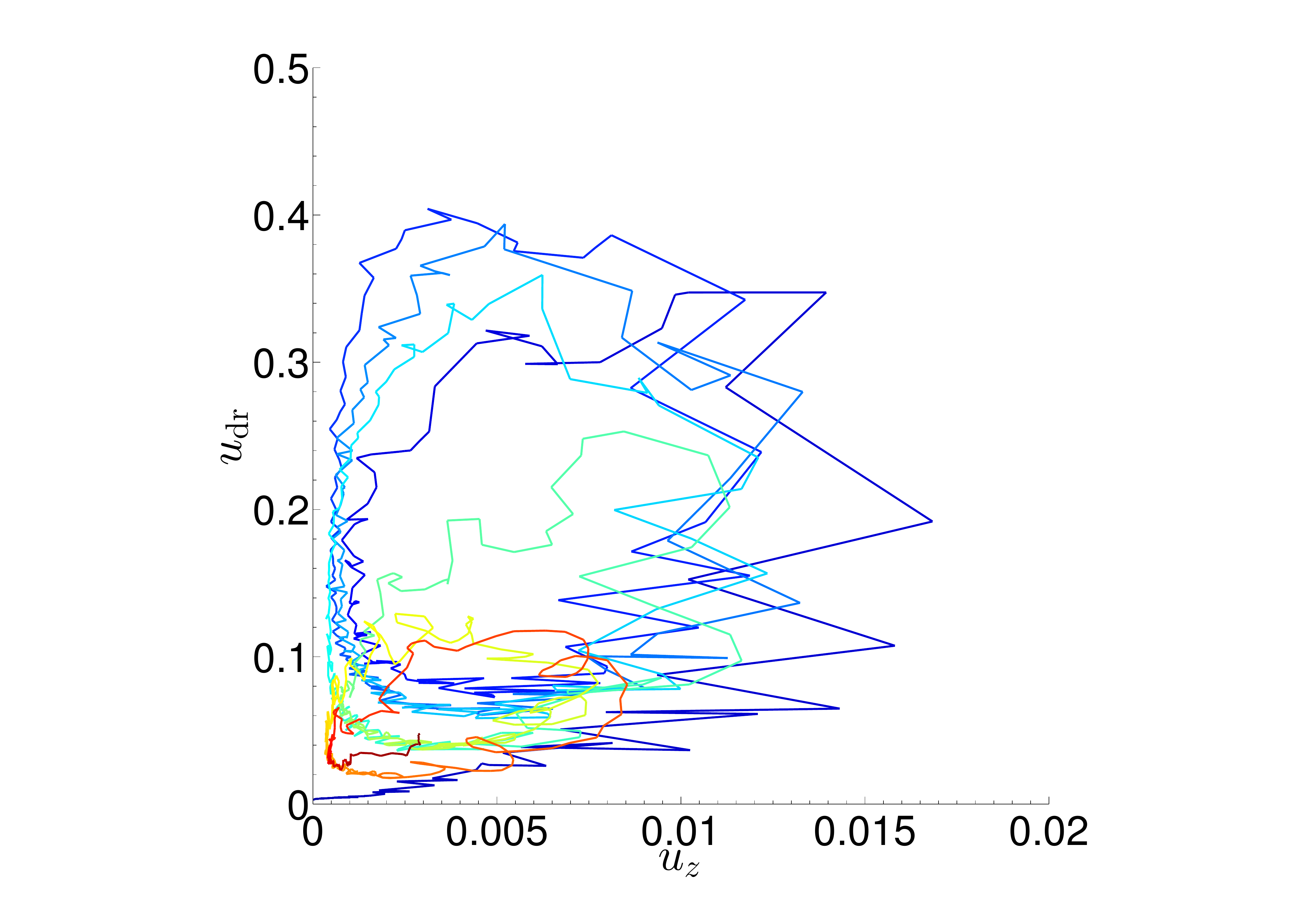} } 
       \end{center}
  \caption{Illustration of zonal flows produced in a simulation with $\Omega=0.2,n=0.01,A=0.05$ and $\nu=3\times 10^{-5}$. Top left: Vertically-averaged azimuthal velocity $\langle u_\phi^{\prime} \rangle_z$ on the $xy$-plane during the initial burst phase. Top right: same during the second burst phase. Bottom left: comparison of the mean zonal flow $\langle u_\phi^{\prime} \rangle_{\phi,z}$ as a function of cylindrical radius $R$ during the first and second burst phases. Bottom right: phase plane plot of $\langle u_z\rangle$ vs $u_\mathrm{dr}$, with the colour representing time (from blue to green to red). This behaviour is reminiscent of predator-prey dynamics (with waves acting as the prey and zonal flows as the predators).}
  \label{3}
\end{figure}

In the bottom right panel of Fig.~\ref{2}, I plot the components of the angular momentum in the inertial frame ($\boldsymbol{L}_0$), which is related to that in the bulge frame ($\boldsymbol{L}_n$) by $\boldsymbol{L}_0=\boldsymbol{L}_n+I \boldsymbol{n}$,
where $I=\frac{8}{15}\pi a b c$ is the moment of inertia of a rigid ellipsoid (for rotations about $z$), and 
\begin{eqnarray}
\boldsymbol{L}_n=\int_V \boldsymbol{x}\times \boldsymbol{u} \,\mathrm{d}V.
\end{eqnarray}
This panel shows that the spin remains aligned with the orbit during the tidal synchronisation process, as we might expect (by the absence of appreciable growth in $L_x$ or $L_y$). 

In order to analyse the cyclic behaviour, I define the differential rotation as follows: firstly, the perturbed velocity is
\begin{eqnarray}
\boldsymbol{u}^{\prime}=\boldsymbol{u}-\boldsymbol{U}(x,y,t),
\end{eqnarray}
where $\boldsymbol{U}(x,y,t)=\langle\gamma (t)\rangle(-\frac{ya}{b},\frac{xb}{a},0)$.
For simplicity, I use the initial geometry when computing the differential rotation, but for all cases considered the results below differ negligibly if I instead use the instantaneous shape of the ellipsoid. I interpolate results from the non-uniformly spaced grid points in Nek5000 to a uniform Cartesian grid containing the entire body (consisting of $50^3$ points), and the vertically-averaged azimuthal perturbation in the $xy$-plane (using cylindrical polar coordinates) is defined by (motivated by similar calculations in \citealt{Favier2015})
\begin{eqnarray}
\langle u^{\prime}_\phi (R,\phi,t) \rangle_z &=&\frac{1}{N_z} \sum_z u^{\prime}_\phi(R,\phi,z,t),
\end{eqnarray}
where $N_z$ is the number of points in $z$ for a given $R$ and $\phi$. The energy in the differential rotation is
\begin{eqnarray}
E_{\mathrm{dr}}(t)&=&\frac{1}{V}\int_V \frac{1}{2}\left[u^{\prime}_\phi (R,\phi,z,t)\right]^2 \, \mathrm{d}V.
\end{eqnarray}

In the top left panel of Fig.~\ref{2}, $E_{\mathrm{dr}}$ and $\langle u_z\rangle$ are plotted as a function of time, where $\langle u_z\rangle$ is a measure of the (square root of the) energy contained in the waves\footnote{I have, somewhat unusually, compared $E_\mathrm{dr}$ with the RMS $u_z$ in this case, because the former is typically much larger than $\frac{1}{2}\langle u_z\rangle^2$ by a factor of approximately $10^{3}$. In addition, this comparison highlights more clearly the time delay between these two quantities than if I had plotted energies on a log-scale.} (in addition to the weak viscously-driven flows). $E_{\mathrm{dr}}$ begins to grow shortly after the initial instability has set in. Once the energy in the differential rotation has increased to a sufficient level, the energy in the waves subsequently decays. Only when the differential rotation has decayed sufficiently due to viscosity\footnote{The viscous timescale is $\sim L^2/\nu\sim 5000$ if $L\sim 0.4$, which is somewhat larger than the observed decay, but matches it to within an $O(1)$ factor.} can the instability grow once more. This leads to cyclic behaviour\footnote{The cyclic behaviour shown in Fig.~\ref{2} is somewhat similar to predator-prey dynamics, in which the waves can be thought of as ``rabbits" and the zonal flows as ``foxes" (e.g.~\citealt{Murray2002}). However, in this case, the energy source feeding the instability dies out as the spin synchronises with the orbit, so we do not observe strictly periodic behaviour because the ``food source runs out" (in addition to more complicated nonlinearities).} during which the instability grows, transfers energy into differential rotation, which then inhibits further growth until the differential rotation is sufficiently damped by viscosity. Differential rotation, in the form of zonal flows, therefore plays an important role in the saturation of the instability. 

I further illustrate this cyclic behaviour in the bottom right panel of Fig.~\ref{3}, where I plot the ``phase plane" $\langle u_z\rangle$ against $u_\mathrm{dr}=\sqrt{2 E_\mathrm{dr}}$ to show the appearance of cyclic behaviour with evolving cycle amplitudes (the colour denotes time, which increases from blue to red). Analogous cyclic behaviour occurs in local hydrodynamical simulations of the elliptical instability, where columnar vortices play the role of zonal flows \citep{BL2013}. It interesting to note that cyclic behaviour has also been observed in the nonlinear evolution of the ``r-mode" instability in neutron stars, where inertial modes are instead driven by gravitational radiation, even in integrations of three-mode couplings that neglect zonal flows \citep{Brink2005,Bondarescu2009}.

\begin{figure}
  \begin{center}
      \subfigure{\includegraphics[trim=0cm 0cm 0cm 0cm, clip=true,width=0.39\textwidth]{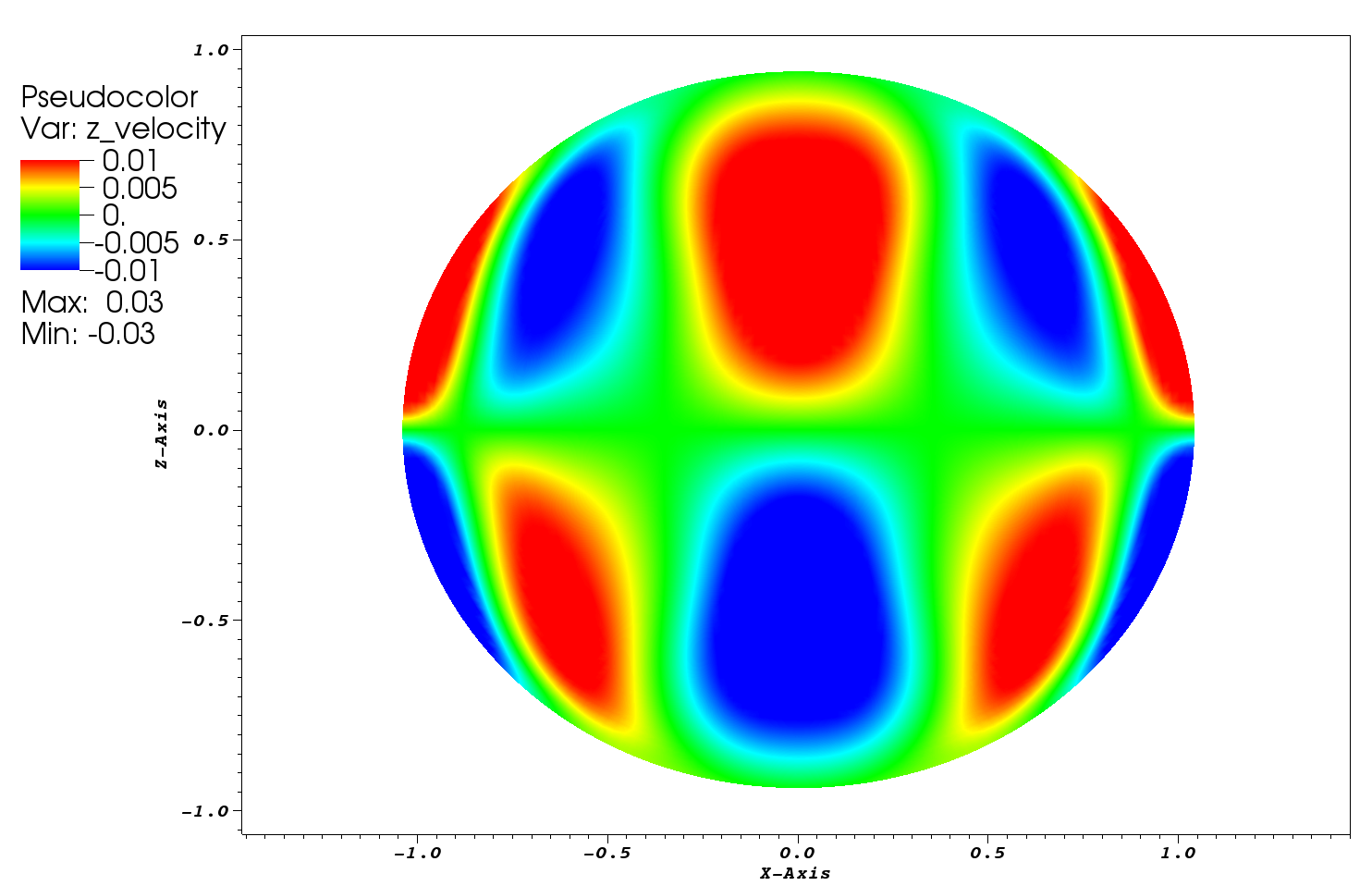} } 
      \subfigure{\includegraphics[trim=0cm 0cm 0cm 0cm, clip=true,width=0.39\textwidth]{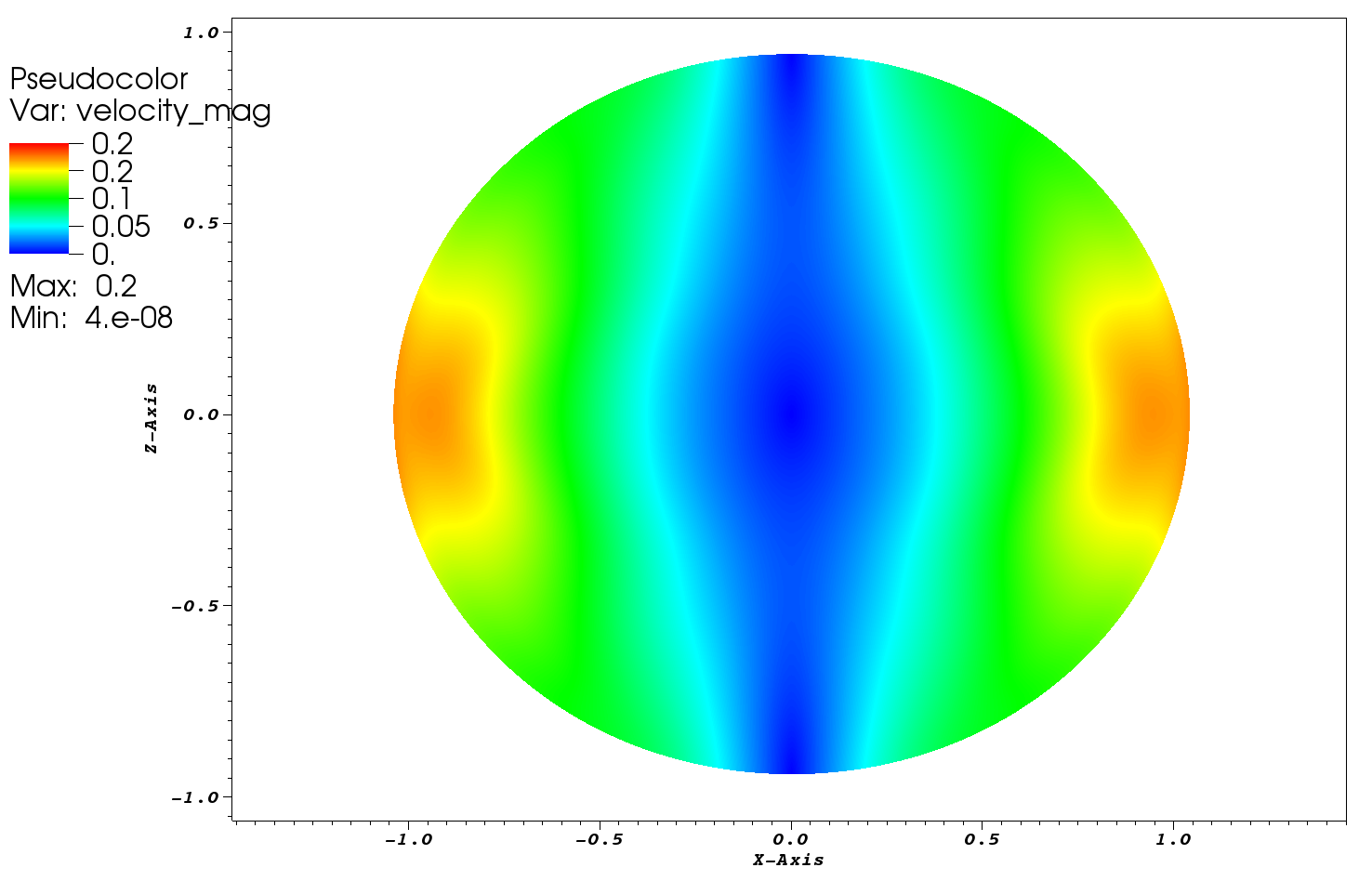} } 
      \subfigure{\includegraphics[trim=0cm 0cm 0cm 0cm, clip=true,width=0.39\textwidth]{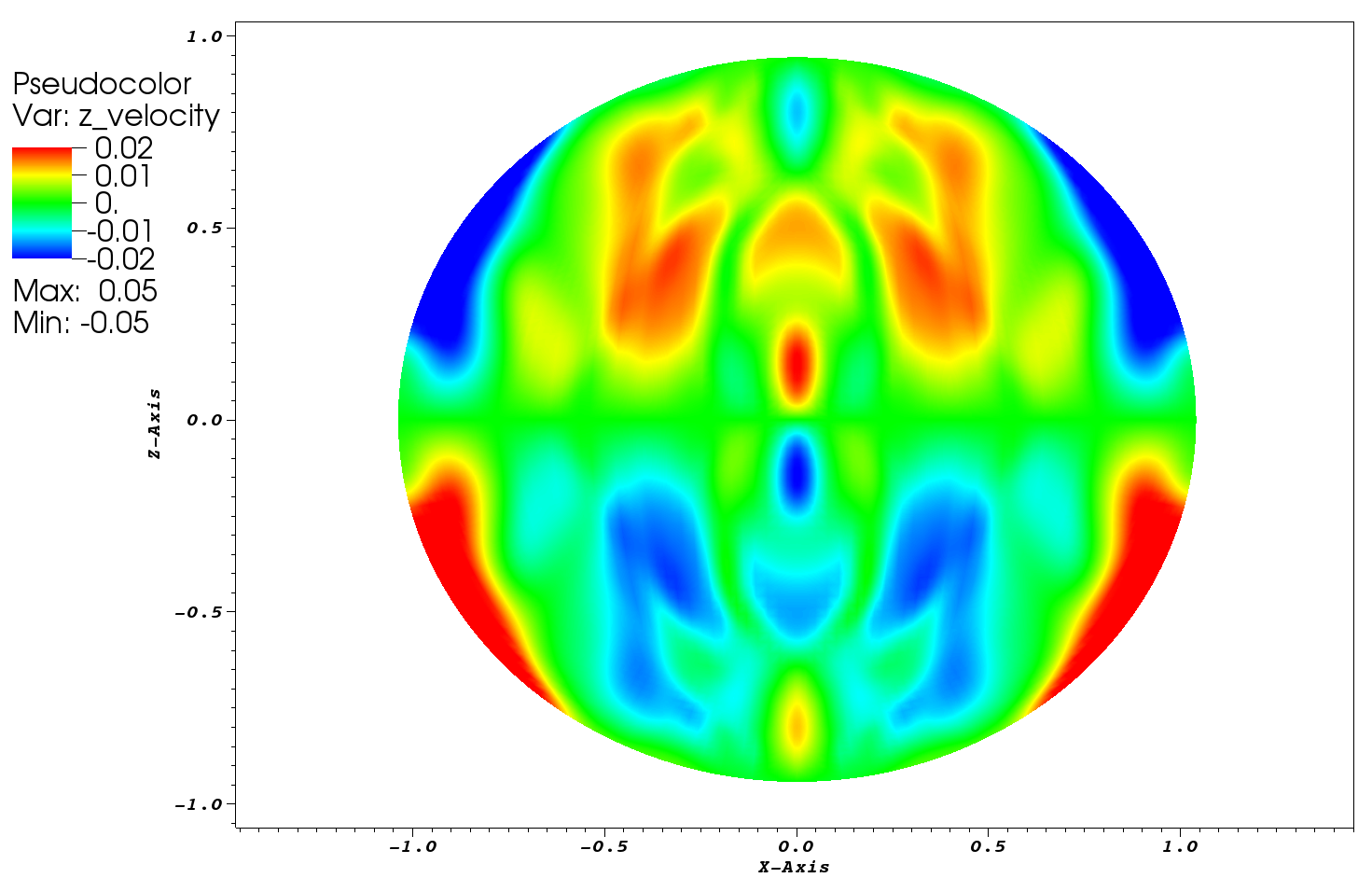} } 
      \subfigure{\includegraphics[trim=0cm 0cm 0cm 0cm, clip=true,width=0.39\textwidth]{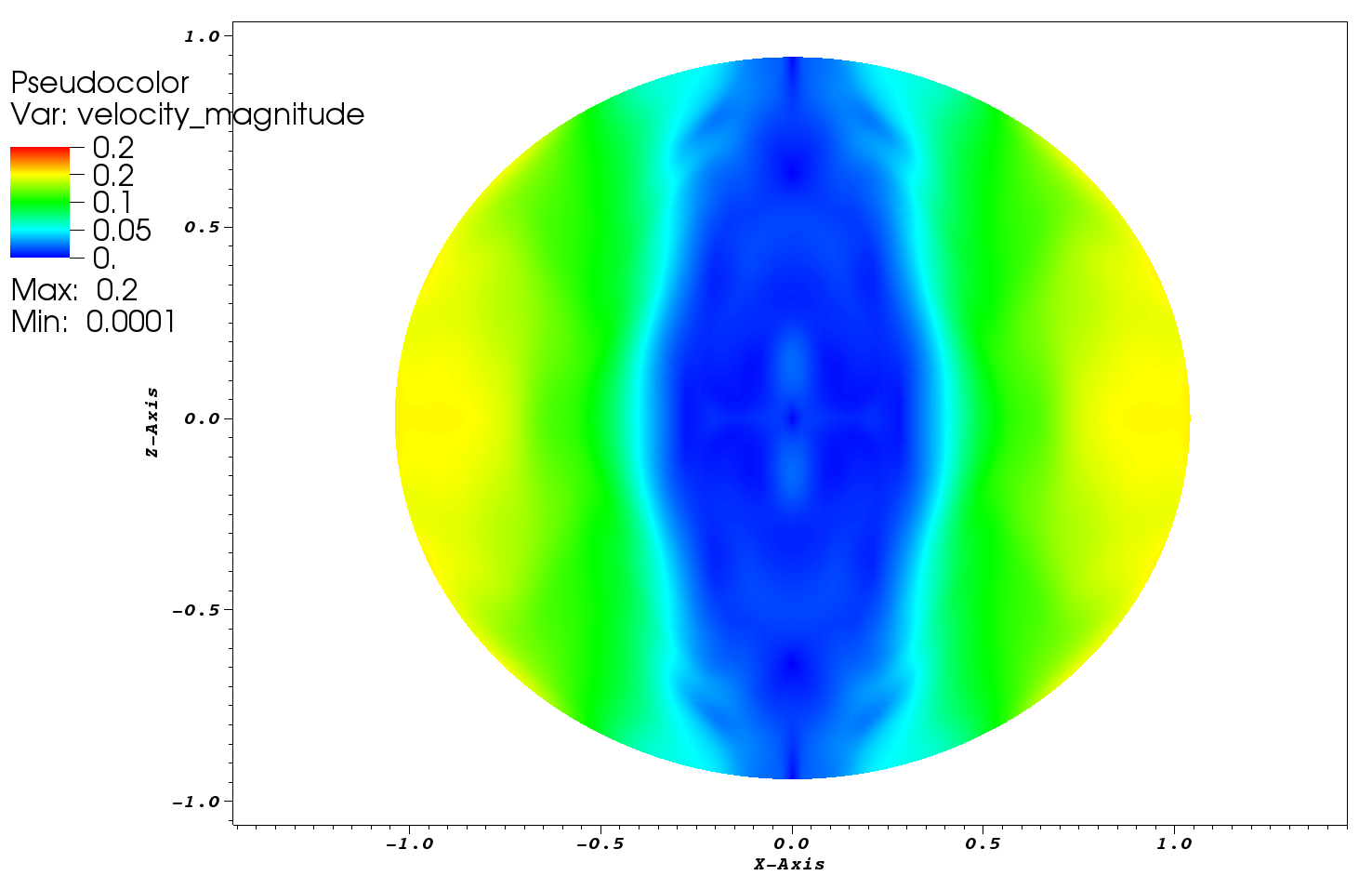} } 
       \end{center}
  \caption{Illustration of the flow in the simulation with $\Omega=0.2,n=0.01,A=0.05$ and $\nu=3\times 10^{-5}$. Top and top middle: instantaneous vertical velocity and $|\boldsymbol{u}|$ during initial linear growth phase at $t=1700$. Bottom middle and bottom: same during initial ``turbulent" phase at $t=1999.7$.}
  \label{3a}
\end{figure}

The vertical fluid velocity and $|\boldsymbol{u}|$ in the $xz$-plane are shown in Fig.~\ref{3a} during the first linear growth phase at $t=1700$ (top two panels), and during the subsequent ``turbulent" burst phase at $t=1999.7$ (bottom two panels). I also illustrate the vertically and temporally averaged azimuthal velocity in the $xy$-plane, during the first two burst phases (for the differential rotation) in the top two panels in Fig.~\ref{3}. In the bottom left panel of Fig.~\ref{3}, I have plotted the azimuthal velocity as a function of cylindrical radius at both of these times, illustrating the radial structure of the zonal flows. The zonal flow reaches velocity amplitudes up to approximately $4\%$ of the fluid rotation during this phase.

The bottom left panel of Fig.~\ref{2} also shows that there are transient periods of tidal desynchronisation (though the net evolution is towards synchronism). This occurs between each turbulent burst, and is caused by viscous and nonlinear damping of the zonal flows, whose net angular momentum is transferred back to the mean rotation of the fluid. Ultimately, the energy source driving the zonal flows comes from the mean asynchronism of the flow, so these transient phases of desynchronisation are not sustained. Nevertheless, this points out the possibility of periods of tidal desynchronisation, even in a system that lacks an additional energy source (cf.~\citealt{OgilvieLesur2012}).

The generation of zonal flows occurs in many rotating fluids (e.g.~\citealt{FBBO2014}), but this is the first example in which it has been observed in global simulations with a free surface. In this example, the flow driven by the elliptical instability was only weakly turbulent because of the relatively large viscosity in relation to the tidal amplitude considered. In the next subsection, I briefly examine two further simulations with a prograde spin, but in which the instability is driven more strongly. In these cases, zonal flows continue to play a role, but the turbulence is less bursty.

\subsection{Two further prograde examples}

\begin{figure}
  \begin{center}
    \subfigure{\includegraphics[trim=5cm 0cm 7cm 1cm, clip=true,width=0.23\textwidth]{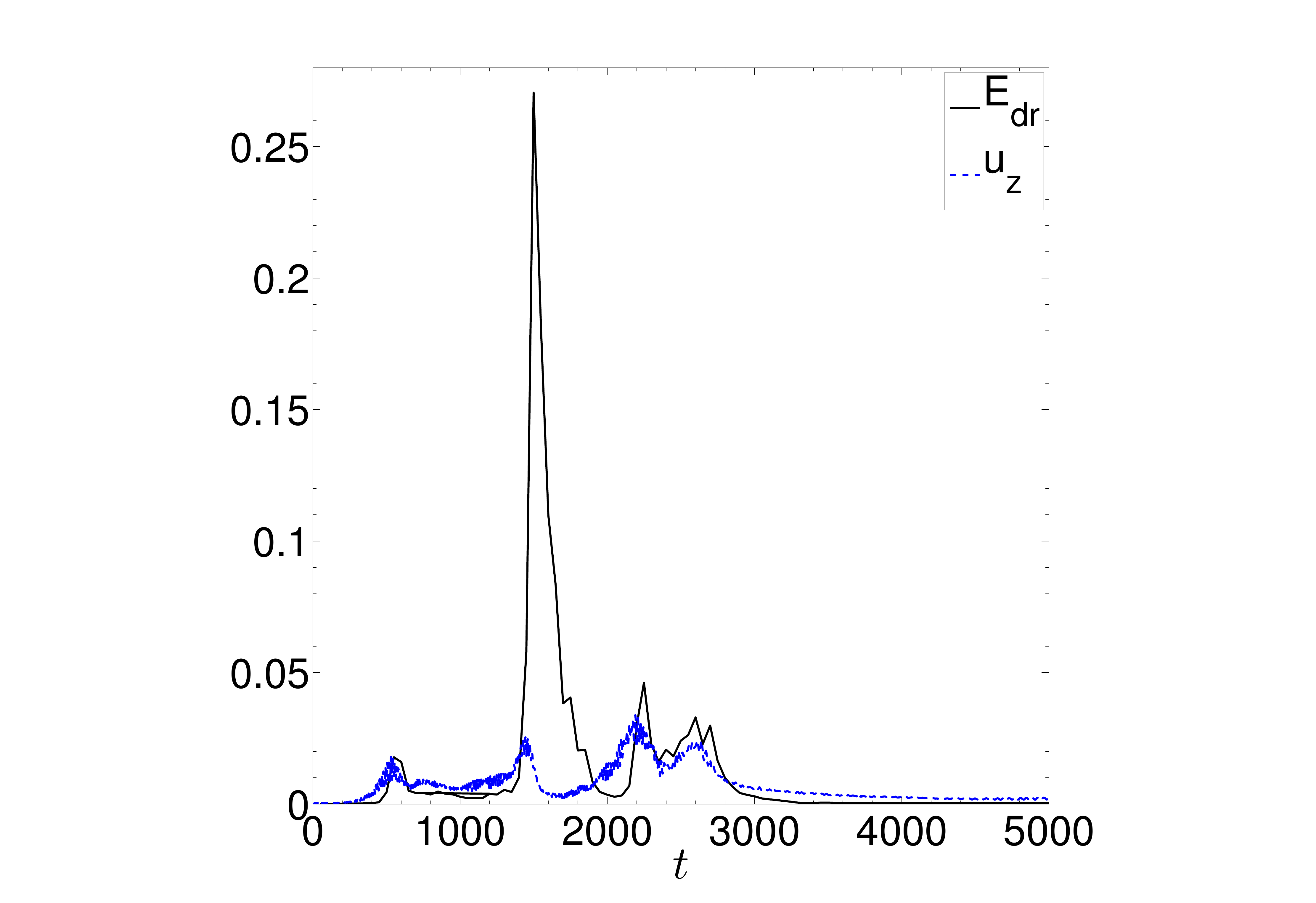} } 
    \subfigure{\includegraphics[trim=5cm 0cm 7cm 1cm, clip=true,width=0.23\textwidth]{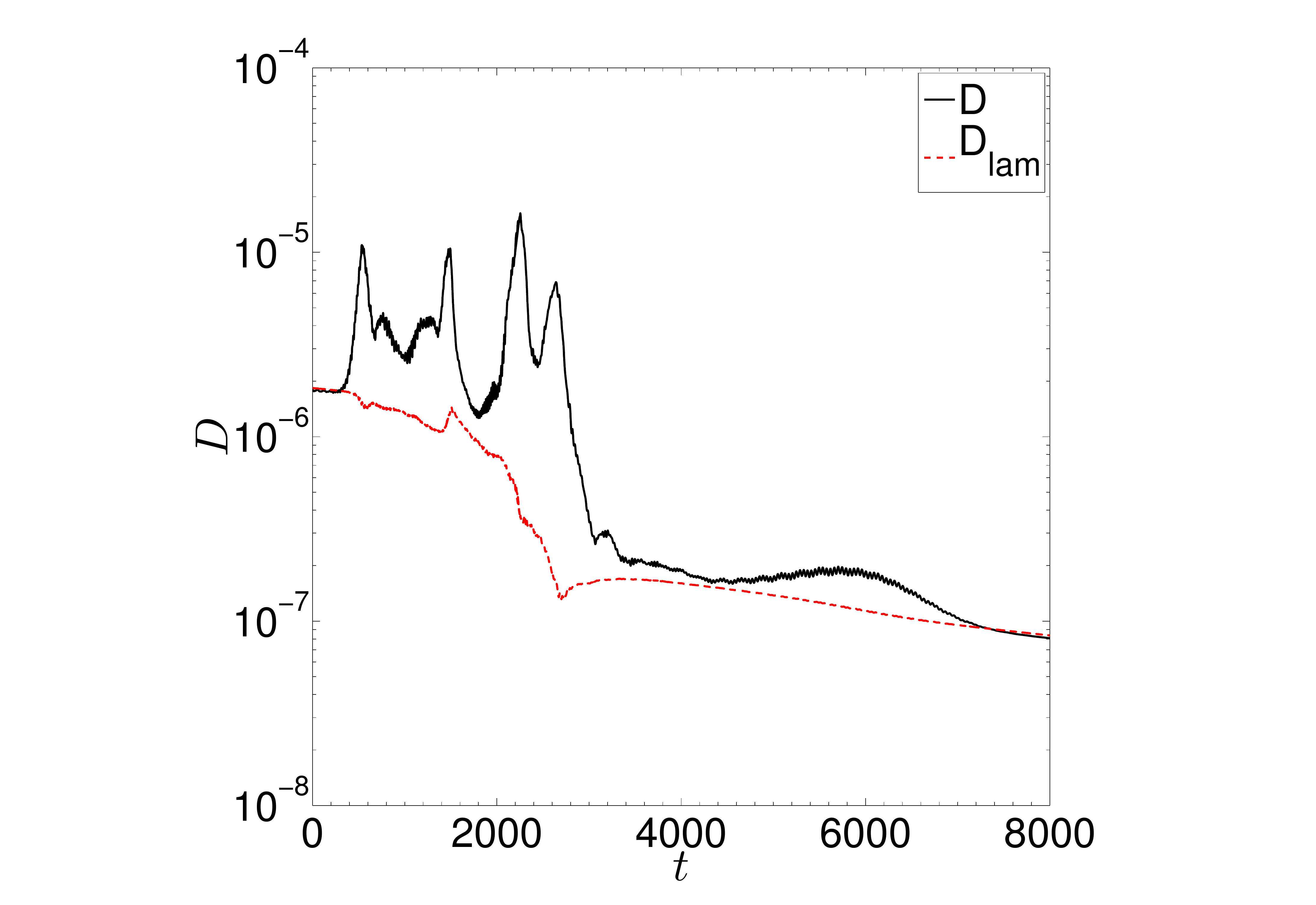} } 
    \subfigure{\includegraphics[trim=5cm 0cm 7cm 1cm, clip=true,width=0.23\textwidth]{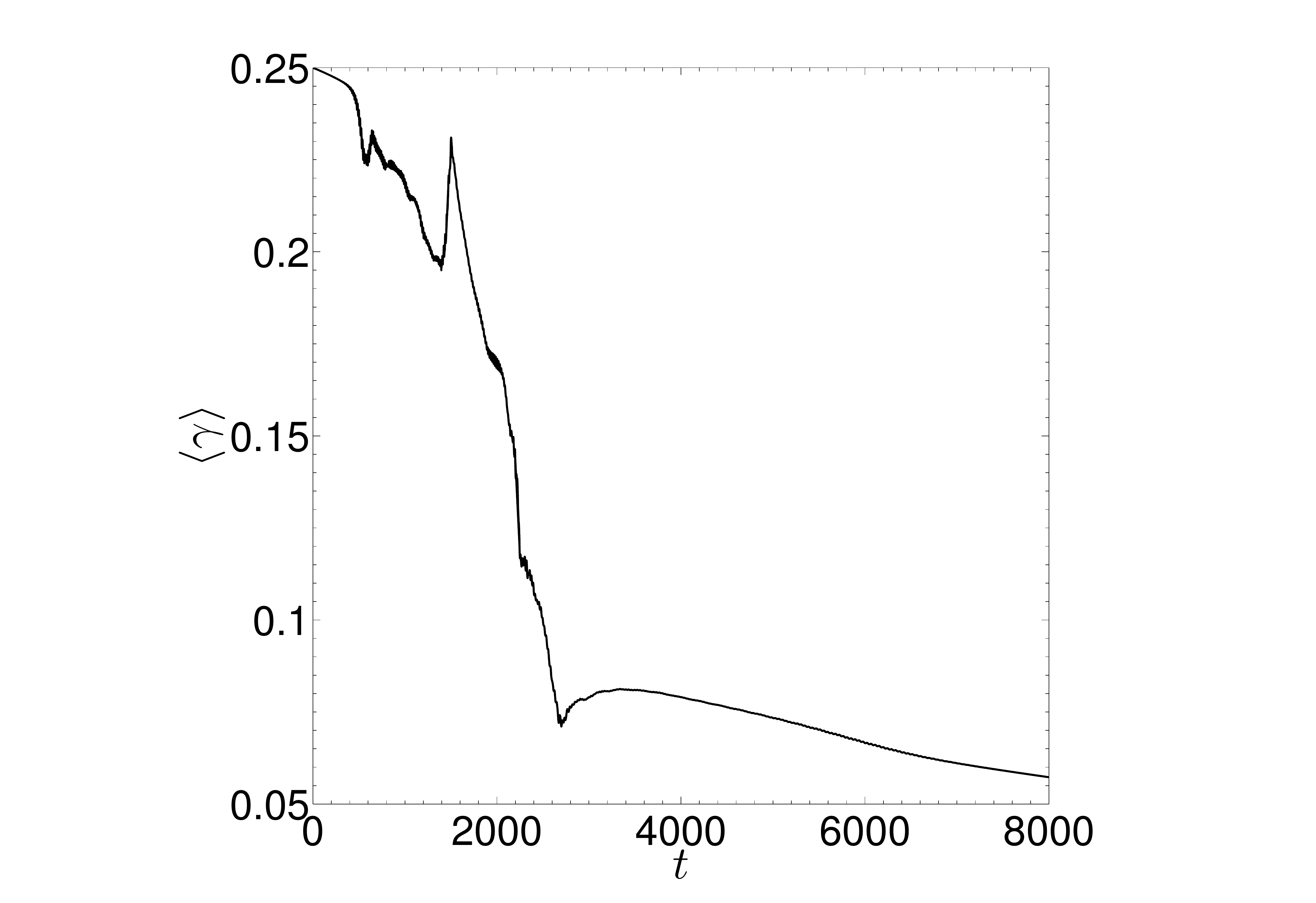} } 
    \subfigure{\includegraphics[trim=5cm 0cm 7cm 1cm, clip=true,width=0.23\textwidth]{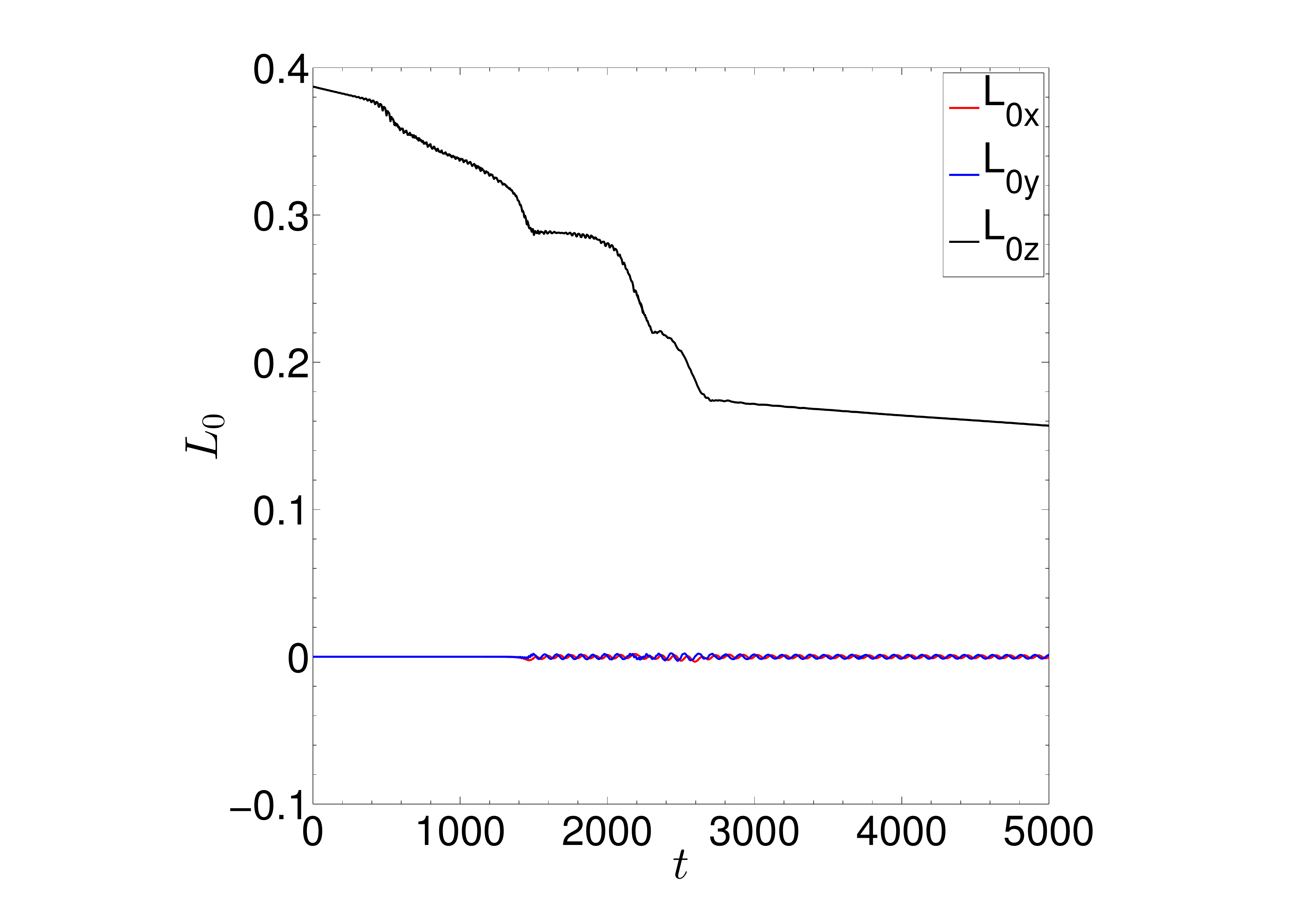} } 
       \end{center}
  \caption{Evolution of various flow quantities with time for a simulation with $\Omega=0.3,n=0.05,A=0.15$ and $\nu=10^{-4}$. Top left: comparison of RMS $u_z$ with the energy in the differential rotation, $E_\mathrm{dr}$. Top right: viscous dissipation rate (black line) and laminar viscous dissipation rate prediction (red line). Bottom left: mean asynchronism of the flow $\langle\gamma\rangle$. Bottom right: Cartesian components of the angular momentum of the fluid in the inertial frame.}
  \label{4}
\end{figure}
\begin{figure}
  \begin{center}
    \subfigure{\includegraphics[trim=5cm 0cm 7cm 2cm, clip=true,width=0.23\textwidth]{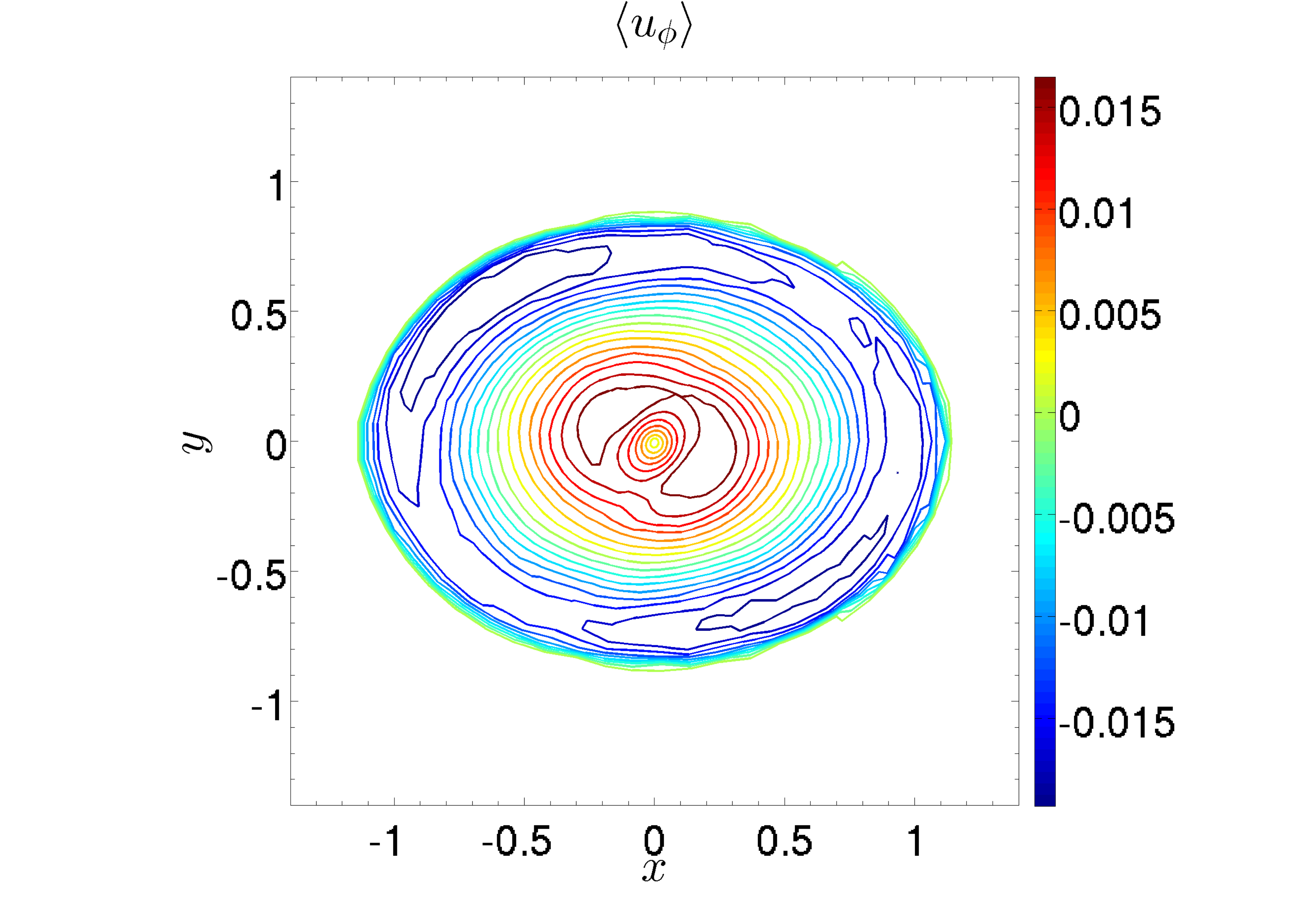} } 
    \subfigure{\includegraphics[trim=4cm 0cm 7cm 1cm, clip=true,width=0.23\textwidth]{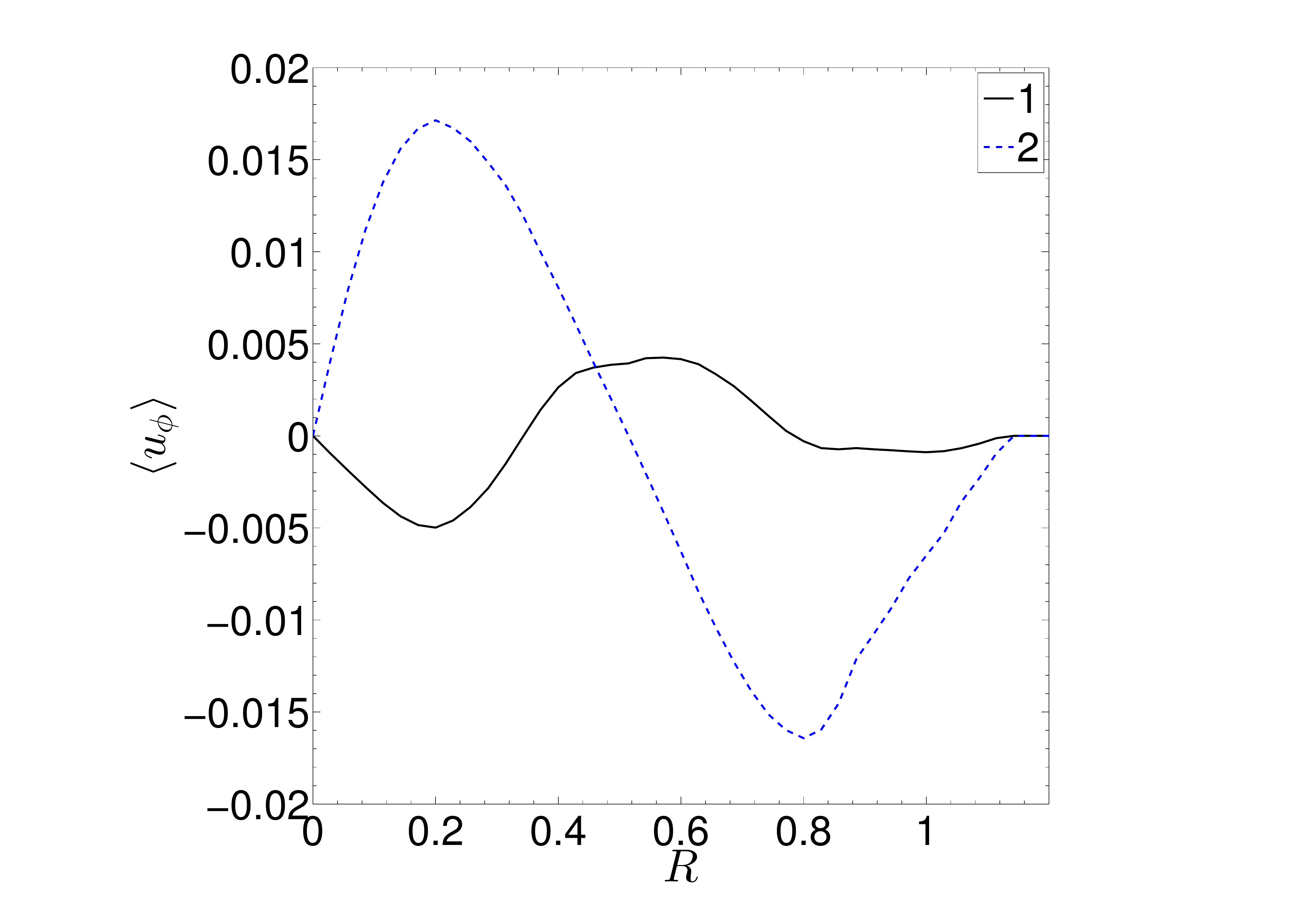} } 
       \end{center}
  \caption{Illustration of zonal flows produced in a simulation with $\Omega=0.3,n=0.05,A=0.15$ and $\nu=10^{-4}$. Left: Vertically-averaged azimuthal velocity $\langle u_\phi^{\prime}  \rangle_z$ on the $xy$-plane during the the strongest (second) burst phase. Right: comparison of the mean zonal flow $\langle u_\phi^{\prime} \rangle_{\phi,z}$ as a function of cylindrical radius $R$ during the first and second burst phases.}
  \label{5}
\end{figure}

A further example with a prograde spin is illustrated in Figs.~\ref{4} and \ref{5}. In this case, $\Omega=0.3$, $n=0.05$, with $A=0.15$ and $\nu=10^{-4}$ (using a resolution of $N=10$). As shown in the top left panel of Fig.~\ref{4}, the initial elliptical instability (shown by the blue dashed line that plots the RMS $u_z$) has grown by $t\sim 500$, leading to the generation of differential rotation in the form of zonal flows (shown by the solid black line, and plotted as a function of $R$ in the right panel of Fig.~\ref{5} as the solid black line). This simulation also exhibits bursty behaviour, though it is much less regular than the example discussed in \S~\ref{example}. A strong zonal flow is produced during the second burst phase at $t\sim 1500$, which is plotted as a function of $R$ in the right panel of Fig.~\ref{5} (blue dashed line, where the black solid line shows the zonal flow in the first burst phase), which subsequently leads to a reduction in wave activity ($\langle u_z\rangle$). I have plotted the mean azimuthal velocity on the $xy$-plane during this phase in the left panel of Fig.~\ref{5}. As this zonal flow is subsequently damped, this produces a period of rapid tidal desynchronisation, in which $\langle \gamma(t) \rangle$ increases sharply until the zonal flow has damped and transferred its angular momentum to the mean rotation (shown in the bottom left panel of Fig.~\ref{4}). Following this there are two further burst phases before the instability has succeeded in partially synchronising the planet with its orbit, such that $\langle \gamma(t) \rangle \lesssim 0.07$, after which the instability is not excited strongly over viscous damping. The enhanced viscous dissipation during the bursts of instability (over the laminar prediction shown as the red dashed line) is shown in the top right panel of Fig~\ref{4}. As with the previous example, the instability preserves the rotation axis of the flow, as indicated by the Cartesian components of the mean angular momentum in the inertial frame in the bottom right panel of Fig.~\ref{4}.

\begin{figure}
  \begin{center}
      \subfigure{\includegraphics[trim=6cm 0cm 7cm 1cm, clip=true,width=0.23\textwidth]{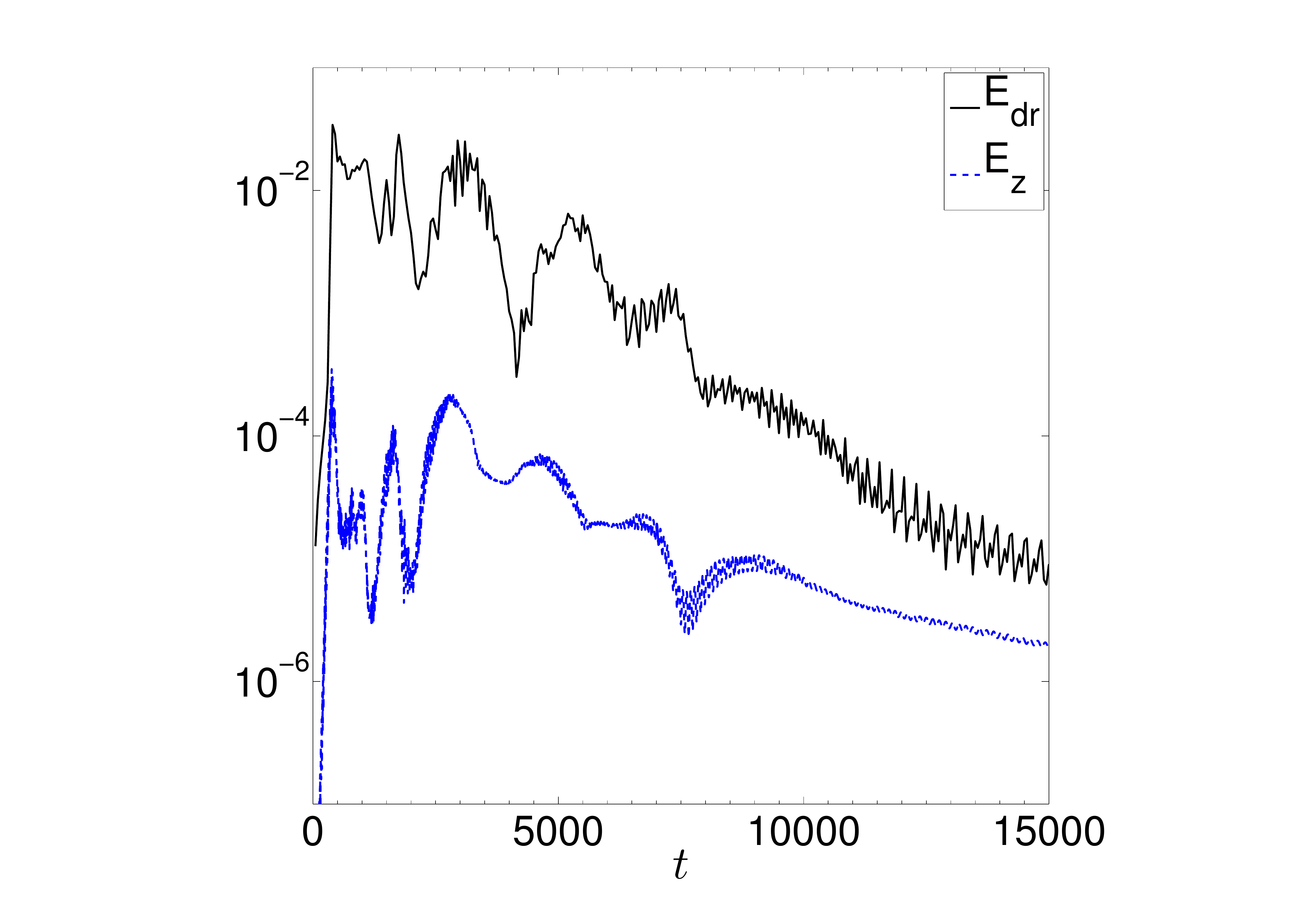} } 
      \subfigure{\includegraphics[trim=6cm 0cm 7cm 1cm, clip=true,width=0.23\textwidth]{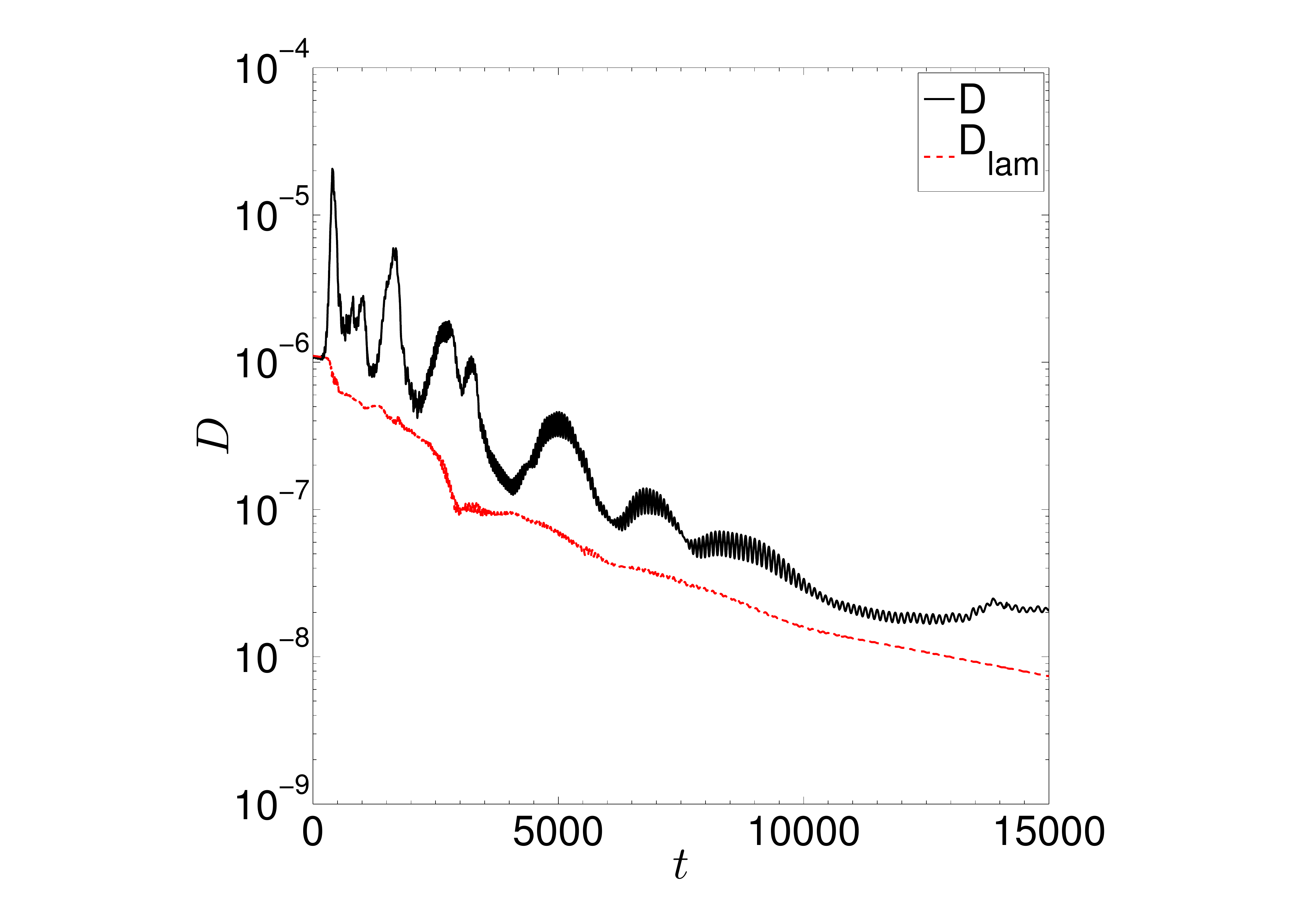} } 
      \subfigure{\includegraphics[trim=6cm 0cm 7cm 1cm, clip=true,width=0.23\textwidth]{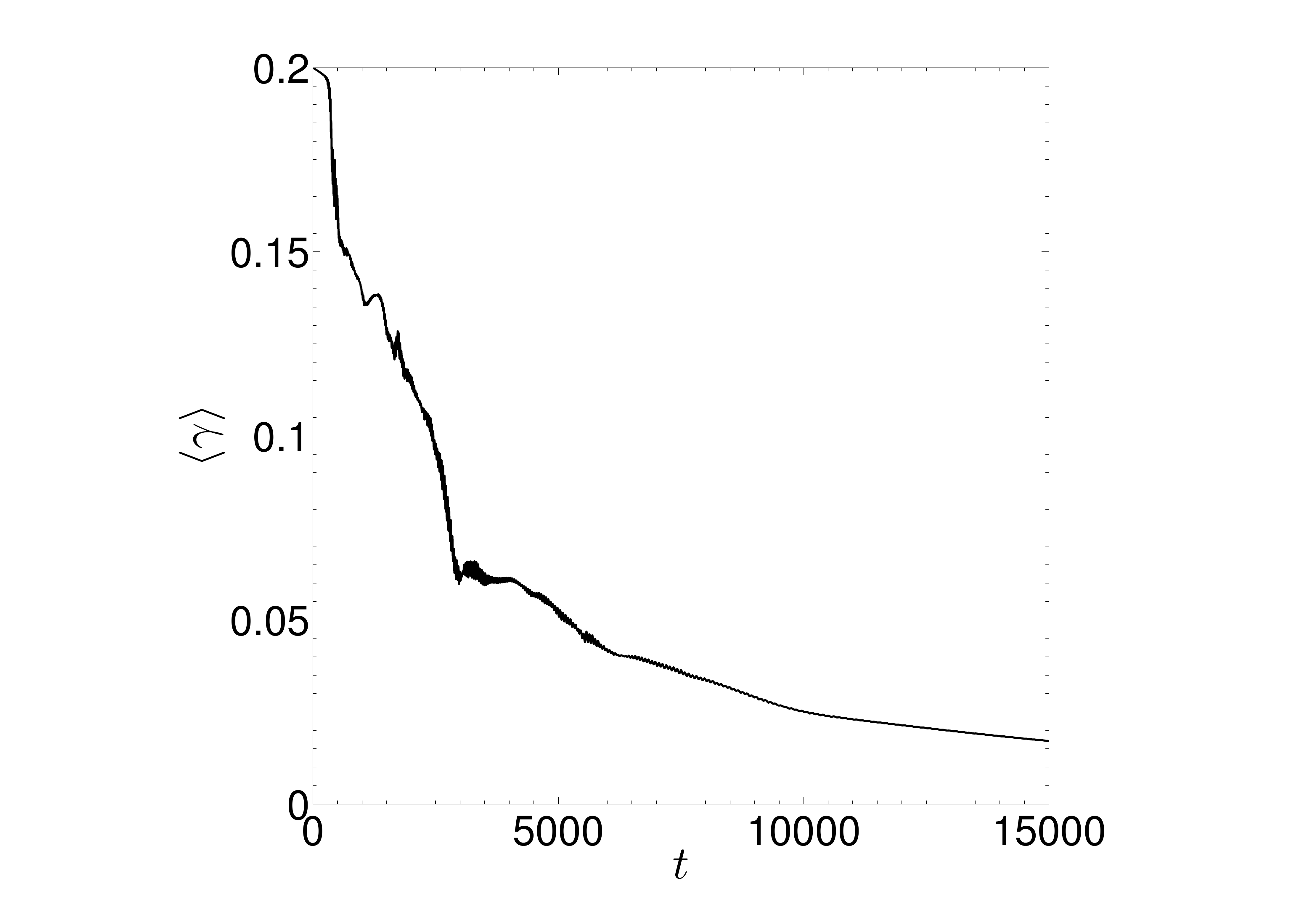} } 
       \end{center}
  \caption{Evolution of various flow quantities with time for a simulation with $\Omega=0.2,n=0,A=0.15$ and $\nu=10^{-4}$. Top left: comparison of $E_z$ with the energy in the differential rotation, $E_\mathrm{dr}$. Top right: viscous dissipation rate (black line) and laminar viscous dissipation rate prediction (red line). Bottom: mean asynchronism of the flow $\langle\gamma\rangle$. Given that there is no reference orbit (as $n=0$), I have not plotted the angular momentum components for this example.}
  \label{10}
\end{figure}

The time-evolution of the various flow quantities for a different example which exhibits qualitatively similar behaviour is plotted in Fig.~\ref{10}.  This example has $\Omega=0.2$, $A=0.15$ and $\nu=10^{-4}$, but with $n=0$  -- i.e. the planet is not moving, which is unphysical but the properties of the flow are worth presenting in this case because the flow is more turbulent. The top left panel again shows similar behaviour to the two previous examples, plotting the $E_z=\frac{1}{2}\langle u_z\rangle^2$ (instead of $\langle u_z\rangle$, which was used for the previous two examples) and $E_\mathrm{dr}$ as a function of time on a log-scale. However, this simulation is less regular in its bursty behaviour. When the zonal flows are strong, the energy in the waves is reduced, until the zonal flows are sufficiently damped -- either by their own shear instabilities, nonlinear energy transfers or by viscosity. This behaviour persists until the planet is mostly synchronised with its orbit, as shown in the bottom panel. After $t\gtrsim 6000$, the elliptical instability is much weaker, leading to mostly laminar evolution, as the wave excitation is balanced by viscous dissipation, at a rate that is an $O(1)$ factor larger than the viscous dissipation of the bulk tidal flow, as is shown in the top right panel of Fig.~\ref{10}.

\subsection{Summary}

In the examples presented so far, the initial flow becomes unstable and global inertial modes are excited by the elliptical instability. These grow and produce zonal flows, which inhibit further growth of the instability, either by changing the flow (by inducing differential rotation) in such a way that resonance with a particular pair of global modes is no longer possible, or by perturbing the phases of the growing modes in such a way that they cannot be coherently driven \citep{BL2013}. In addition, the gradual evolution of the bulk rotation of the fluid means that different resonances can be excited as the system evolves. Global modes are again driven when the zonal flows are sufficiently damped by viscosity, or by their own shear instabilities or nonlinear energy transfers, and this leads to ``bursty" cyclic behaviour, somewhat reminiscent of predator-prey dynamics in some cases. Whenever it is excited, the elliptical instability produces enhanced tidal dissipation that leads to partial synchronisation of the spin of the planet with its orbit.

So far, I have only discussed cases in which the initial spin was prograde/aligned with the orbit. In \S~\ref{retrogradecases}, I present the results of simulations in which the planet has a retrograde/anti-aligned spin. This allows the action of the elliptical instability on the spin-orbit angle to be analysed. In Appendix~\ref{rigidsims}, I compare these results with those obtained in simulations that adopt a rigid container.

\section{Nonlinear simulations with retrograde spin: spin-orbit alignment driven by the elliptical-instability}
\label{retrogradecases}

Until now, I have only discussed examples with a prograde (relative to the orbit; or with $n=0$) planetary spin that is initially aligned with its orbit. In these simulations, the elliptical instability led to gradual spin-synchronisation, with the planetary spin remaining prograde and approximately aligned with the orbit. In this section, I present simulations of the elliptical instability in planets with an initially purely retrograde (anti-aligned) spin, with a particular emphasis on studying tidal spin-orbit alignment driven by the elliptical instability.
In Fig.~\ref{8} (see also \citealt{Barker2015a}), I have demonstrated that the strongest instabilities indeed occur for retrograde spins i.e.~when $n \leq0$, and that instability is also possible when $\frac{n}{\Omega}\lesssim -1$, if $A$ is sufficiently large, where the elliptical instability is usually thought to be absent.

The spin-orbit evolution can be analysed by considering the spin-orbit angle $\psi$, defined by 
\begin{eqnarray}
\cos \psi = \hat{\boldsymbol{n}} \cdot \hat{\boldsymbol{L}}_0 =\frac{\mathrm{sgn}(n)L_{0,z}}{\sqrt{L_{0,x}^2+L_{0,y}^2+L_{0,z}^2}},
\end{eqnarray}
where $\boldsymbol{L}_0$ is the fluid angular momentum in the inertial frame and $\hat{\boldsymbol{n}}$ is the orbit's unit normal vector.
Starting from a purely anti-aligned spin, I expect the spin to align (and synchronise) with the orbit so that this angle will evolve from $\cos\psi=-1$ to $\cos\psi=1$. In this paper, I assume $\boldsymbol{n}$ is constant, i.e.~the orbit is fixed in space. This is only appropriate in the limit of very large orbital angular momentum, which is a reasonable approximation when considering the spin evolution of a hot Jupiter (in reality, $\boldsymbol{n}$ will also evolve due to tides in the star). In this limit, we expect the planet's spin to align with its orbit on a timescale that is comparable with (but not necessarily identical to) the spin-synchronisation timescale.

\subsection{Violent instability when $\frac{n}{\Omega}\lesssim -1$}
\label{violent}

\begin{figure}
  \begin{center}
\subfigure{\includegraphics[trim=5cm 0cm 7cm 1cm, clip=true,width=0.23\textwidth]{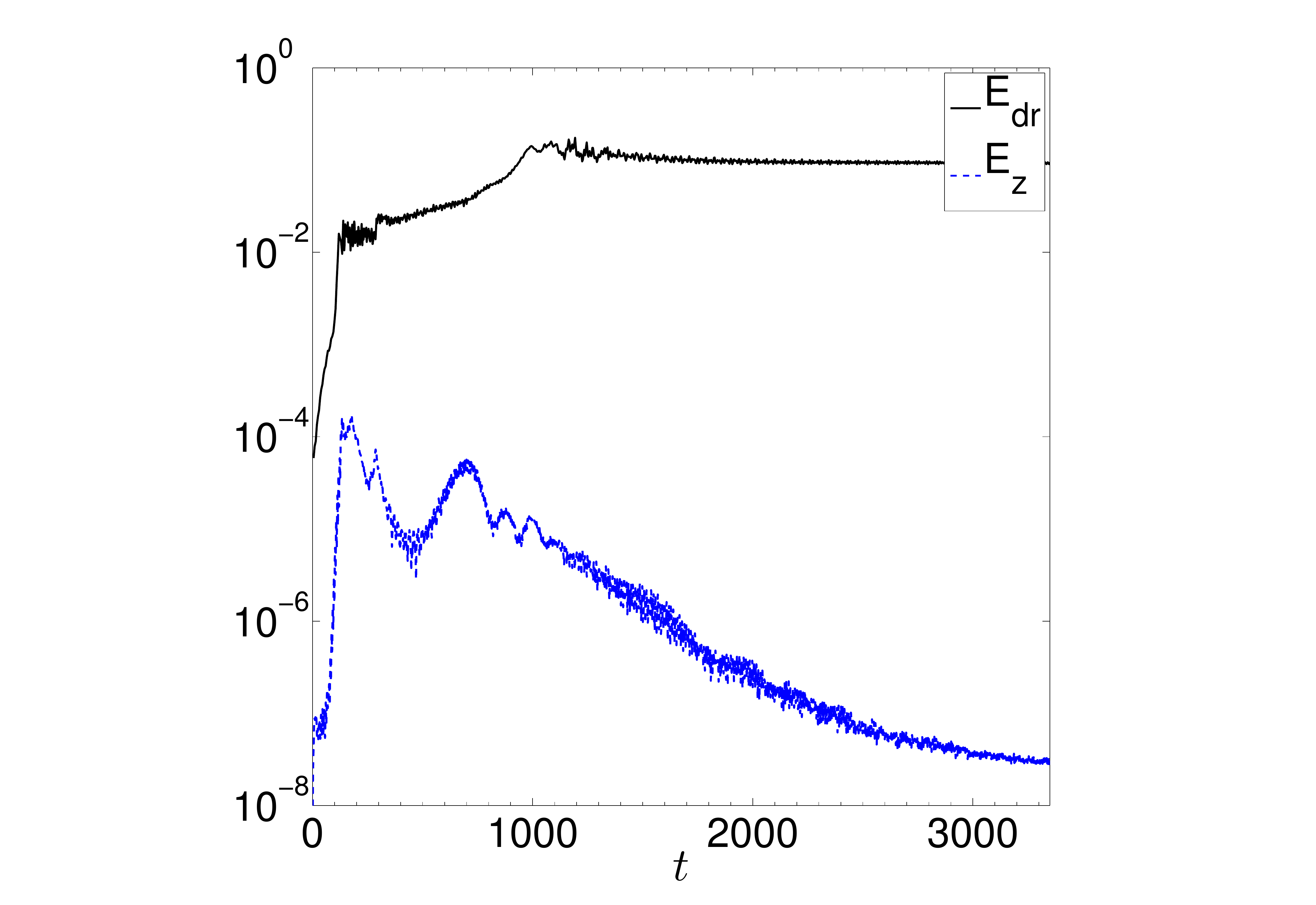} } 
\subfigure{\includegraphics[trim=5cm 0cm 7cm 1cm, clip=true,width=0.23\textwidth]{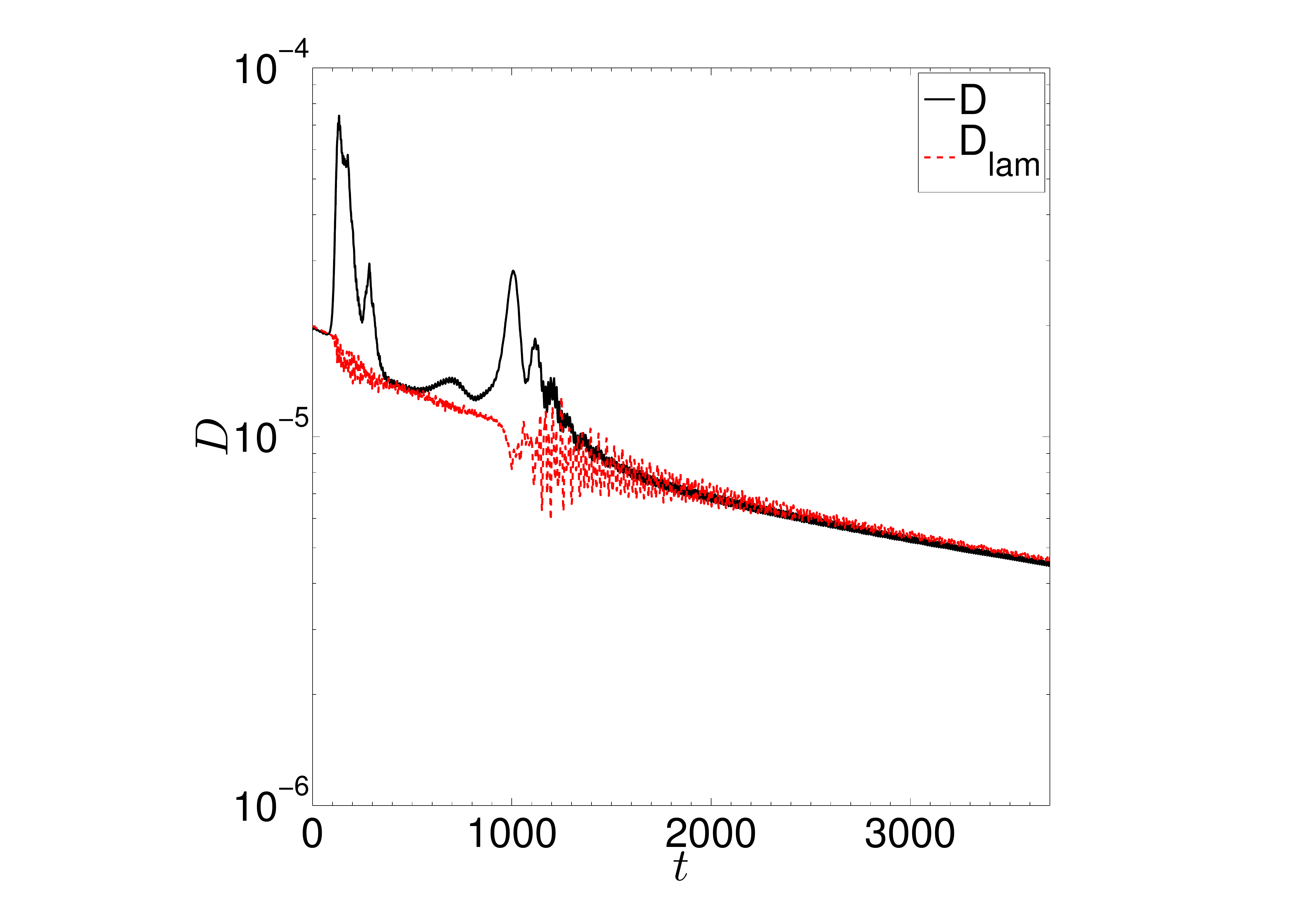} } 
\subfigure{\includegraphics[trim=5cm 0cm 7cm 1cm, clip=true,width=0.23\textwidth]{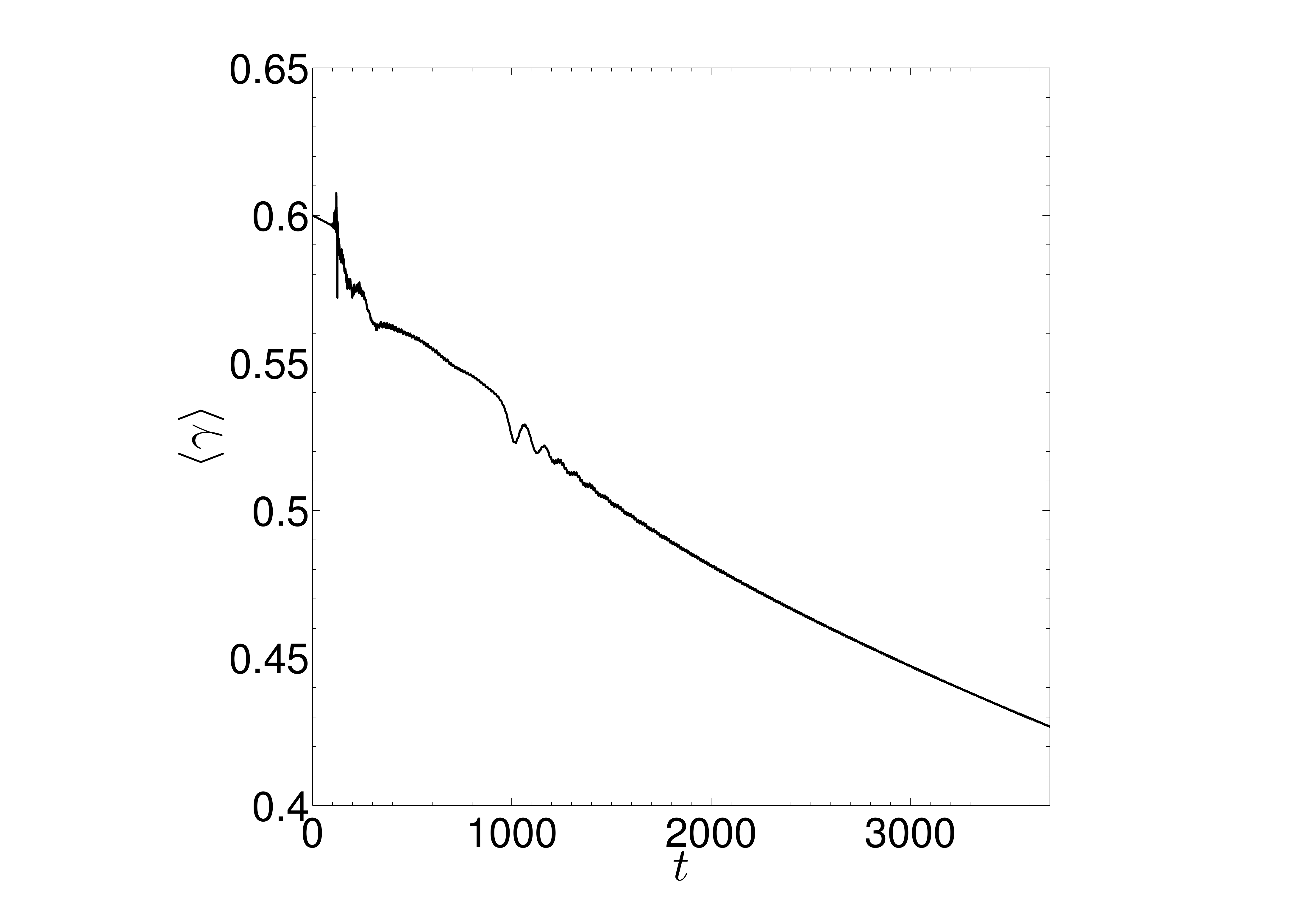} } 
\subfigure{\includegraphics[trim=5cm 0cm 7cm 1cm, clip=true,width=0.23\textwidth]{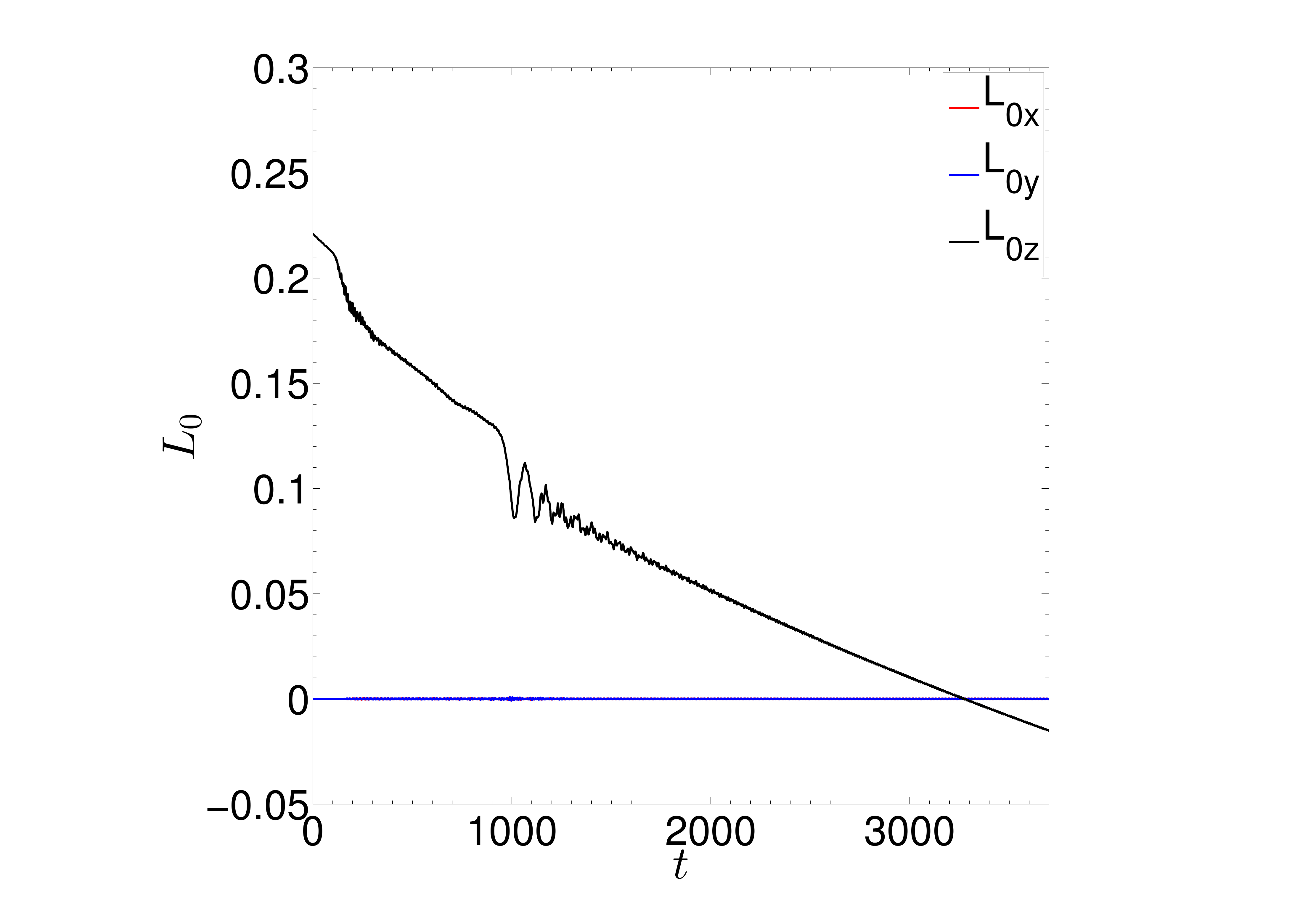} } 
\subfigure{\includegraphics[trim=5cm 0cm 7cm 1cm, clip=true,width=0.23\textwidth]{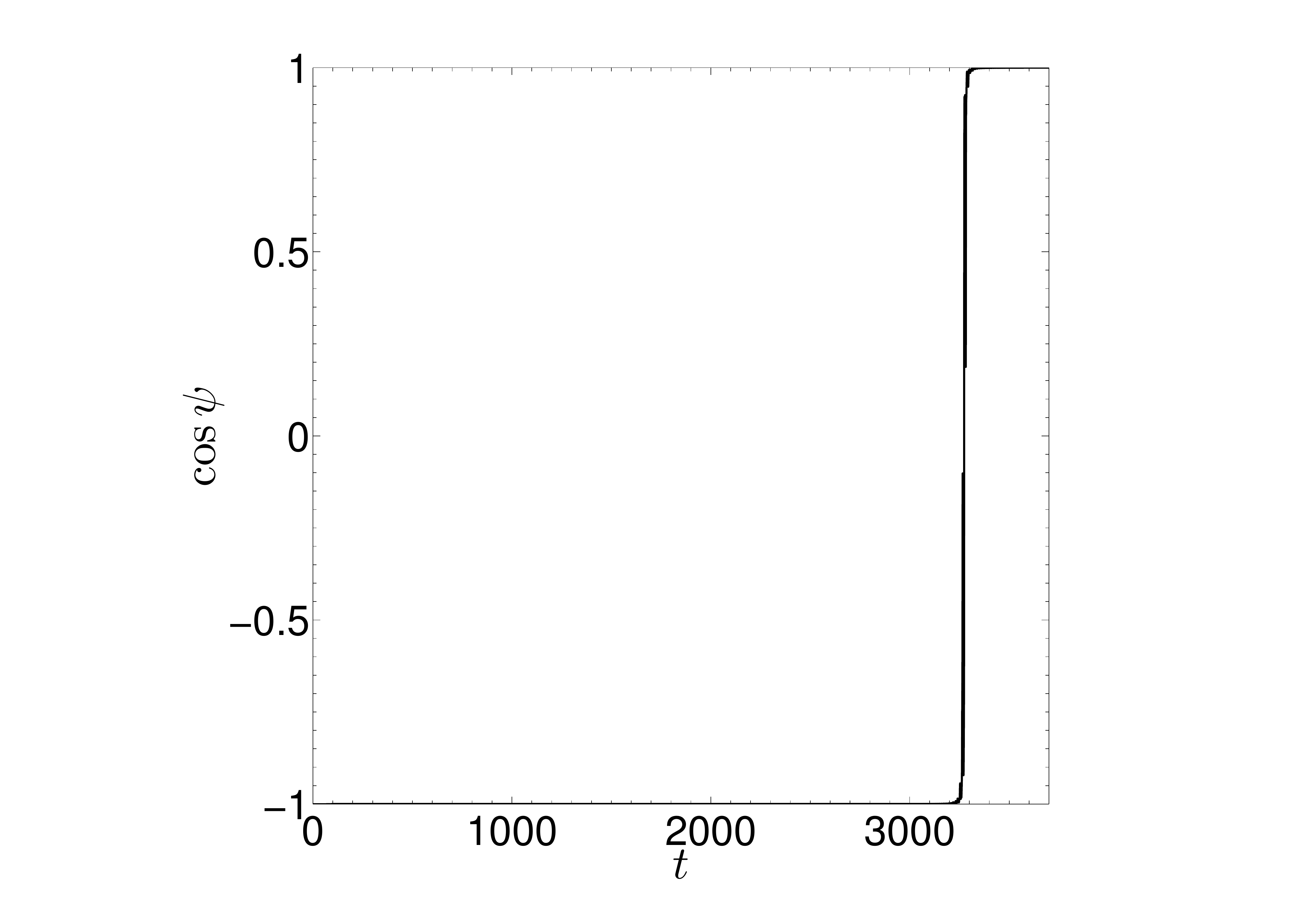} } 
   \end{center}
  \caption{Evolution of various flow quantities with time for an initially anti-aligned simulation with $\Omega=0.2,n=-0.4,A=0.1$ and $\nu=10^{-4}$, for which the elliptical instability is not usually thought to be excited. Top left: comparison of $E_z$ with the energy in the differential rotation, $E_\mathrm{dr}$. Top right: viscous dissipation rate (black line) and laminar viscous dissipation rate prediction (red dashed line). Middle left: mean asynchronism of the flow $\langle\gamma\rangle$. Middle right: Cartesian components of the angular momentum in the inertial frame. Bottom: cosine of the spin-orbit angle $\psi$, which exhibits rapid alignment at $t\sim 3200$, due to viscous dissipation of the laminar tidal flow, which preserves the rotation axis of the flow.}
  \label{5b}
\end{figure}

\begin{figure}
  \begin{center}
     \subfigure{\includegraphics[trim=0cm 0cm 0cm 0cm, clip=true,width=0.35\textwidth]{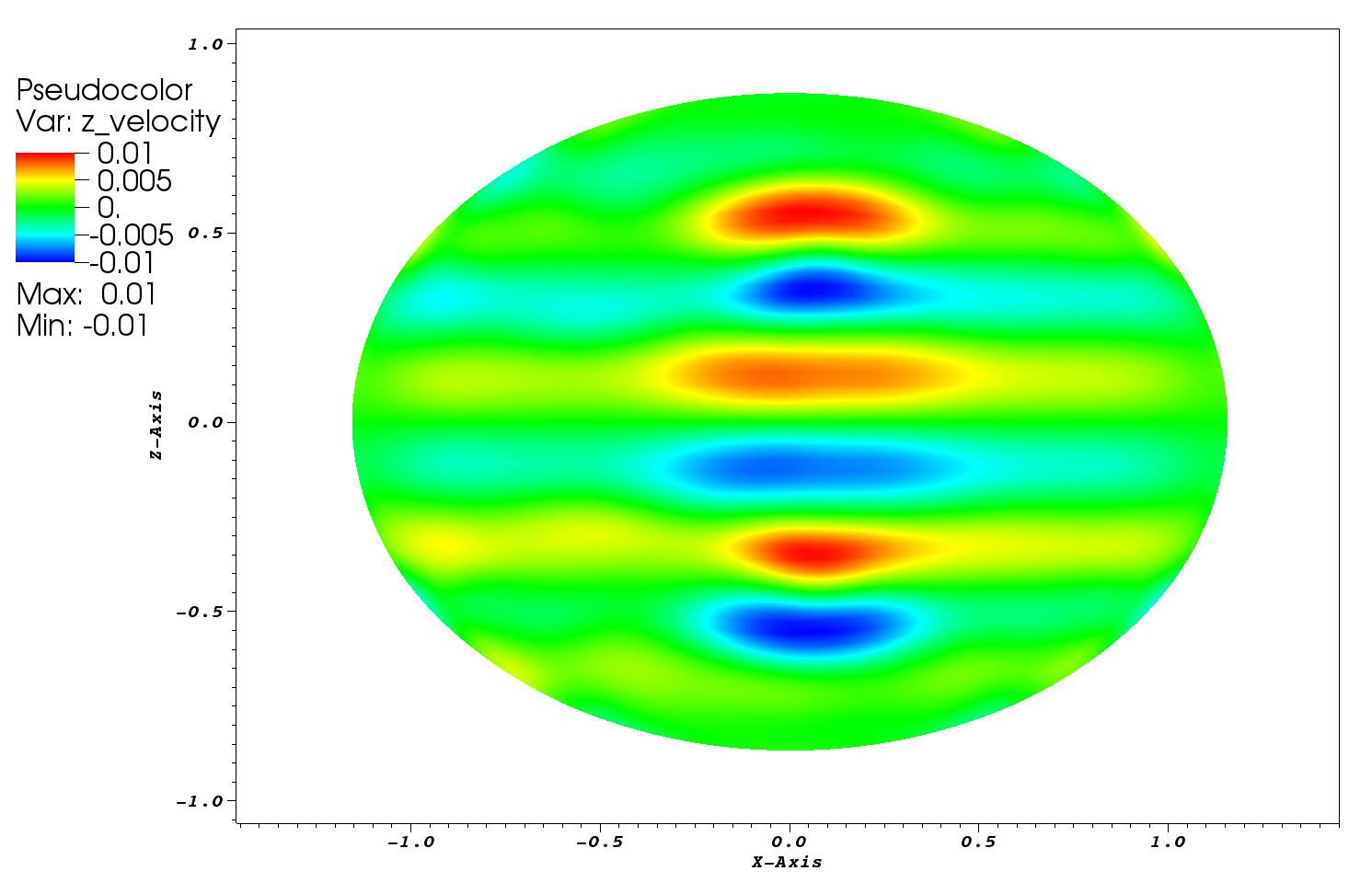} } 
     \subfigure{\includegraphics[trim=0cm 0cm 0cm 0cm, clip=true,width=0.35\textwidth]{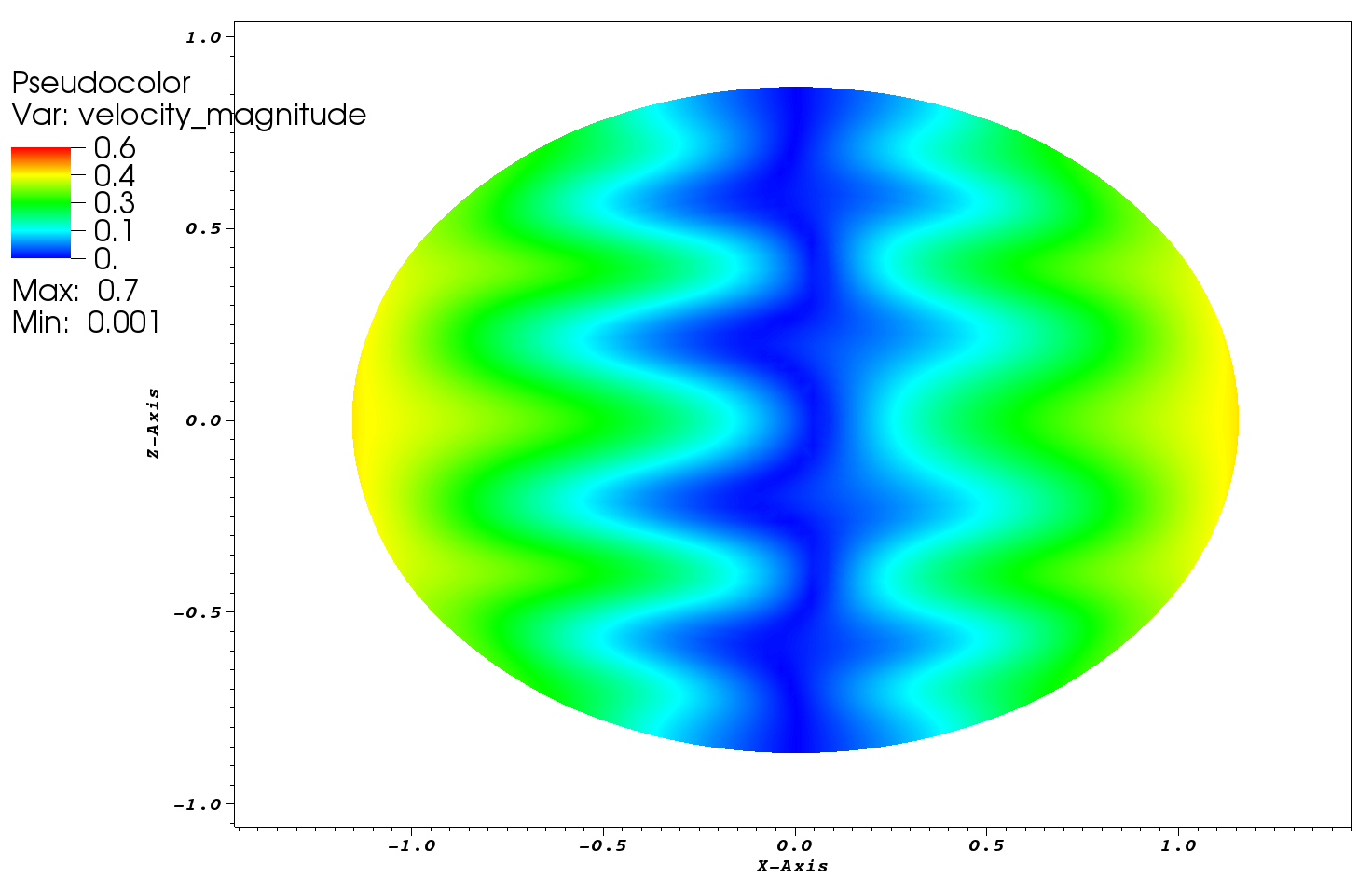} }
     \subfigure{\includegraphics[trim=0cm 0cm 0cm 0cm, clip=true,width=0.35\textwidth]{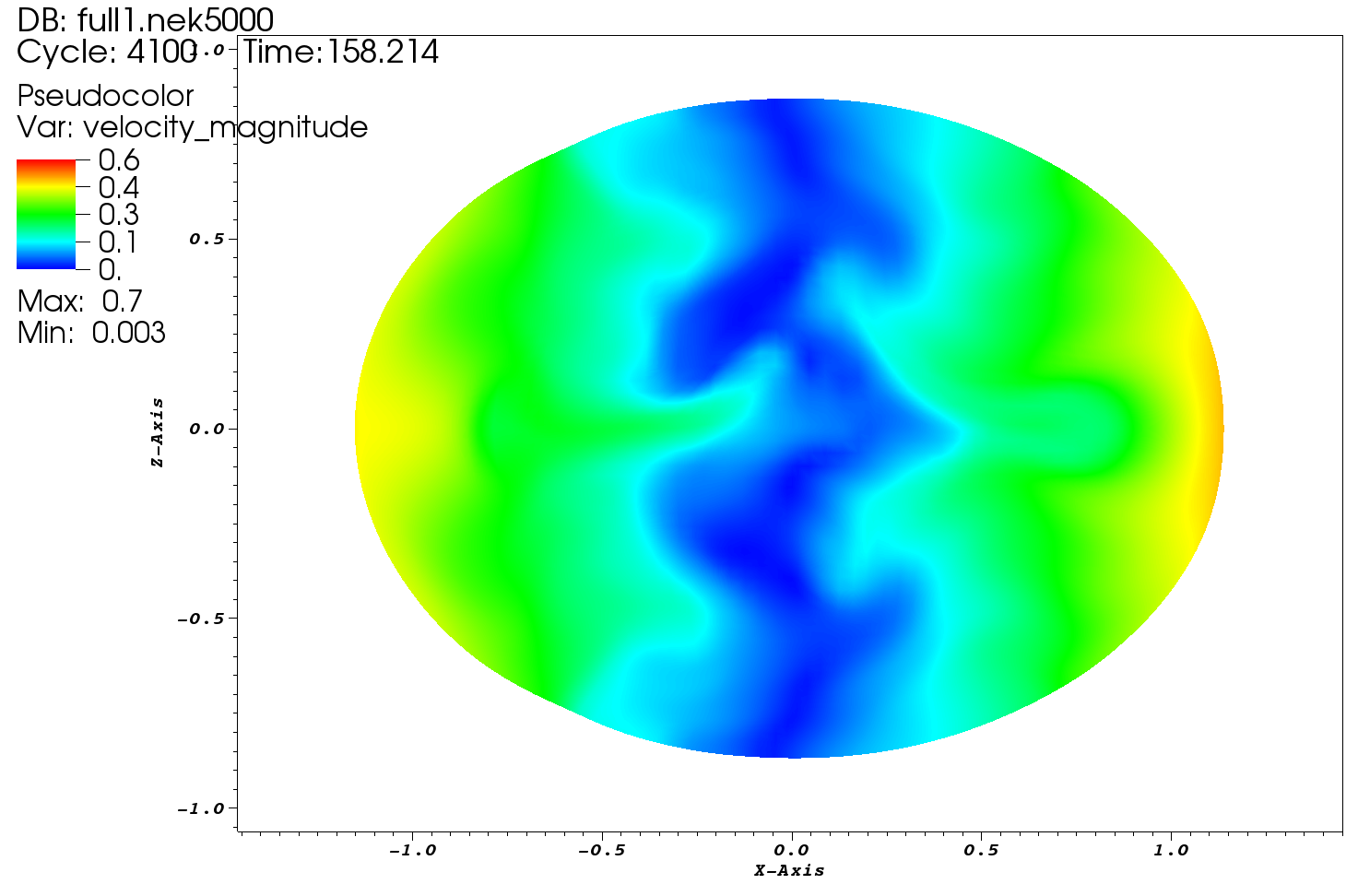} }  
       \end{center}
  \caption{Illustration of the flow on the $xz$-plane at two times in the simulation with $\Omega=0.2,n=-0.4,A=0.1$, and $\nu=10^{-4}$. Top and Middle: $u_z$ and $|\boldsymbol{u}|$ at $t=114.73$, respectively, both during the linear growth phase. Bottom: $|\boldsymbol{u}|$ during initial turbulent phase at $t=158.21$. This instability is a global version of the ``stack of pancakes" instability (involving primarily horizontal epicyclic motions) that arises in a local analysis of the elliptical instability (\citealt{Barker2015a}; see also \citealt{Craik1989,LL1996a}).}
  \label{3b}
\end{figure}

In  Fig.~\ref{8}, I have demonstrated that elliptical instability can occur when $\frac{n}{\Omega} \lesssim -1$ if $A$ is sufficiently large -- as predicted by the global (and local) stability analysis \citep{Barker2015a} -- which is outside the frequency range in which it is usually thought to operate. Here I analyse the nonlinear outcome of a simulation with $\Omega=0.2$, $n=-0.4$, $A=0.1$ and $\nu=10^{-4}$, whose initial growth rate has been plotted in the left panel of Fig.~\ref{8}. In Fig.~\ref{5b}, I plot the temporal evolution of various flow quantities during the nonlinear evolution of this simulation, and in Fig.~\ref{3b}, I plot the vertical velocity and $|\boldsymbol{u}|$ for the growing mode at $t=114.73$, and $|\boldsymbol{u}|$ during the initial turbulent phase at $t=158.21$. 

The initial instability is violent, causing a rapid partial-synchronisation of the planetary spin at $t\sim 100$. Zonal flows (quantified by $E_\mathrm{dr}$) are produced as the instability saturates. A second burst of instability occurs at $t\sim 1000$, after which there is no further burst of instability and waves are no longer efficiently driven (i.e.~$\langle u_z\rangle$ is small), but a zonal flow persists. During the subsequent evolution, the spin of the planet very gradually synchronises with its orbit due to viscous dissipation of the zonal flows and the basic tidal flow. 
Viscous torques preserve the rotation axis of the flow, so that there is a sudden transition in $\cos\psi$ (i.e.~the rotation axis flips by $180^\circ$) when $L_{0z}$ passes through zero\footnote{Note that this occurs shortly before $\Omega$ passes through zero itself, i.e. before $\langle \gamma\rangle =\Omega-n=0.4$, because I have defined $\cos\psi$ using the angular momentum unit vector, and this differs from the angular velocity unit vector because the body is highly non-symmetric. I would observe the same behaviour if the spin-orbit angle was defined using the angular velocity components but at a slightly later time.}. This behaviour would be expected based on the simplest models of tidal dissipation, such as the constant lag-time model -- if $\psi=180^{\circ}$ initially, and the orbit is fixed \citep{Hut1981,Eggleton1998,Barker2009}, then $\psi$ should not evolve, except when $L_{0z}$ passes through zero. 

\subsection{Spin-orbit alignment when the ``spin-over" mode is excited}
\label{spinover}

In this section I collect together two examples in which the spin is initially anti-aligned with the orbit, where a secondary elliptical instability excites the ``spin-over" mode. The spin-over mode in a sphere is the only inertial mode with harmonic degree $\ell=2$ (with azimuthal wavenumber $m=\pm 1$), and represents a rigid tilt of the planet's rotation axis. This mode is excited by the elliptical instability for a narrow range of $\Omega$ and $n<0$ (see Fig.~\ref{8} and \citealt{Barker2015a}), but is not excited when $n>0$ because the phase velocity of the mode must match that of the orbit \citep{Kerswell1994}. It occurs when the polar axis ($c$) becomes the middle axis ($b<c<a$), and is related to the ``middle moment of inertia instability" of a rigid body. Previous laboratory experiments and numerical simulations have emphasised the importance of this mode \citep{Lacaze2004,LeBars2007,LeBars2010,Cebron2010,Cebron2013}.

\begin{figure}
  \begin{center}
  \subfigure{\includegraphics[trim=5cm 0cm 8cm 1cm, clip=true,width=0.23\textwidth]{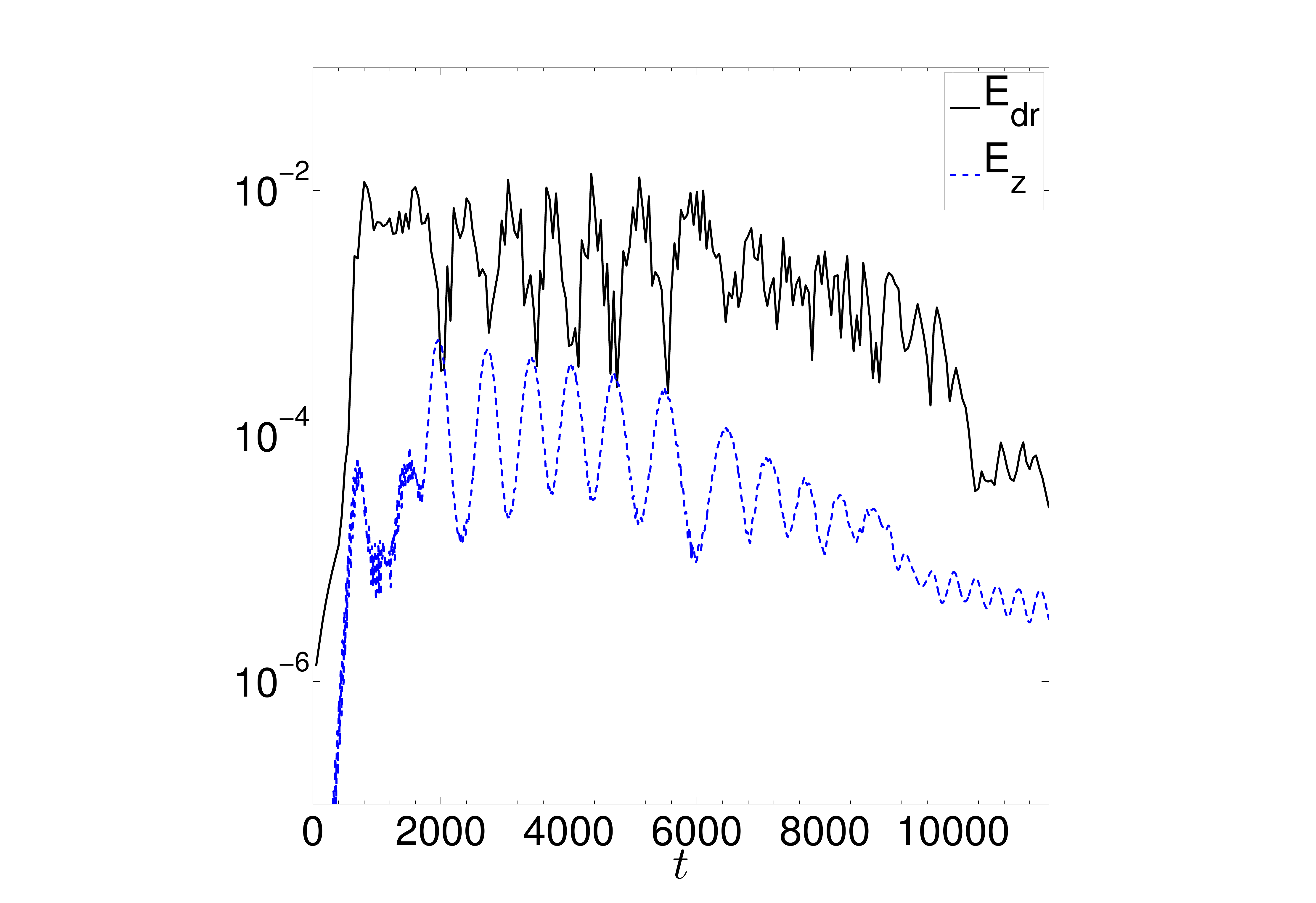} }
  \subfigure{\includegraphics[trim=5cm 0cm 8cm 1cm, clip=true,width=0.23\textwidth]{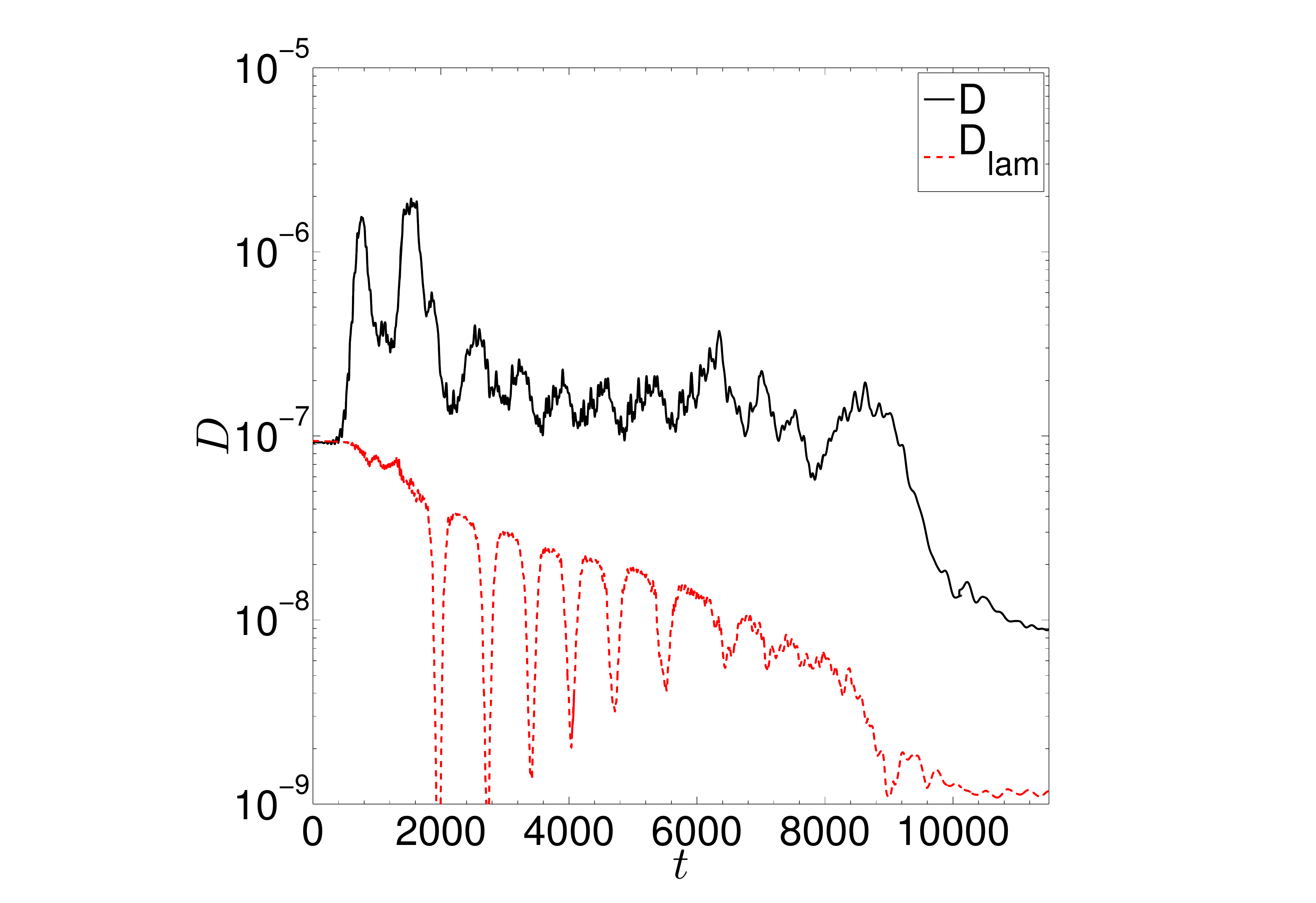} }
  \subfigure{\includegraphics[trim=5cm 0cm 8cm 1cm, clip=true,width=0.23\textwidth]{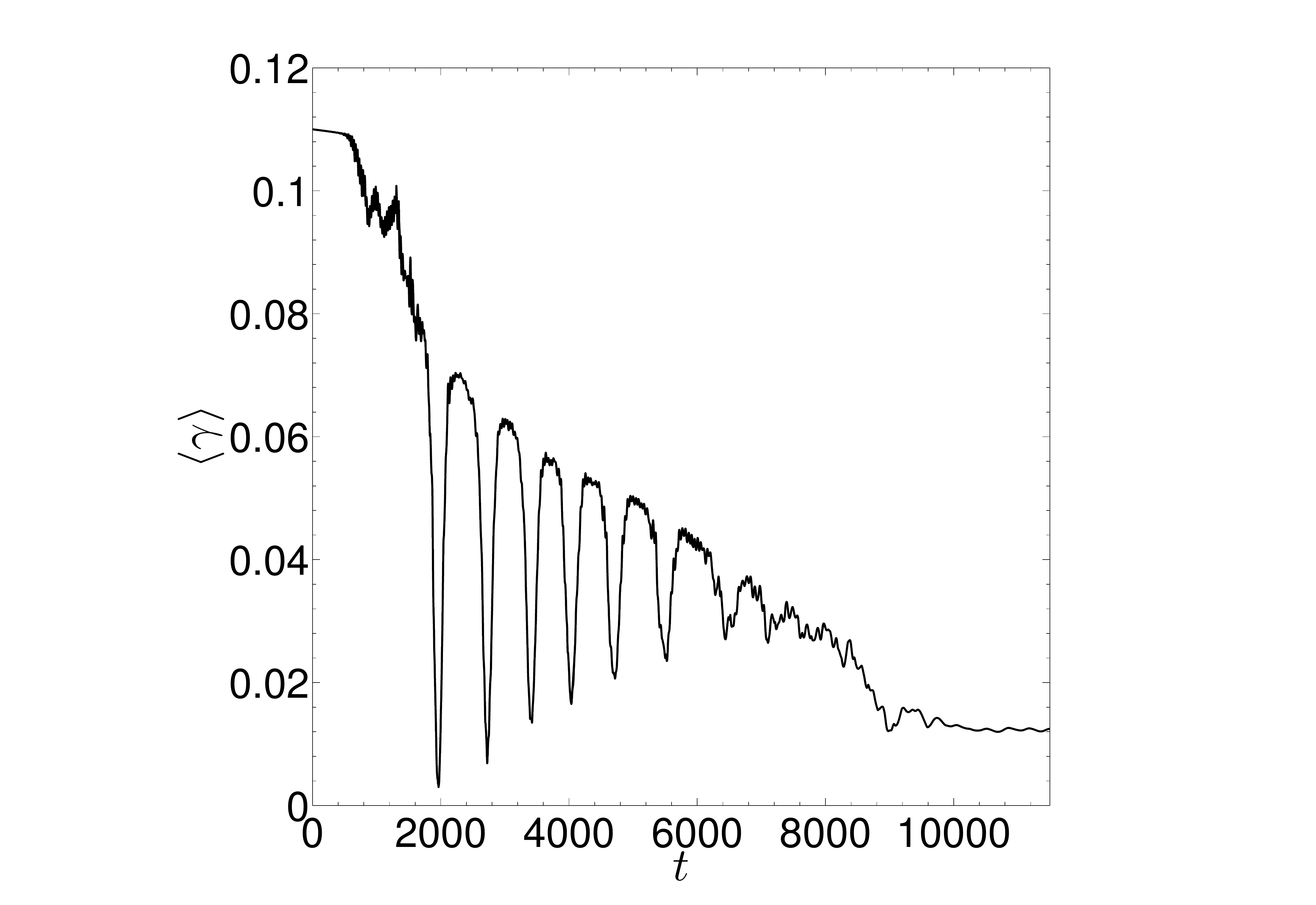} }
  \subfigure{\includegraphics[trim=5cm 0cm 8cm 1cm, clip=true,width=0.23\textwidth]{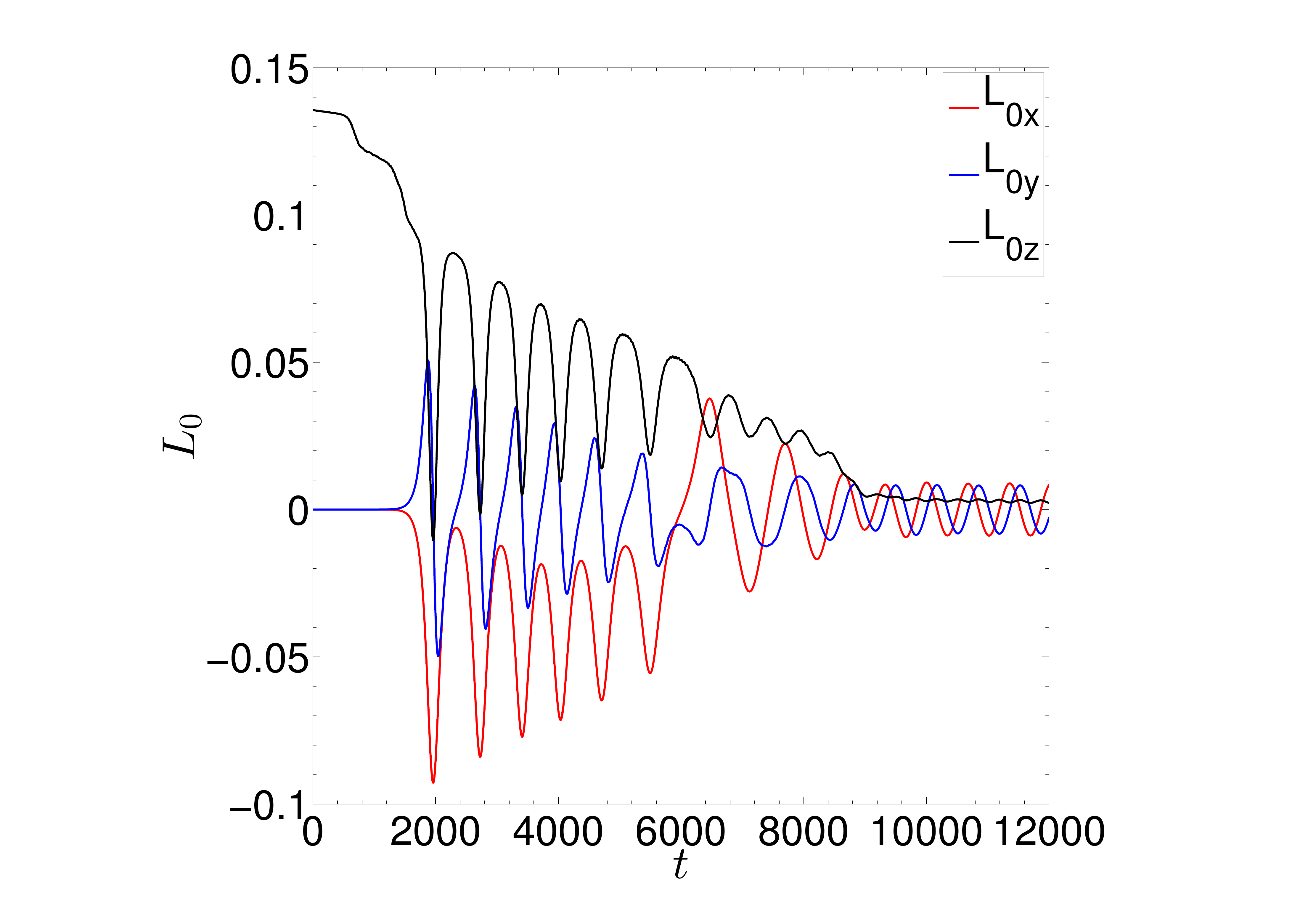} }
  \subfigure{\includegraphics[trim=5cm 0cm 8cm 1cm, clip=true,width=0.23\textwidth]{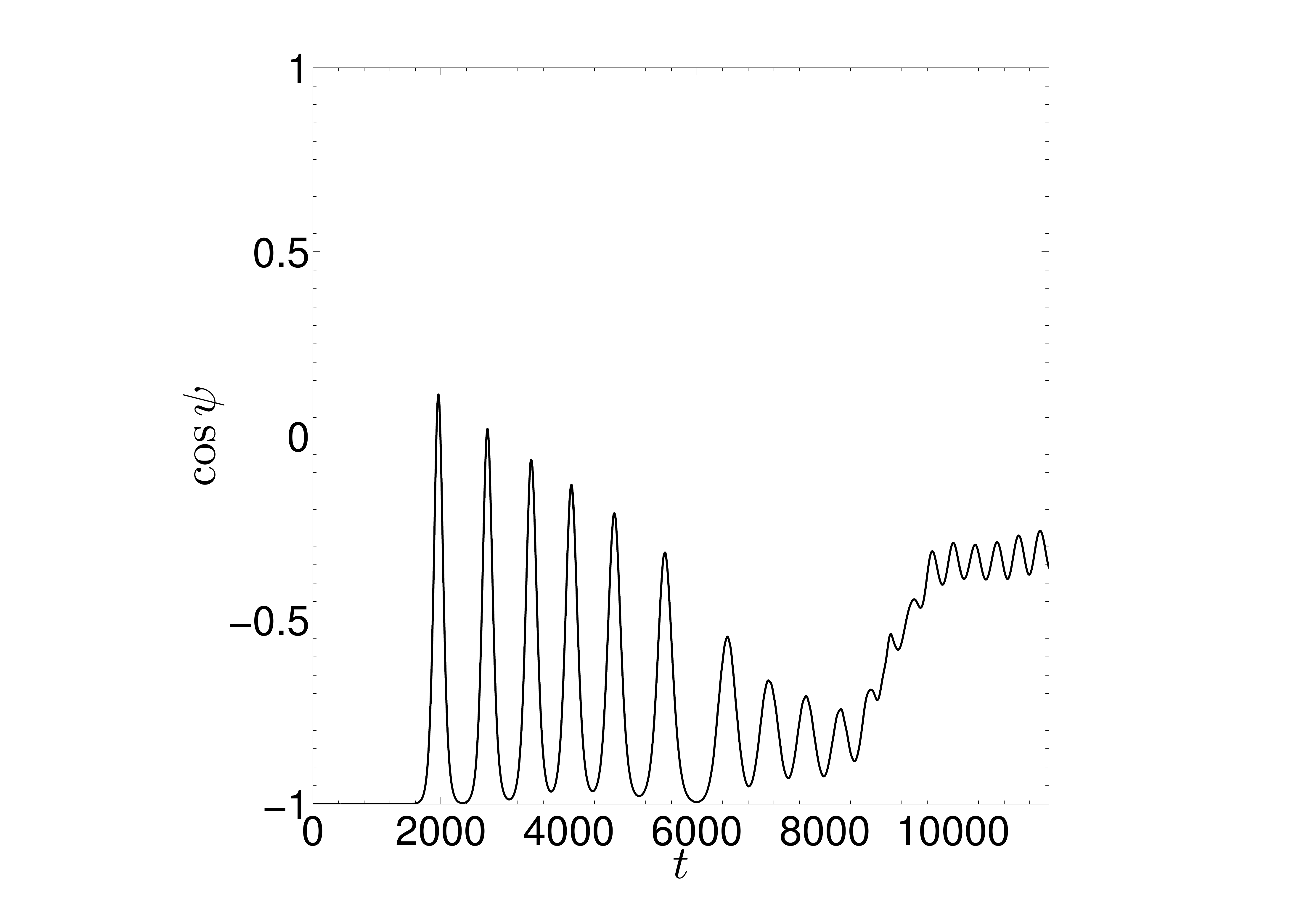} } 
       \end{center}
  \caption{Evolution of various flow quantities with time for an initially anti-aligned simulation with $\Omega=0.1,n=-0.01,A=0.15$ and $\nu=3\times10^{-5}$. Top left: comparison of $E_z$ with the energy in the differential rotation, $E_\mathrm{dr}$. Top right: viscous dissipation rate (black line) and laminar viscous dissipation rate prediction (red dashed line). Middle left: mean asynchronism of the flow $\langle\gamma\rangle$. Middle right: Cartesian components of the angular momentum in the inertial frame. Bottom: cosine of the spin-orbit angle $\psi$. The spin-over mode is excited when $t\sim 2200$, leading to precessional motion that is gradually damped.}
 \label{15a}
\end{figure}

\begin{figure}
  \begin{center}
     \subfigure{\includegraphics[trim=0cm 0cm 0cm 0cm, clip=true,width=0.35\textwidth]{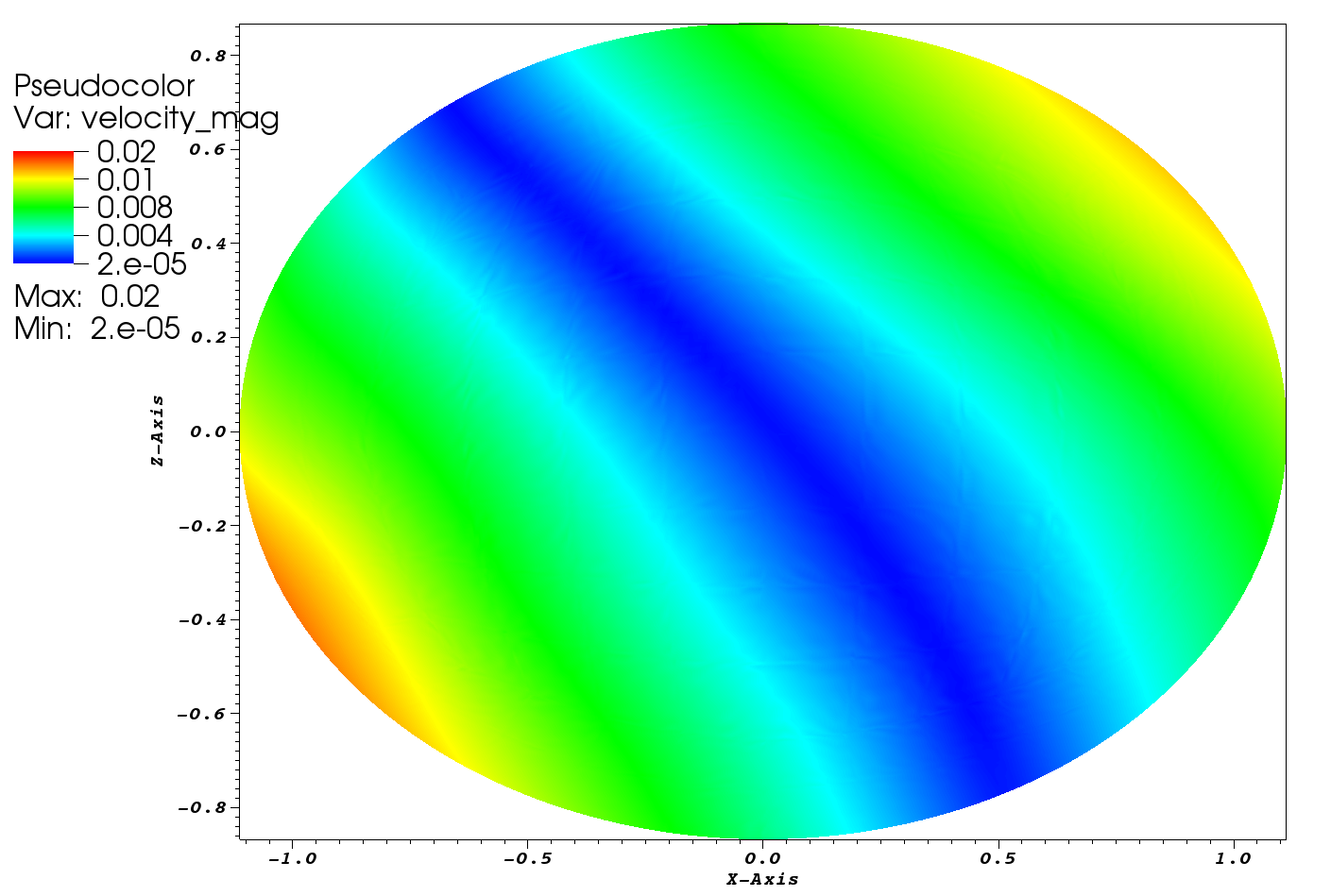} } 
       \end{center}
  \caption{Illustration of $|\boldsymbol{u}|$ on the $xz$-plane at $t=11697.8$ in the simulation with $\Omega=0.1,n=-0.01,A=0.15$, and $\nu=3\times10^{-5}$. During this phase, the rotation axis of the fluid precesses about $z$ due to the tidal torque.}
  \label{15aa}
\end{figure}

The first example has $\Omega=0.1$, $n=-0.01$, $A=0.15$ and $\nu=3\times10^{-5}$, for which the temporal evolution of various flow quantities are plotted in Fig.~\ref{15a}. In this case, the initial elliptical instability saturates by $t\sim 500$, by which time it has produced partial synchronisation of the spin and orbit. The spin remains anti-aligned with the orbit during this phase, so that this initial instability preserves the rotation axis of the flow. However, a secondary elliptical instability subsequently grows after $t\sim 1500$, which corresponds with the excitation of the spin-over mode (I have confirmed that this occurs when the container shape satisfies $b\lesssim c\lesssim a$). This tilts the planet's spin axis away from the $z$-direction, so that the non-dissipative tidal torque acting on the equatorial bulge causes the spin axis to precess. This is illustrated by oscillations in the $x$ and $y$ components of the angular momentum, plotted in the middle right panel in the figure. I also plot $|\boldsymbol{u}|$ for the spin-over mode at $t=11697.8$ in Fig.~\ref{15aa}.

The precessional motion is gradually damped by its own instabilities (e.g.~\citealt{Kerswell1993,LorenzaniTilgner2003,Cebron2010a,Lin2015}), and by viscosity, causing gradual spin-orbit alignment. The timescale for this process appears initially to be faster than the laminar viscous timescale for $t\lesssim 9000$ (i.e.~$D>D_\mathrm{lam}$), presumably due to instabilities of the precessional flow or to the generation and damping of non-negligible differential rotation in the fluid, whose presence is indicated by the solid black line in the top left panel of Fig.~\ref{15a}. By comparing the left middle and bottom panels, the timescale for spin-orbit alignment is similar to that in which the planet's spin synchronises, but there is an initial phase in which the spin evolves towards anti-alignment. This transient spin-orbit evolution towards anti-alignment was discussed by \cite{Lai2012}, but the eventual evolution, which includes the ``equilibrium tide" viscous torque, is towards alignment (see also \citealt{RogersLin2013} and \citealt{LiWinn2015}).

\begin{figure}
 \begin{center}
\subfigure{\includegraphics[trim=5cm 0cm 8cm 1cm, clip=true,width=0.23\textwidth]{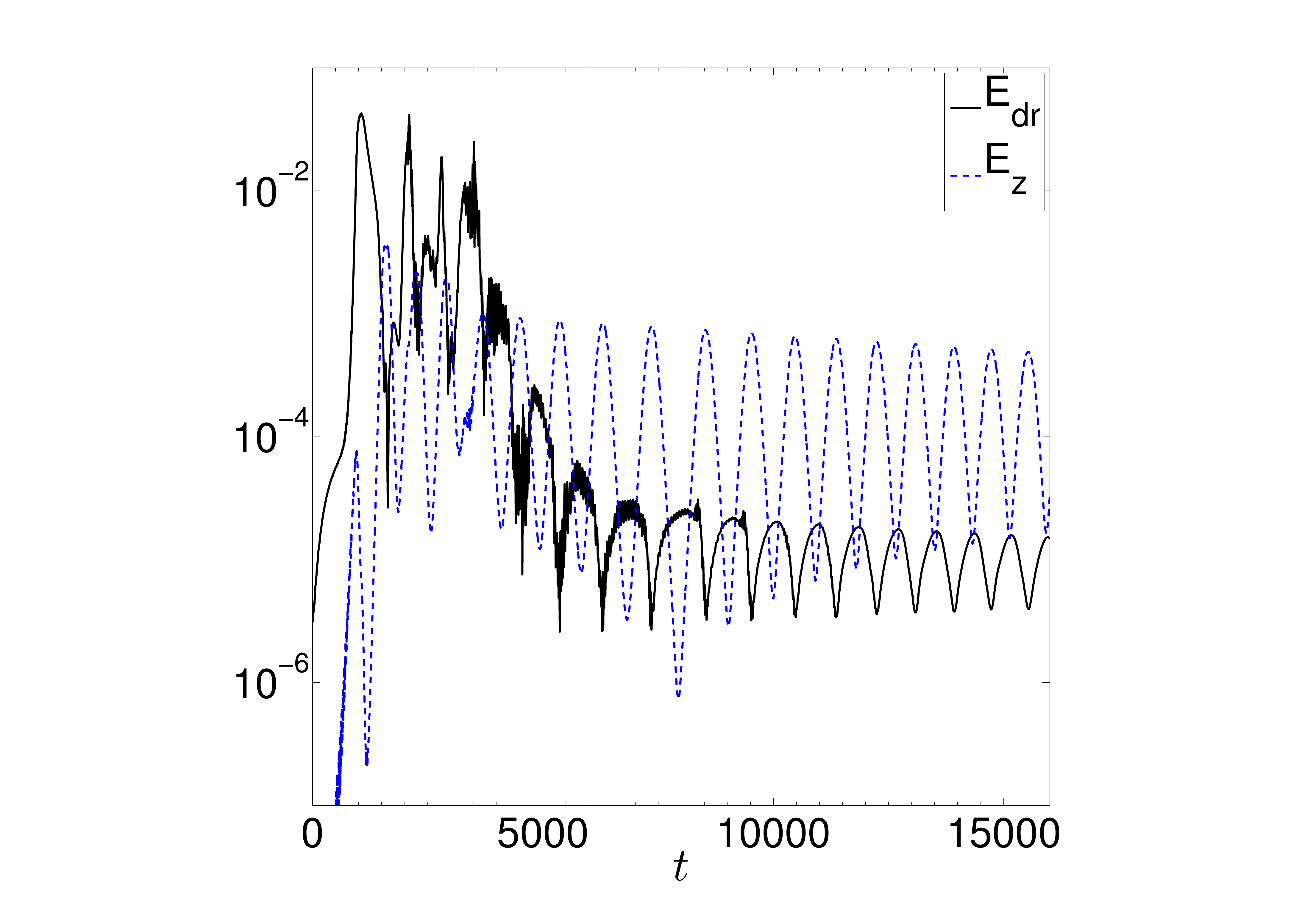} }
\subfigure{\includegraphics[trim=5cm 0cm 8cm 1cm, clip=true,width=0.23\textwidth]{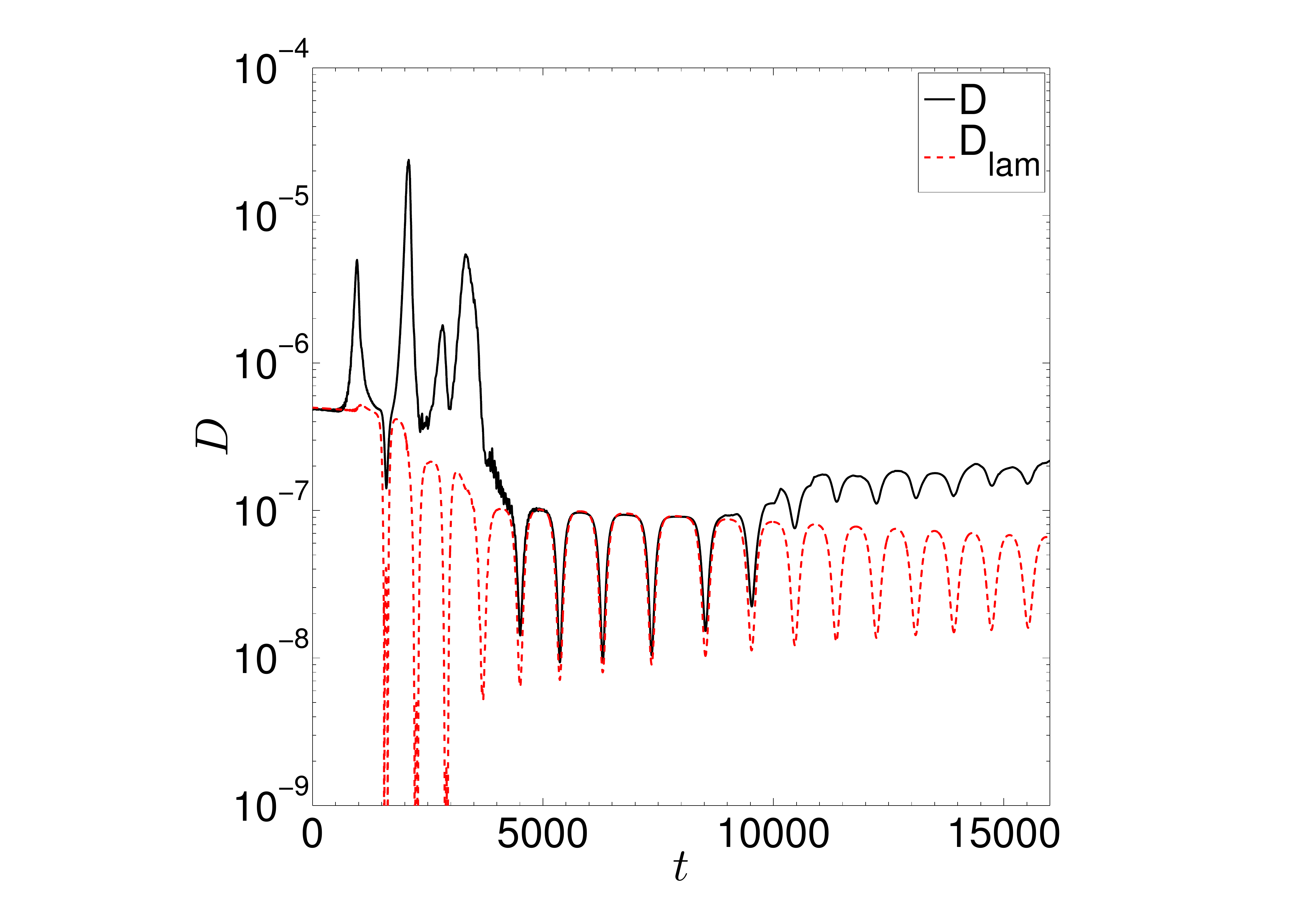} } 
\subfigure{\includegraphics[trim=5cm 0cm 8cm 1cm, clip=true,width=0.23\textwidth]{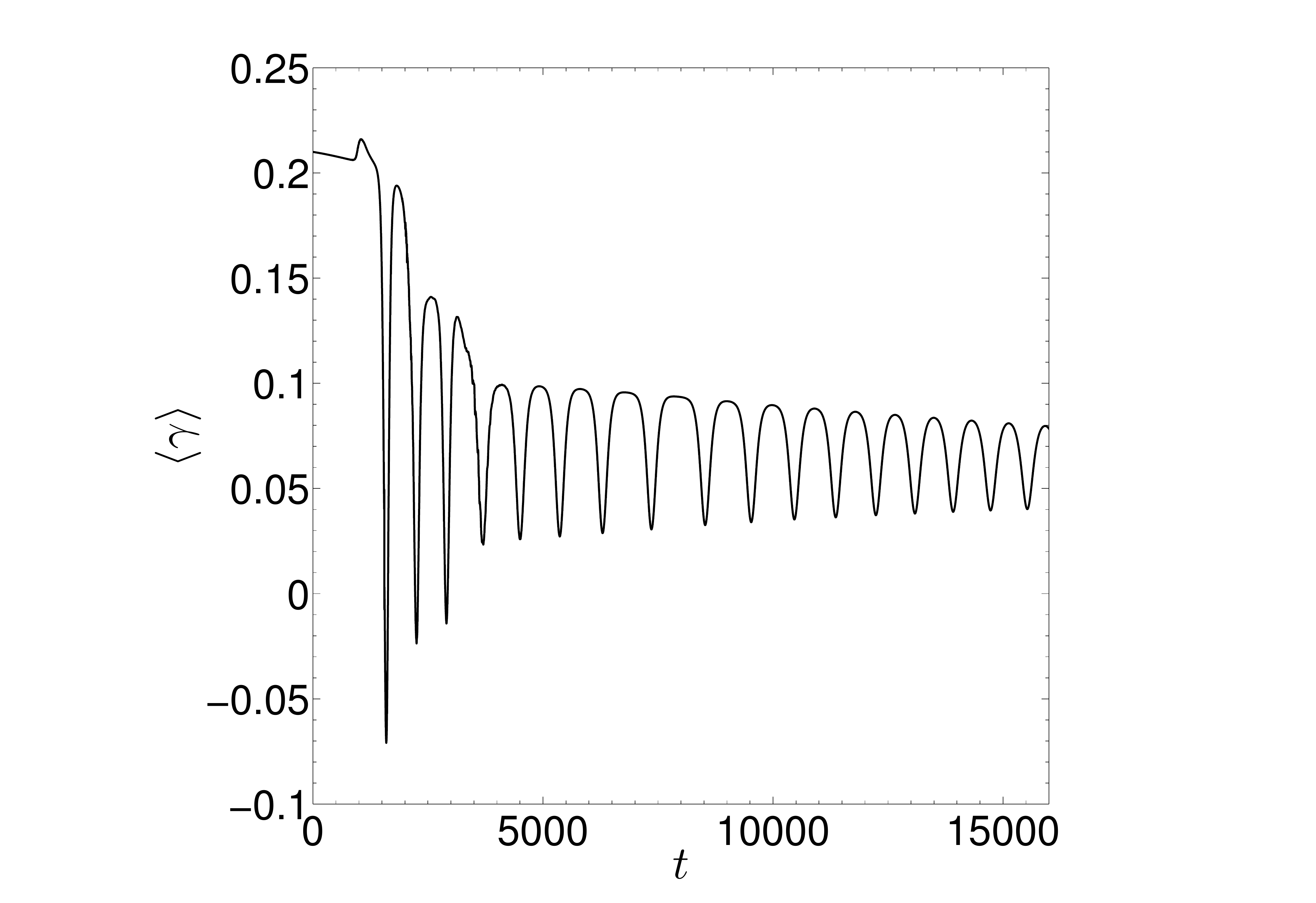} } 
\subfigure{\includegraphics[trim=5cm 0cm 7cm 1cm, clip=true,width=0.23\textwidth]{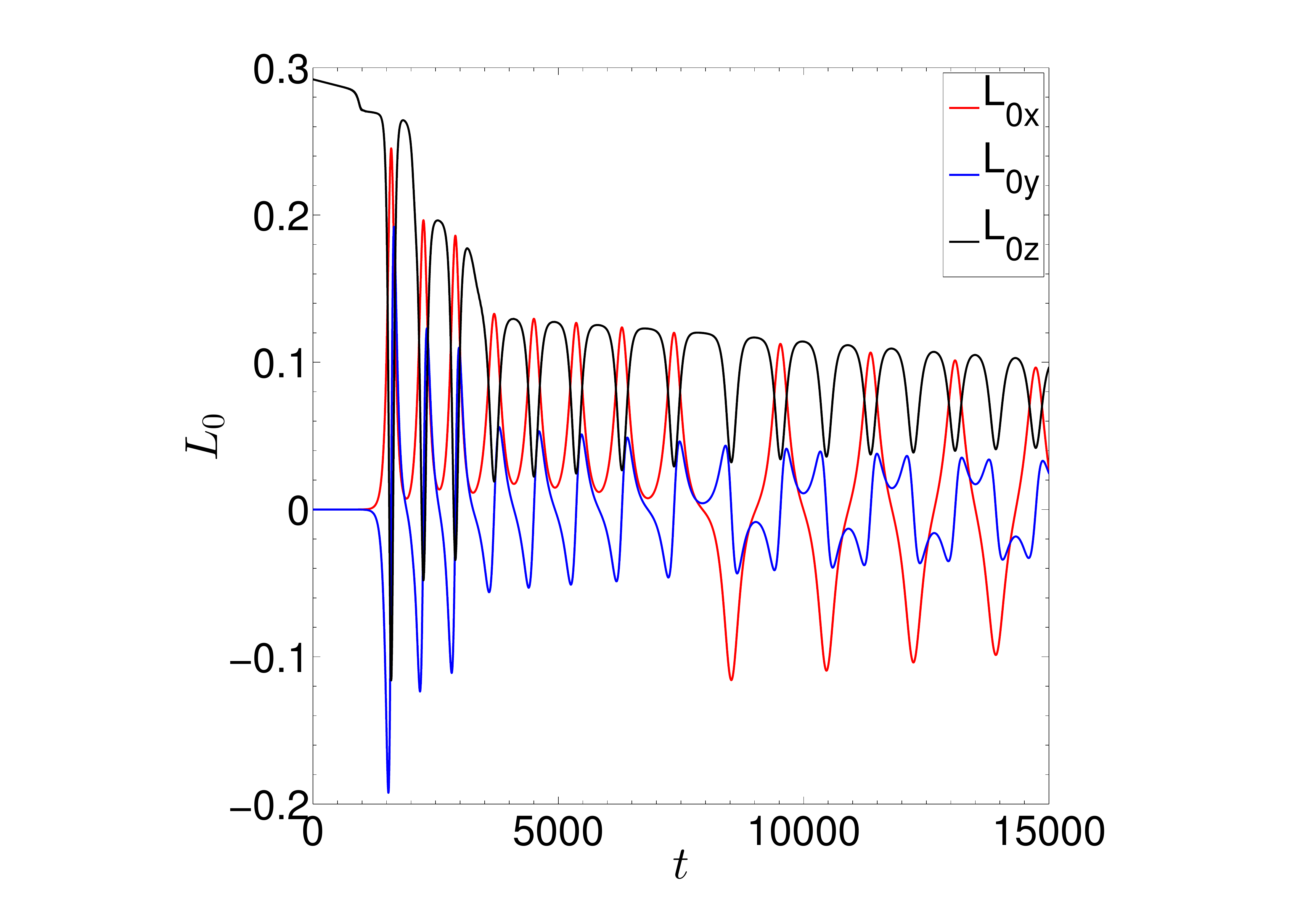} } 
\subfigure{\includegraphics[trim=5cm 0cm 8cm 1cm, clip=true,width=0.23\textwidth]{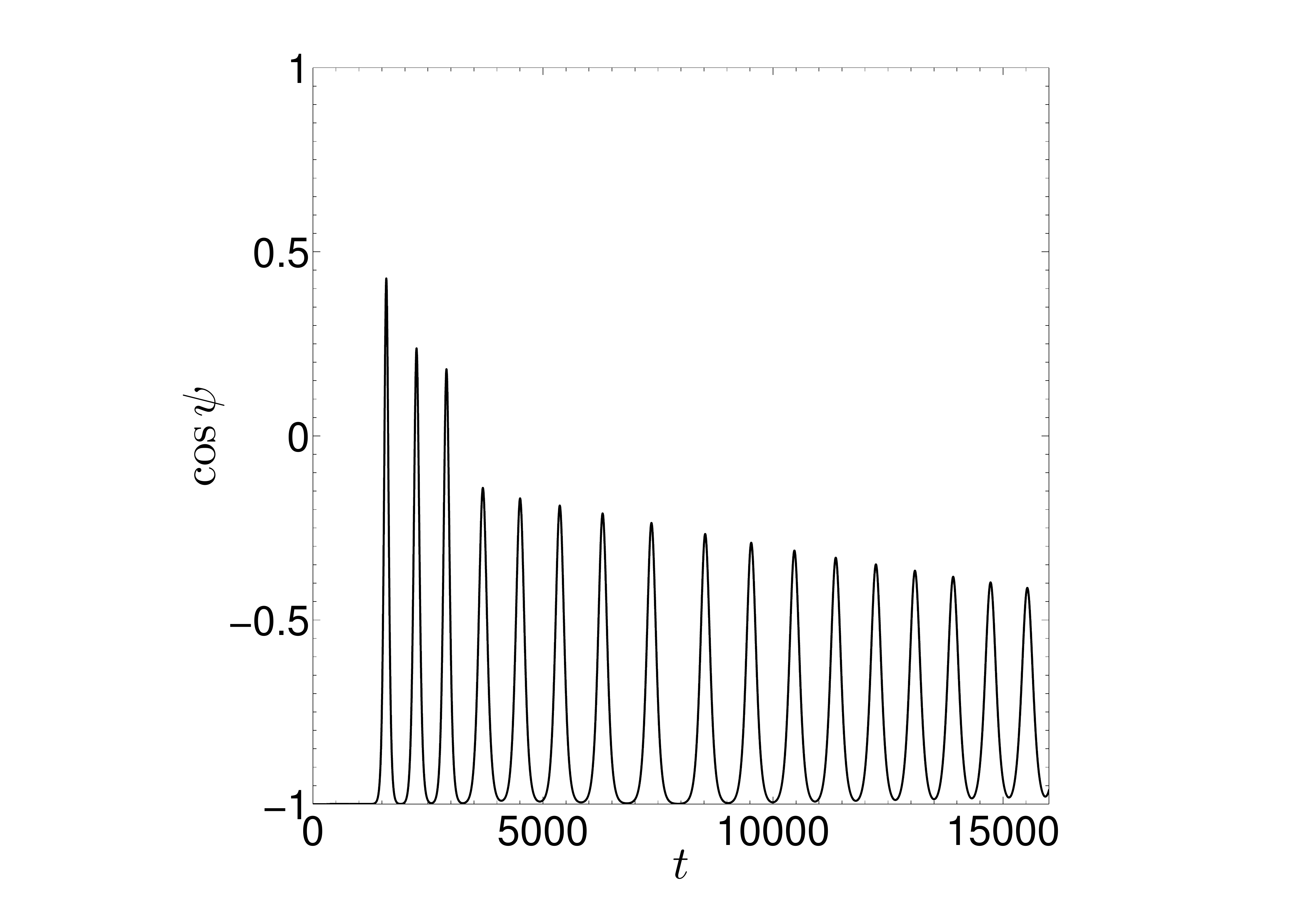} } 
  \end{center}
  \caption{Evolution of various flow quantities with time for an initially anti-aligned simulation with $\Omega=0.2,n=-0.01,A=0.1$ and $\nu=10^{-4}$. Top left: comparison of $E_z$ with the energy in the differential rotation, $E_\mathrm{dr}$. Top right: viscous dissipation rate (black line) and laminar viscous dissipation rate prediction (red dashed line). Middle left: mean asynchronism of the flow $\langle\gamma\rangle$. Middle right: Cartesian components of the angular momentum in the inertial frame. Bottom: cosine of the spin-orbit angle $\psi$. The spin-over mode is excited when $t\sim 1500$, leading to precessional motion that is gradually damped.}
 \label{15b}
\end{figure}

\begin{figure}
  \begin{center}
     \subfigure{\includegraphics[trim=0cm 0cm 0cm 0cm, clip=true,width=0.35\textwidth]{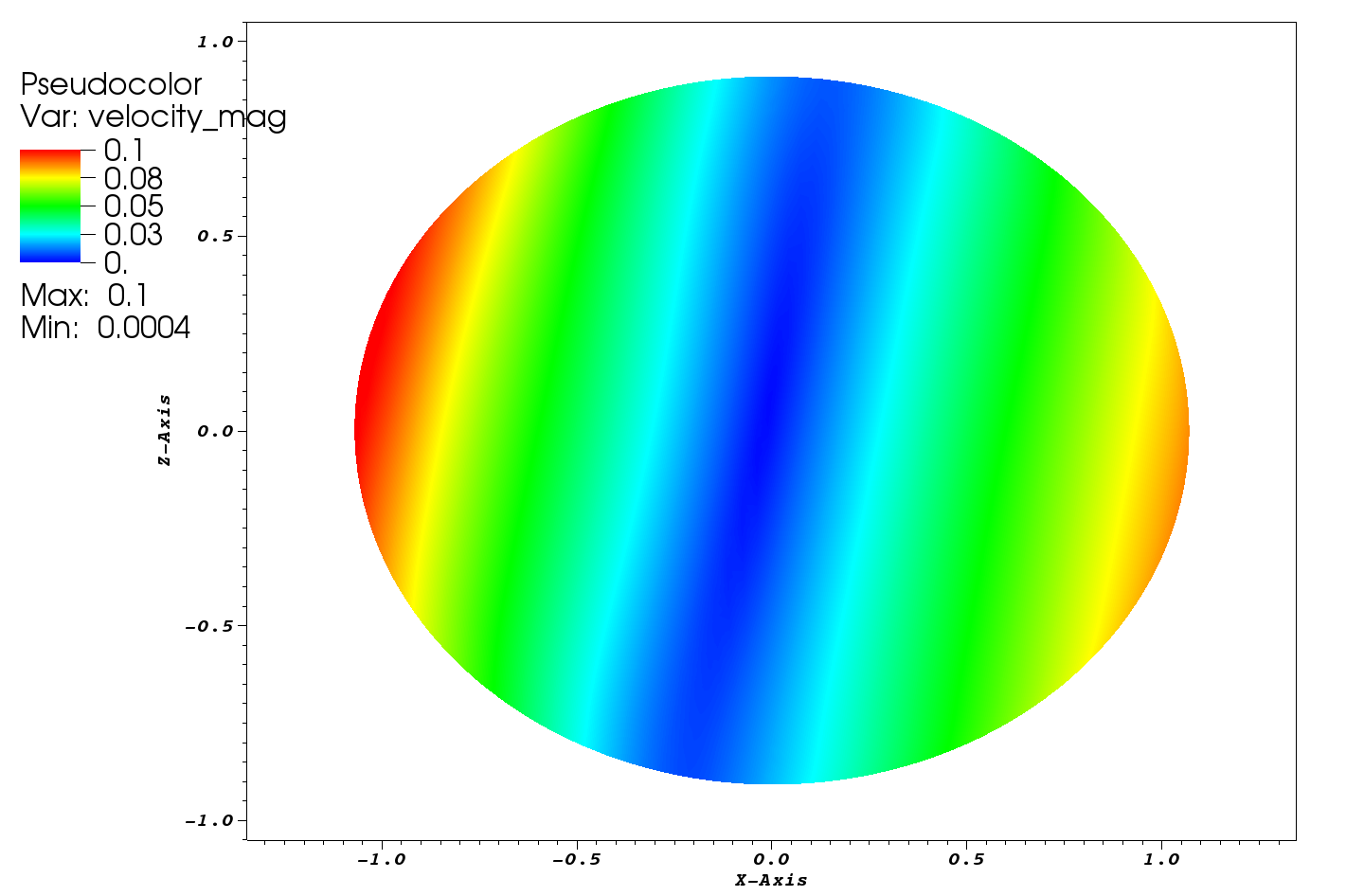} } 
       \end{center}
  \caption{Illustration of $|\boldsymbol{u}|$ on the $xz$-plane at $t=4174.5$ in the simulation with $\Omega=0.2,n=-0.01,A=0.1$, and $\nu=10^{-4}$. This illustrates the flow after the spin-over mode has been excited, when the rotation axis of the fluid precesses about $z$ due to the tidal torque.}
  \label{15bb}
\end{figure}

A second example in which the spin-over mode is excited is plotted in Fig.~\ref{15b}. This simulation has $\Omega=0.2$, $n=-0.01$, $A=0.1$ and $\nu=10^{-4}$. The initial elliptical instability again preserves the anti-alignment of the spin and orbit until $t\sim 1500$, when the spin-over mode is subsequently excited (at this time I observe $b\approx c\lesssim a$). This tilts the spin axis of the fluid, which precesses about the $z$-axis due to the (non-dissipative) tidal torque. This precessional motion is gradually damped on a similar timescale to that of the spin-synchronisation, due to a combination of laminar viscous dissipation (which explains the damping from $t\approx5000$ until $t\approx10000$), in combination with additional instabilities of this precessional flow. I plot $|\boldsymbol{u}|$ for the spin-over mode at $t=4174.5$ in Fig.~\ref{15bb}. In this simulation, the damping of the precessional flow drives evolution of the spin-orbit angle towards anti-alignment. This evolution could have been predicted \citep{Lai2012} -- however, if the simulation was run for longer, we would expect it to evolve towards alignment as a result of the laminar viscous tidal torque.

In both simulations, the spin-over mode is excited when $b\lesssim c\lesssim a$, corresponding with an instability when the axis of rotation is the middle axis, as expected \citep{Kerswell1994}. Similar spin-orbit alignment is observed in cases with a rigid outer boundary (see Fig.~\ref{15c}). The spin-over mode is not observed in cases with an aligned spin and orbit, as predicted by the global stability analysis \citep{Barker2015a}, which I have confirmed in the simulations performed for this work (discussed in \S \ref{Nonlinear}). The excitation of the spin-over mode, and the gradual damping of the resulting precessional motion, is not captured in the simplest models of tidal dissipation, such as the constant time-lag model \citep{Hut1981,Eggleton1998,ML2002,Barker2009}. However, it can be captured by considering the different components of the tidal response to be damped at different rates (e.g.~\citealt{Lai2012,Ogilvie2014}). In particular, if the $\ell=2, |m|=1, \omega=-\Omega$ component of the tidal response is damped at a rate that is enhanced by an $O(1)$ factor over that of the other components. However, I do not find evidence that this component could be damped \textit{much} more efficiently than the basic tidal flow (which would be required in order to explain tidal spin-orbit alignment for stars hosting hot Jupiters without planetary inspiral, for example).

Whether the spin-over mode would be exited in reality by the elliptical instability inside a planet is unclear. Hot Jupiters are likely to begin their lives rapidly rotating, to be subsequently spun-down by tides. Basing my intuition on Eqs.~\ref{shape}, I expect the body to be oblate ($c<b$), where this mode would not be excited if the spin is aligned or anti-aligned with its orbit. However, this mode could potentially be excited if the planet is slowly rotating (retrogradely with respect to the orbit), but it is then moved to a very-short period ($P\ll 1$ d) orbit so that it has $b<c$, where this instability could operate. In addition, this instability could be excited when the spin and orbit is already significantly misaligned.

For a different application, it does not seem likely that this mode could be excited by the elliptical instability inside a solar-type star hosting a short-period planet on a retrograde orbit (cf.~\citealt{Cebron2013}). This is because the tidal deformation is likely to be much too weak to allow $b<c$ for realistic stellar rotation rates. Nevertheless, further work to study the excitation of this mode and to understand the damping of precessional motions in planets and stars more generally would certainly be worthwhile (e.g.~\citealt{PapPringle1982}).

\section{Discussion}
\label{discussionresults}

The global nonlinear evolution of the elliptical instability shares many properties with its local Cartesian counterpart \citep{BL2013,BL2014}. The instability in both cases leads to ``bursty", cyclic behaviour associated with the formation of coherent structures in the flow, e.g., we can compare the top panel of Fig.~\ref{2} with the same panel in Fig.~4 of \cite{BL2013}. In the local model, this cyclic behaviour was related to the formation of vertically-aligned columnar vortices, which subsequently inhibited energy injection into the flow by the elliptical instability until these vortices had been sufficiently damped by viscosity. In this work I have found zonal flows to play an analogous role to the columnar vortices obtained in the local model. (A difference to note in the setup is that the local model treated the tidal flow as a fixed background flow, whereas here I allow it to evolve self-consistently.)

\begin{figure}
  \begin{center}
      \subfigure{\includegraphics[trim=5cm 0cm 5cm 0cm, clip=true,width=0.46\textwidth]{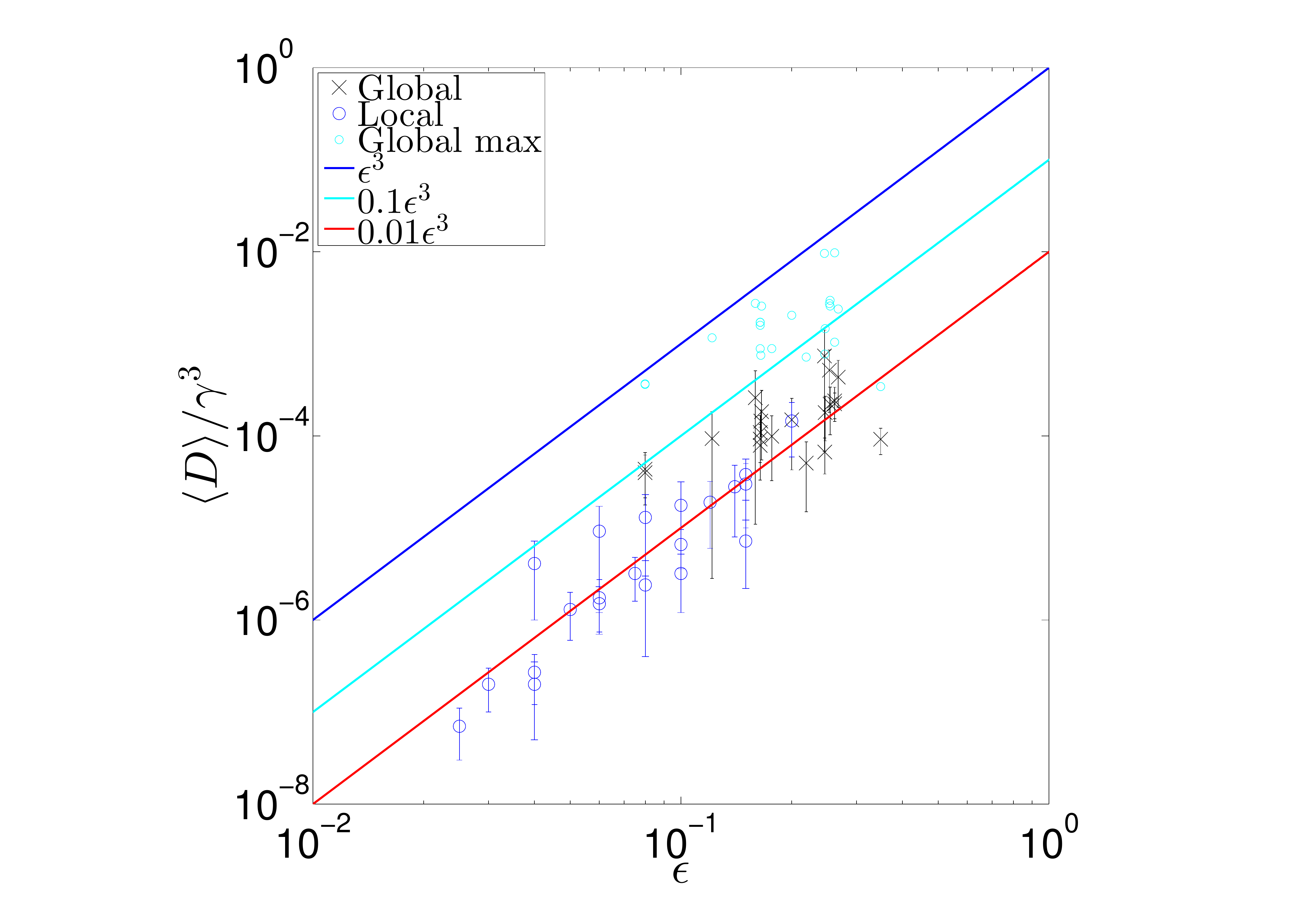} }
      \subfigure{\includegraphics[trim=5cm 0cm 5cm 0cm, clip=true,width=0.46\textwidth]{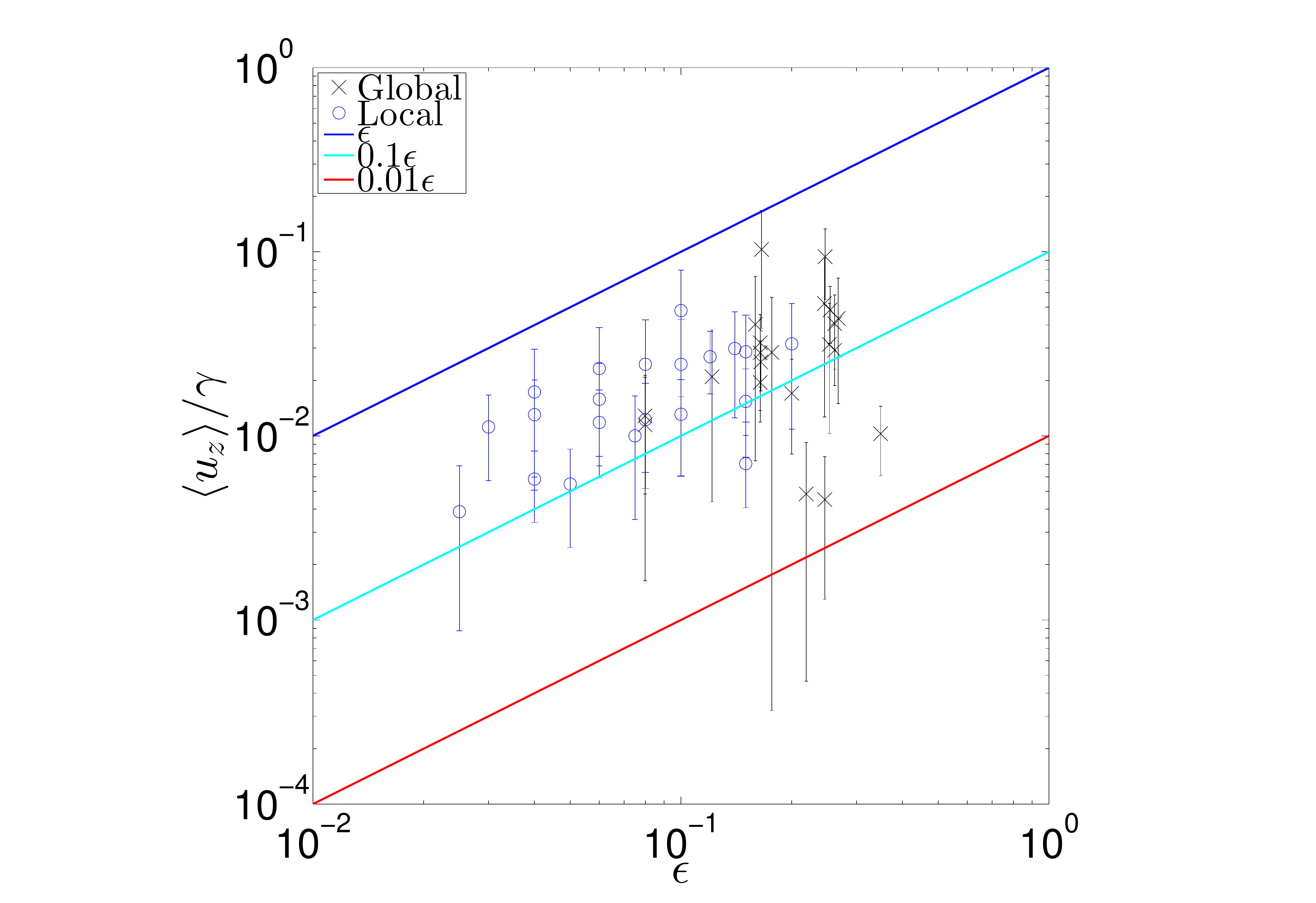} }
       \end{center}
  \caption{Collection of results plotting mean dissipation and mean RMS vertical velocity (normalised by the appropriate power of $\gamma$) against $\epsilon$ for the global simulations with a free surface (black crosses), with RMS errors identified by the error bars. Also added are the results from the local model. The magnitudes of these quantities are consistent with the scalings from the local model. The maximum dissipation is somewhat larger than the mean dissipation, but it is only attained for a short duration. This demonstrates that the RMS turbulent velocity is consistent with being $O(\epsilon)$ and the turbulent dissipation with being $O(\epsilon^3)$ in the limit of small $\epsilon$ (see \S~\ref{astrophysical} for an alternative possible scaling).}
  \label{dissplot}
\end{figure}

In \cite{BL2013}, we provided crude arguments to estimate the turbulent dissipation resulting from the elliptical instability. We assumed that an unstable mode (with a velocity amplitude $u$ and a wavelength $\lambda$) grows until secondary shear instabilities (with growth rates scaling with $u/\lambda$) become strong enough to prevent further amplitude growth (when the primary elliptical instability growth rate $\sigma \sim u/\lambda$), leading to a saturated velocity amplitude that can be written
\begin{eqnarray}
\label{umlt}
u\sim u_z\equiv C \lambda \sigma, 
\end{eqnarray}
where $C$ is assumed to be independent of $\epsilon$ and $\nu$. If this is the case, we have a turbulent cascade with an ``outer scale" given by $\lambda$, and turbulent dissipation $D\sim u^3/\lambda\sim \sigma^3\lambda^2$. Given that $\sigma \sim \epsilon\gamma$, we can write
\begin{eqnarray}
\label{dmlt}
D \equiv \chi \epsilon^3 \gamma^3 \lambda^2,
\end{eqnarray}
where $\chi$ is an efficiency factor (which is assumed to be independent of $\epsilon$ and $\nu$, though may vary to some extent with $\Omega$ and $n$). For simplicity, I will also consider $\lambda=R_p$, the size of the equivalent spherical planet, so that the wavelength of the dominant scale is taken into account with $\chi$. In \S~\ref{astrophysical} we discuss the possibility that $\lambda$ may in fact depend on $\epsilon\gamma$ when $\epsilon\gamma\ll1$.

In the local model (with weak magnetic fields; \citealt{BL2014}), we provided evidence to support Eqs.~\ref{umlt} \& \ref{dmlt}, with $\chi\sim 0.01$ and $C\sim 0.2$ (which appeared to be approximately independent of viscosity, where $\lambda$ was taken to be the size of the box), at least over the limited range of $\epsilon$ (and $\nu$) that we could study numerically. To conclusively confirm or refute the scalings of Eqs.~\ref{umlt} \& \ref{dmlt} would require global simulations over a much wider range of $\epsilon$ than I have considered here. This is not possible with current computational resources. Instead, I will compare the global results with the scalings obtained from the local model. While I have not observed sustained turbulence in all of the global simulations (unlike the local model with magnetic fields), it is nevertheless worthwhile to verify whether the results are consistent with these scalings for the computed parameter range, given that my ultimate aim is to determine the astrophysical importance of the instability. (However, it should be noted that the global simulations have not been demonstrated to exhibit mean dissipation rates that are independent of the viscosity -- unlike the local simulations, at least as far as we can probe this numerically.)

I compute a time average of $D$ and (RMS) $\langle u_z\rangle $ over each simulation and plot these quantities in Fig.~\ref{dissplot}, after normalising by the appropriate power of $\gamma$. Several caveats that should be kept in mind is that I do not remove the viscous decay of the basic flow (which can be substantial in some cases, as listed at $t=0$ in Table \ref{table2}), and I also include the whole simulation (including the linear growth phase; except for the laminar phases at late times in cases where the turbulence does not persist). Nevertheless, results presented here are representative of those obtained in the whole simulation. In the top panel I also plot the maximum value of $D$, to illustrate an absolute upper bound on the dissipation obtained in the simulations (however, these values are only attained for a very short time interval).

The top two panels of Fig.~\ref{dissplot} illustrate that the mean dissipation and mean RMS vertical velocity obtained in the global model are roughly consistent with the scalings that describe the local results for a given $\epsilon$. While there is significant scatter (due to differences in $\Omega,n,\nu$), these results are broadly consistent (to within an $O(1)$ factor) if $\chi\sim 0.01-0.1$ and $C\sim 0.1$ over this range of $\epsilon$  (the peak dissipation is somewhat stronger, but is attained only for a short duration). This demonstrates that the dissipation and turbulent velocities are quantitatively similar to those obtained from the local model over the observed range of $\epsilon$. Note that $\chi$ is controlled by which mode is driven unstable, which depends on $n$ and $\Omega$, in addition to $\epsilon$, which may explain some of the scatter. Fig.~\ref{dissplot} suggests that we can consider $\chi \lesssim 0.1$ to provide an upper limit for the mean dissipation resulting from the elliptical instability.

In the local model, magnetic stresses were found to significantly modify the hydrodynamical evolution by preventing the maintenance of large-scale coherent vortices, thereby allowing sustained (as opposed to cyclic) turbulence \citep{BL2014}. In addition, the elliptical instability was found to drive a small-scale dynamo. Recent global simulations have shown that the elliptical instability could act as a ``system-scale" dynamo \citep{Cebron2014}. However, it remains to be seen how magnetic fields would modify the global evolution of the elliptical instability discussed in this work.

\subsection{Astrophysical implications}
\label{astrophysical}

I have simulated the elliptical instability for a range of values of $A,n$ and $\Omega$. Which values are realistic?
The shortest-period observed hot Jupiters, such as WASP-19 b \citep{Hebb2010} or WASP-121 b \citep{WASP121} have $A\sim 0.05$ and $|n|\sim 0.2$, which is at the lower end of the tidal amplitudes, and within the considered range of $|n|$, that I have simulated. Unfortunately, we do not have constraints on the rotation rates (or axes) of these planets, therefore $\Omega$ (and the sign of $n$) is not determined. Planets in wider orbits typically have smaller $A$, which I have not directly studied, requiring us to resort to scaling laws such as Eq.~\ref{dmlt} to apply  the results to these planets.

We can estimate the role of the elliptical instability for tidal circularisation and synchronisation by assuming Eq.~\ref{dmlt} to be valid for both circularisation and synchronisation -- for the synchronisation problem, this is consistent with the local model and compatible with the global simulation results, and I expect it to remain valid for the circularisation problem because the corresponding linear instability has similar properties \citep{KerswellMalkus1998}. I define the circularisation period $P_e$ to be the maximum orbital period for which an initially eccentric planetary orbit can be circularised within $1$ Gyr. The synchronisation period $P_\Omega$ is defined similarly. Using the equations listed in Appendix \ref{timescales} (based on Eqs.~24 and 25 of \citealt{BL2014}), I obtain 
\begin{eqnarray}
P_{e}\approx 2.8 \;\mathrm{d} \left(\frac{\chi}{0.1}\right)^{\frac{3}{25}}\left(\frac{m_p}{1 M_J}\right)^{\frac{2}{25}}\left(\frac{m_\star}{1 M_\odot}\right)^{-\frac{2}{25}}\left(\frac{P_{\mathrm{dyn}}}{3.6 \mathrm{hr}}\right)^{\frac{22}{25}},
\end{eqnarray}
where $P_\mathrm{dyn}=2\pi/\omega_d$. Similarly, I obtain
\begin{eqnarray}
\label{estimate2}
P_{\Omega}\approx 14.7 \;\mathrm{d} \left(\frac{\chi}{0.1}\right)^{\frac{1}{6}}\left(\frac{P_{\mathrm{dyn}}}{3.6 \mathrm{hr}}\right) \left(\frac{1 \mathrm{d}}{P_{\mathrm{rot}}}\right)^{\frac{1}{6}},
\end{eqnarray}
if the tidal period is $P_{\mathrm{rot}}/2$ (which would be appropriate if $\Omega\gg n$ -- note that we obtained a different estimate in \citealt{BL2014}, where we took the tidal period to be $P$). I have assumed the star to be solar-like and the planet to have Jupiter's mass, radius and radius of gyration for these estimates. The simulations (Fig.~\ref{dissplot}) suggest these to provide an upper limit on $P_e$ and $P_\Omega$ (note also that the estimates in \citealt{BL2014} assumed $\chi=10^{-2}$). 

These estimates are not strongly sensitive to $\chi$ (as long as $D$ scales approximately as $\epsilon^{3}$), so the scatter in Fig.~\ref{dissplot} is unlikely to change these predictions significantly. However, the radius of the planet does significantly affect these quantities, since $P_e\propto R_p^{\frac{33}{25}}$ and $P_\Omega \propto R_p^{\frac{3}{2}}$, so much greater dissipation would be expected early in the life of the system, when the planet had a larger radius, or alternatively if the planet can remain inflated. If I instead take $R_p=1.5 R_J$ for a significantly inflated (or very young) hot Jupiter, then $P_e=4.8$ d and $P_\Omega=27$ d. Note also that the timescale for spin-orbit alignment will be comparable with the spin synchronisation timescale (because the angular momentum in the orbit is typically much less than that in the planetary spin), so the spins of hot Jupiters should therefore be aligned with their orbits if their orbital periods are shorter than approximately $P_\Omega$.

I conclude that the circular orbits of hot Jupiters inside about 3 days may be explained by the elliptical instability. In addition, I predict the spin synchronisation (and spin-orbit alignment) of these planets with their orbits out to about 10-15 days. These estimates may be revised somewhat if the planet can remain inflated, or if consideration of the coupled orbital and thermal evolution of these planets can modify this picture. However, it appears necessary to invoke other mechanisms to explain tidal circularisation for longer orbital periods e.g. (linear) excitation and (linear or nonlinear) dissipation of inertial waves in a planet with a core \citep{Gio2004,GoodmanLackner2009,Ogilvie2013,FBBO2014}, or dissipation in the core itself \citep{Remus2012,Storch2014}.

If we turn to the related problem of explaining the observed circularisation and synchronisation of close binary stars \citep{Mazeh2008}, we find $P_e\approx 3.8$ d and $P_\Omega\approx 8.5$ d -- we have considered both stars to have the Sun's current mass and radius and $P_\mathrm{rot}=$ 10 d. The elliptical instability is therefore unlikely to be the primary explanation of the observed circularisation and synchronisation of close solar-type binary stars, but it could play an important role at short orbital periods.

It was pointed out by the referee that there is an alternative possible scaling for the dissipation rate when $\epsilon\gamma \ll 1$ to the one given by Eq.~\ref{dmlt} with $\lambda=R_p$. One reason is that elliptical instability occurs in frequency bands of width $O(\epsilon\gamma)$ around exact resonance. Since there are only a finite number of global modes with $\lambda\sim R_p$, the probability that a planetary-scale mode would be excited becomes very small as $\epsilon\gamma\rightarrow 0$. Resonances will always be found on small-enough scales, since the number of modes with a given maximum wavelength $\lambda$ scales as $\lambda^{-3}$. This suggests that the ``outer-scale" $\lambda$ may scale as $(\epsilon\gamma)^{\frac{1}{3}}$. If we otherwise follow the above arguments, we would obtain $D\propto (\epsilon\gamma)^{\frac{11}{3}}$, i.e., $\chi\propto \left(\epsilon\gamma\right)^{\frac{2}{3}}$. Over the range of $\epsilon$ that I have simulated such a scaling is also consistent with the local model data (with fixed $\gamma$) plotted in Fig.~\ref{dissplot} if $D\approx 0.1\epsilon^{\frac{11}{3}}$ and $ u_z\approx 0.4\epsilon^{\frac{4}{3}}$ (this is not plotted on Fig.~\ref{dissplot} because not all global simulations have the same $\gamma$). This would predict weaker dissipation (than Eq.~\ref{dmlt} with $\lambda= R_p$) when $\epsilon\gamma\ll 1$, and therefore somewhat smaller values of $P_e$ and $P_\Omega$ -- in particular I estimate that these scalings would predict $P_e\approx 2$ d and $P_{\Omega}\approx 8$ d for hot Jupiters, instead of $3$ d and $15$ d, respectively. This alternative scaling would also suggest that tidal evolution driven by the elliptical instability would become somewhat less efficient, than predicted by Eq.~\ref{dmlt} with $\lambda=R_p$, as we approach exact synchronism or circularity. The currently available data do not allow us to distinguish between $D\propto \epsilon^3$ or $D\propto \epsilon^\frac{11}{3}$ (at fixed $\gamma$). As a result, the values of $P_e$ and $P_\Omega$ quoted above should probably be regarded as upper limits.

A final point to remember is that these simulations have been forced to adopt values of the kinematic viscosity $\nu$ that were very much larger, by at least 10 orders of magnitude, than the values expected in a giant planet interior \citep{Jupiter2007}. It is hoped that these global simulations capture the dominant ``outer scales" of elliptical-instability driven turbulence, and that the mean flow quantities (such as the mean dissipation rates) are not strongly dependent on resolving much smaller scales. But whether this is true is difficult to test numerically. Further simulations that probe more deeply into the $\nu\ll 1$ regime (either using a rigid boundary or a free surface) would be worthwhile.

\section{Conclusions}

I have presented results from the first global simulations of the elliptical instability in a rotating and tidally deformed gaseous planet (or star) with a free surface. My primary motivation was to study tides inside the shortest-period hot Jupiters. The tides in these planets have large enough amplitudes that consideration of nonlinear tidal effects is likely be essential. In particular, the large-scale tidal flow in these planets is probably subject to the elliptical instability, which could play an important role in circularising, synchronising and aligning the spins of the shortest-period hot Jupiters.

The simulations were designed to study the nonlinear evolution of the elliptical instability, to determine its outcome and astrophysical relevance. I have adopted an intentionally simplified model consisting of a rotating, homogenous, and viscous fluid planet subjected to tidal gravity. In a companion paper \citep{Barker2015a}, the global modes and instabilities of such a planet were studied. 
In the simulations, I have observed the elliptical instability to produce turbulence in the planetary interior, but this is bursty, and leads to temporally-variable dissipation and synchronisation of the spin and orbit. 

Angular momentum is deposited non-uniformly throughout the planetary interior, and this leads to the development of differential rotation in the form of zonal flows. These zonal flows play an important role in the saturation of the elliptical instability, leading to bursty evolution that is reminiscent of predator-prey\footnote{The zonal flows can be thought of as the ``foxes" and the instability-driven inertial waves can be thought of as the ``rabbits".} dynamics in some cases. These zonal flows, and their interaction with the elliptical instability, may be responsible for the collapses observed in previous laboratory experiments and numerical simulations (e.g.~\citealt{Malkus1989,LeBarsReview2015}). In addition, we have previously observed similar bursty behaviour in the local model of \cite{BL2013}, but where columnar vortices played the role of zonal flows. These results highlight the ubiquity of zonal flows in tidally forced rotating planets, demonstrating that these are generated even when realistic boundary conditions are adopted.

I have demonstrated that a violent elliptical instability is observed when $\frac{n}{\Omega}\lesssim -1$, as predicted in the companion paper \citep{Barker2015a}, which is outside the frequency range in which it is usually thought to operate. This occurs for retrograde spins if the tidal amplitude is sufficiently large, so that inertial waves can be excited when not exactly in resonance. This could occur during the early stages in the life of hot Jupiters, if the planet is kicked into a very short-period orbit but possesses a retrograde spin.

I have also simulated the instability in a planet in which the surface is modelled as a rigid (but stress-free) boundary rather than a free surface. I have found qualitative and broad quantitative agreement for both the linear properties of the instability \citep{Barker2015a}, as well as its nonlinear evolution (Appendix \ref{rigidsims}). This is promising, because numerical simulations with a rigid (but stress-free) boundary are much less expensive computationally (e.g.~\citealt{Cebron2013}).

In simulations with an initially anti-aligned spin and orbit, the elliptical instability is observed to drive spin-orbit alignment. In all cases, the timescale for spin-orbit alignment is found to be similar (but not the same) as that of the spin synchronisation. In some cases, the ``spin-over" mode is excited (effectively a rigid tilting of the spin axis of the planet), which precesses due to the (non-dissipative) tidal torque. This precessional motion is gradually damped, leading to spin-orbit alignment that is not purely captured using the simplest models of tidal dissipation, such as the constant time-lag model \citep{Hut1981,Eggleton1998,ML2002,Barker2009}, where all components of the tide damp at the same rate \citep{Lai2012,Ogilvie2014}. Further work is required to study in more detail the damping of this precessional flow, and in particular to determine whether or not the alignment could occur before spin synchronisation (or planetary inspiral), which may have relevance to the spin-orbit alignment of hot Jupiter host stars (e.g.~\citealt{Albrecht2012}).

I have quantified the tidal dissipation resulting from the elliptical instability, and I suggest that it could explain the circular orbits of the shortest-period hot Jupiters inside about 3 days. However, it seems necessary to invoke other mechanisms to explain tidal circularisation for longer orbital periods (e.g.~linear excitation of inertial waves in a planet with a core, or dissipation in the core itself). I also predict the spin synchronisation and spin-orbit alignment of hot Jupiters with orbital periods shorter than about 10 (or perhaps 15) days as a result of this mechanism.

Future work is required to adopt more realistic interior models, including the presence of an inner core, as well as realistic density and entropy profiles, in addition to the possible presence of magnetic fields. It would also be worthwhile to probe more deeply into the regime of small viscosities, allowing even smaller values of $\epsilon$ to be simulated -- perhaps over a sufficient range to allow us to distinguish between the two possible scaling laws for the dissipation that are consistent with the data in \S~\ref{astrophysical}. Finally, spin-orbit alignment should be studied for more general (rather than initially anti-aligned) configurations, also taking into account the evolution of the orbit.

\section*{Acknowledgements}
I would like to thank Paul Fischer for the help he provided with Nek5000 during the early stages of this project, and for providing the spherical mesh that was used in this work. I would like to thank Harry Braviner, Benjamin Favier, Yufeng Lin and Gordon Ogilvie for discussions 
at various stages in the project, Jeremy Goodman for a useful and thought-provoking referee report, and Pavel Ivanov for helpful comments. This work was supported by the Leverhulme Trust and Isaac Newton Trust through the award of an Early Career Fellowship, but the early stages were supported by STFC through grants ST/J001570/1 and ST/L000636/1. Some of the simulations reported here used the DiRAC Complexity system, operated by the University of Leicester IT Services, which forms part of the STFC DiRAC HPC Facility (www.dirac.ac.uk). This equipment is funded by BIS National E-Infrastructure capital grant ST/K000373/1 and  STFC DiRAC Operations grant ST/K0003259/1. DiRAC is part of the National E-Infrastructure.

\appendix

\section{Table of simulations}
\label{AppendixTable}
A list of the simulations performed for this work (and presented in Fig.~\ref{dissplot}) is given in Table \ref{table2}.
\begin{table*}
{\renewcommand{\arraystretch}{1}}
\setlength{\tabcolsep}{0.1cm}
\begin{tabular}{c c c c c c | c c c | c  }
\hline
$\Omega$ & $n$ & $\gamma$ & $A$ & $-\log_{10}\nu$ & $N$ & $\langle D/\gamma^3\rangle$ & $\langle u_z/\gamma\rangle$ & $D_\mathrm{lam}(t=0)$ & Comments \\
  \hline
  0.1 & -0.01 & 0.11 & 0.15 & 4.52 & 10 & $1.79\times 10^{-4}$ & $9.4\times 10^{-2}$ & $9.40\times 10^{-8}$ & spin-over $u_z$ \\
  0.1 & 0.01 & 0.09 & 0.1 & 4.52 & 10 & $2.60\times 10^{-4}$ & $4.03\times 10^{-2}$ & $2.53\times 10^{-8}$ & \\
  0.1 & 0.01 & 0.09 & 0.15 & 4.52 & 10 & $7.39\times 10^{-4}$ & $5.24\times 10^{-2}$ & $6.23\times 10^{-8}$ & \\
  0.2 & -0.4 & 0.6 & 0.1 & 4 & 10 & $9.18\times 10^{-5}$ & $1.03\times 10^{-2}$ & $2.0\times 10^{-5}$ & \\
  0.2 & -0.3 & 0.5 & 0.1 & 4 & 10 & $6.67\times 10^{-5}$ & $4.5\times 10^{-3}$ & $6.44\times 10^{-6}$ & \\
  0.2 & -0.25 & 0.45 & 0.1 & 4 & 10 & $5.06\times 10^{-5}$ & $4.8\times 10^{-3}$ & $4.08\times 10^{-6}$ & \\
  0.2 & -0.2 & 0.4 & 0.1 & 4 & 10 & $1.49\times 10^{-4}$ & $1.7\times 10^{-2}$ & $2.67\times 10^{-6}$ & \\
  0.2 & -0.01 & 0.21 & 0.1 & 4 & 8 & $1.83\times 10^{-4}$ & $1.03\times 10^{-1}$ & $4.97\times 10^{-7}$ & spin-over $u_z$ \\
  0.2 & 0 & 0.2 & 0.1 & 4 & 8 & $1.45\times 10^{-4}$ & $2.52\times 10^{-2}$ & $4.47\times 10^{-7}$ & \\
  0.2 & 0 & 0.2 & 0.15 & 4 & 10 & $2.21\times 10^{-4}$ & $4.84\times 10^{-2}$ & $1.11\times 10^{-6}$ & \\
  0.2 & 0.01 & 0.19 & 0.05 & 4.52 & 10 & $3.98\times 10^{-5}$ & $1.28\times 10^{-2}$ & $2.78\times 10^{-8}$ & \\
  0.2 & 0.01 & 0.19 & 0.05 & 4.70 & 10 & $4.35\times 10^{-5}$ & $1.15\times 10^{-2}$ & $1.86\times 10^{-8}$ & \\
  0.2 & 0.01 & 0.19 & 0.075 & 4.70 & 10 & $9.34\times 10^{-5}$ & $2.1\times 10^{-2}$ & $4.32\times 10^{-8}$ & \\
  0.2 & 0.01 & 0.19 & 0.1 & 4.52 & 10 & $7.87\times 10^{-5}$ & $2.83\times 10^{-2}$ & $1.20\times 10^{-7}$ & \\
  0.2 & 0.01 & 0.19 & 0.1 & 4 & 10 & $9.23\times 10^{-5}$ & $1.95\times 10^{-2}$ & $4.0\times 10^{-7}$ & \\
  0.2 & 0.01 & 0.19 & 0.1 & 4.70 & 10 & $1.09\times 10^{-4}$ & $3.2\times 10^{-2}$ & $8.0\times 10^{-8}$ & \\
  0.2 & 0.01 & 0.19 & 0.15 & 4 & 8 & $5.17\times 10^{-4}$ & $3.14\times 10^{-2}$ & $9.89\times 10^{-7}$ & \\
  0.3 & 0.01 & 0.29 & 0.15 & 3.52 & 8 & $4.35\times 10^{-4}$ & $4.34\times 10^{-2}$ & $7.78\times 10^{-6}$ & \\
  0.3 & 0.05 & 0.25 & 0.15 & 4 & 10 & $2.40\times 10^{-4}$ & $4.07\times 10^{-2}$ & $1.84\times 10^{-6}$ & \\
  0.3 & 0.05 & 0.25 & 0.15 & 4 & 8 & $2.23\times 10^{-4}$ & $2.92\times 10^{-2}$ & $1.84\times 10^{-6}$ & \\
  0.4 & 0.1 & 0.3 & 0.15 & 4 & 10 & $9.9\times 10^{-5}$ & $2.84\times 10^{-2}$ & $1.16\times 10^{-6}$ & \\
\end{tabular}
\caption{Table of free surface simulation results. $N+1$ is the resolution, i.e., the polynomial order within each element. $D_\mathrm{lam}(t=0)$ is the numerically computed viscous dissipation rate at the beginning of the simulation, which is always found to accurately agree with Eq.~\ref{disspred}. Note that the case mentioned ``spin-over $u_z$" presents $\langle u_z/\gamma\rangle$ even though the flow is not turbulent during later stages, when it is dominated by the spin-over mode.}
\label{table2}
\end{table*}

\section{Comparison simulations in a rigid ellipsoidal container}
\label{rigidsims}

\begin{figure}
  \begin{center}
 \subfigure{\includegraphics[trim=5.5cm 1cm 5cm 1cm, clip=true,width=0.225\textwidth]{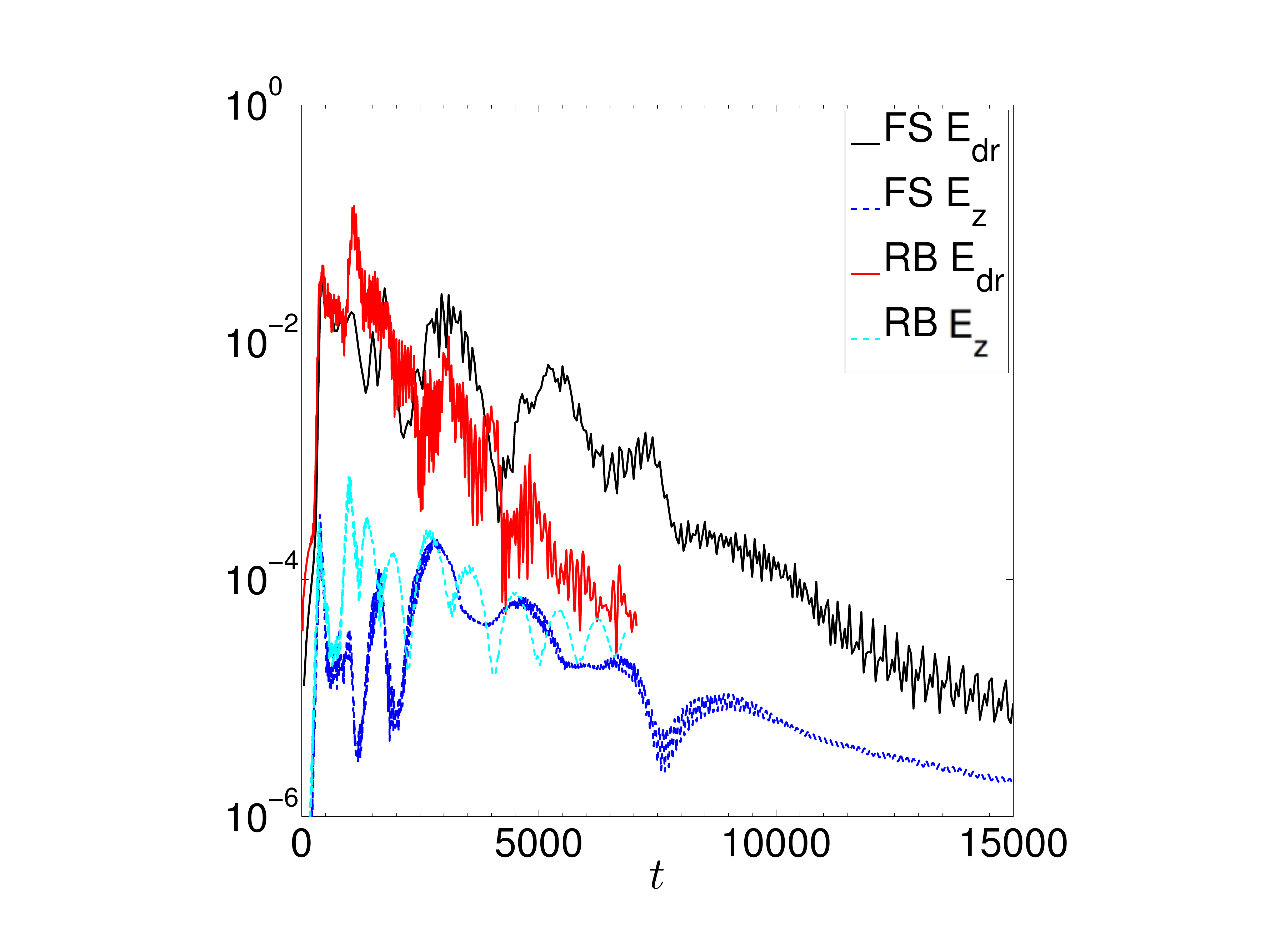} } 
  \subfigure{\includegraphics[trim=5.5cm 0cm 7cm 1cm, clip=true,width=0.23\textwidth]{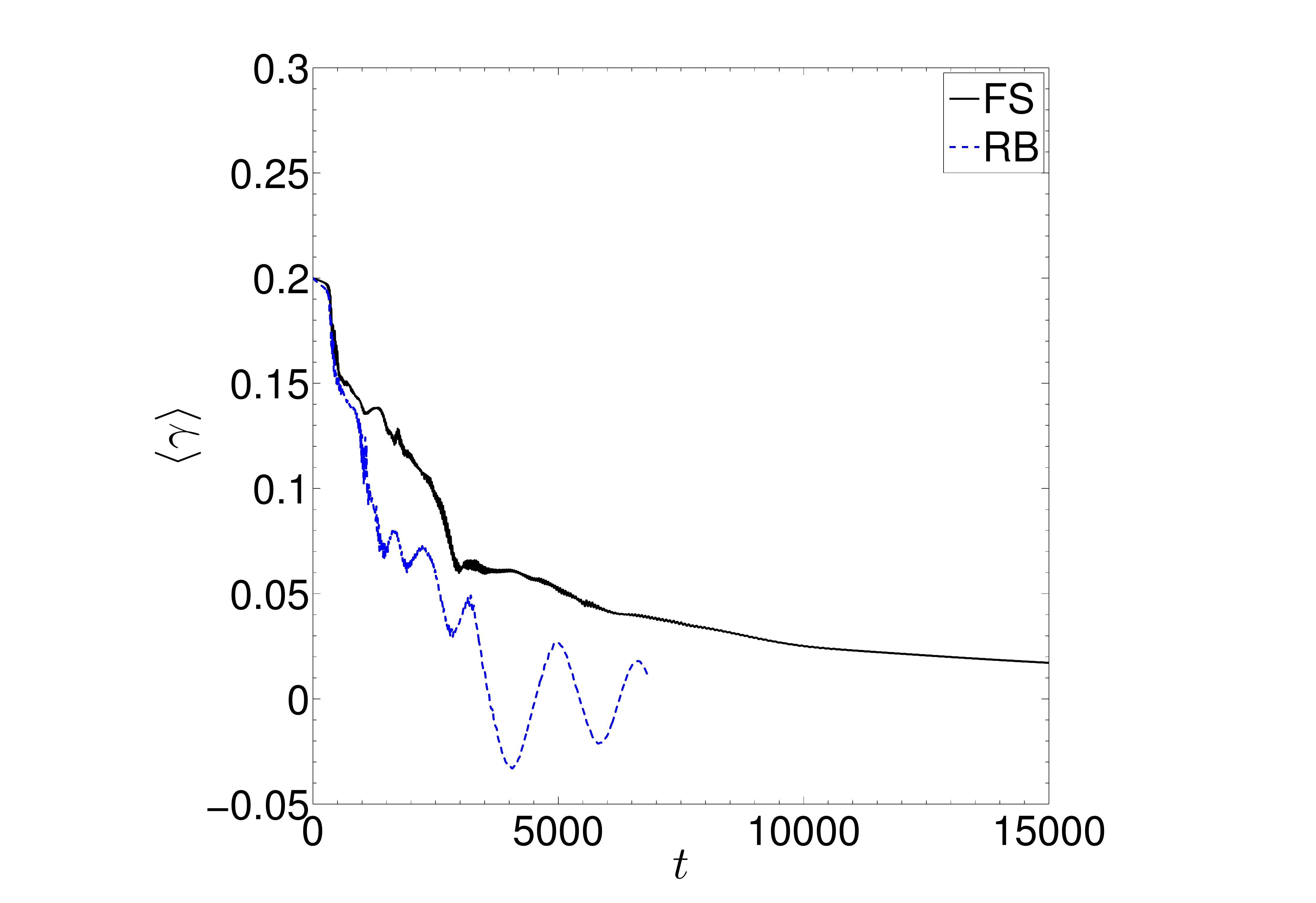} } 
   \end{center}
  \caption{Comparison of simulations with a free surface and a rigid boundary for a pair of simulations with $\Omega=0.2,n=0,A=0.15$ and $\nu=10^{-4}$. Left: mean asynchronism of the flow $\langle\gamma\rangle$ for free surface (black solid line) and rigid boundary (blue dashed line cases. Right: comparison of RMS $u_z$ with the energy in the differential rotation, $E_\mathrm{dr}$ for free surface and rigid boundary cases. This demonstrates the qualitative and broad quantitative agreement of rigid boundary and free surface simulations.}
  \label{6}
\end{figure}

\begin{figure}
  \begin{center}
 \subfigure{\includegraphics[trim=6cm 3cm 6cm 0cm, clip=true,width=0.22\textwidth]{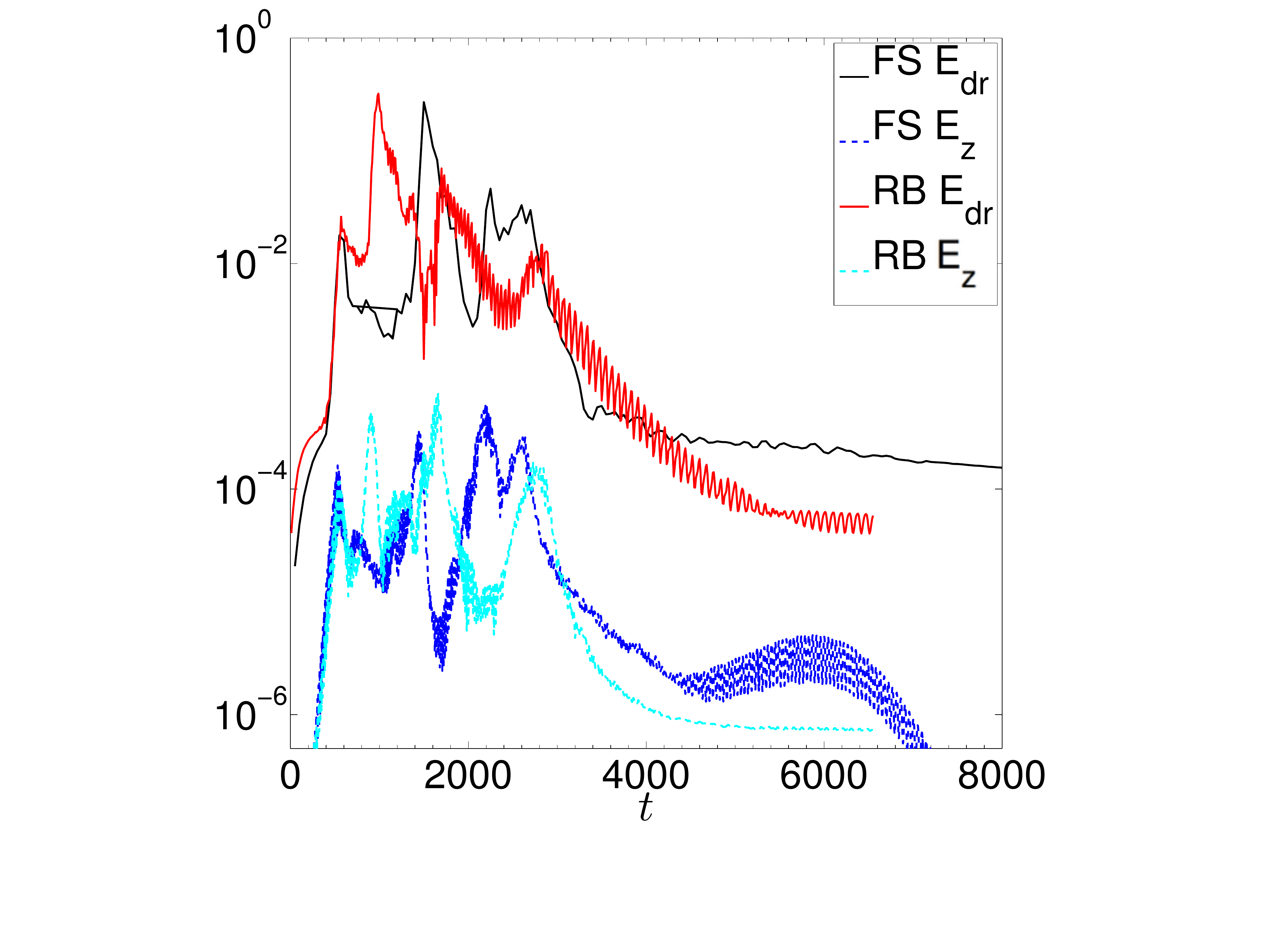} } 
  \subfigure{\includegraphics[trim=6cm 0cm 7cm 1cm, clip=true,width=0.23\textwidth]{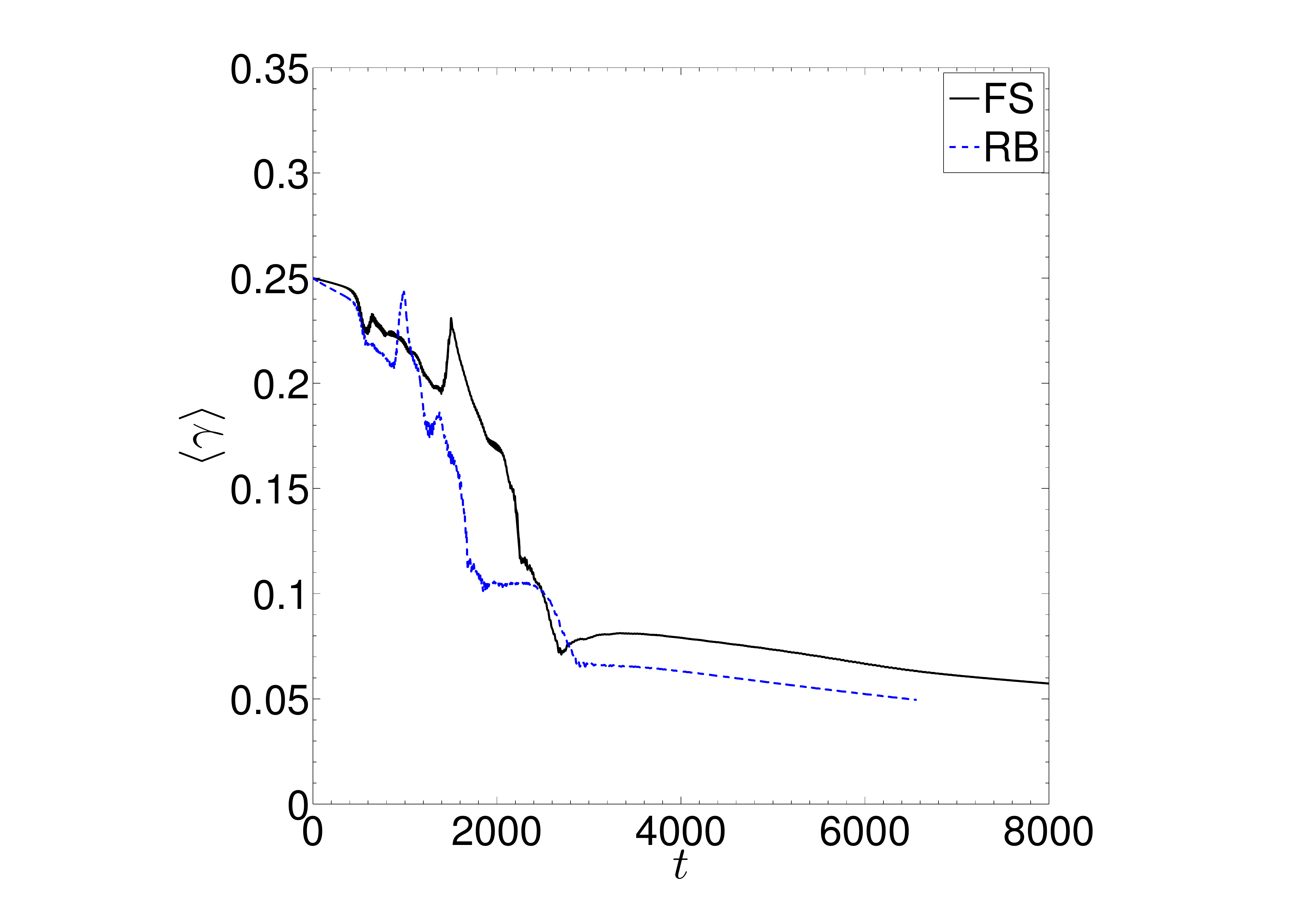} } 
   \end{center}
  \caption{Same as Fig.~\ref{6} but for a pair of simulations with $\Omega=0.3,n=0.05, A=0.15$ and $\nu=10^{-4}$.}
  \label{7}
\end{figure}

\begin{figure}
 \begin{center}
     \subfigure{\includegraphics[trim=6cm 0cm 4cm 2cm, clip=true,width=0.23\textwidth]{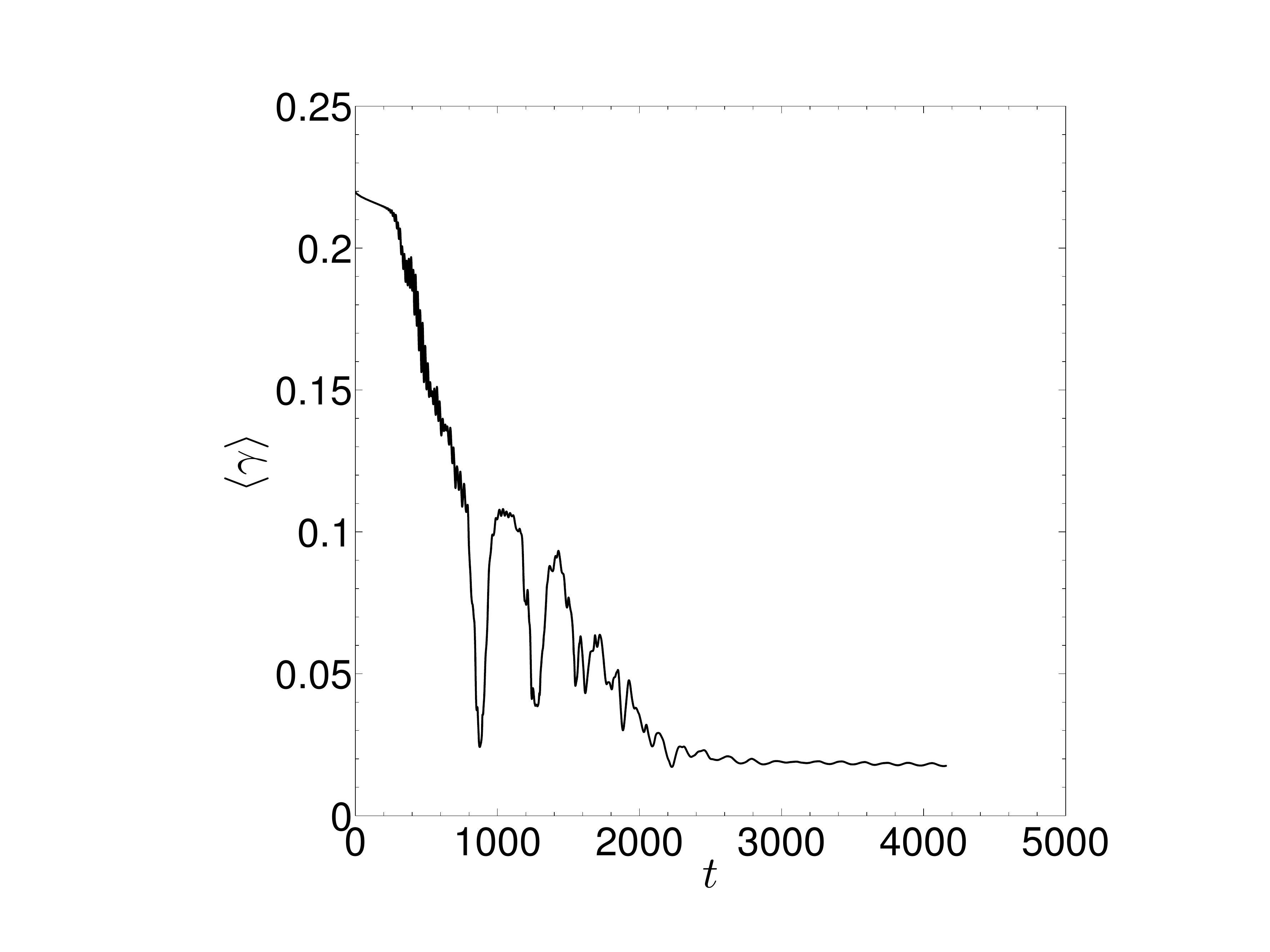} }
     \subfigure{\includegraphics[trim=6cm 0cm 7cm 1cm, clip=true,width=0.23\textwidth]{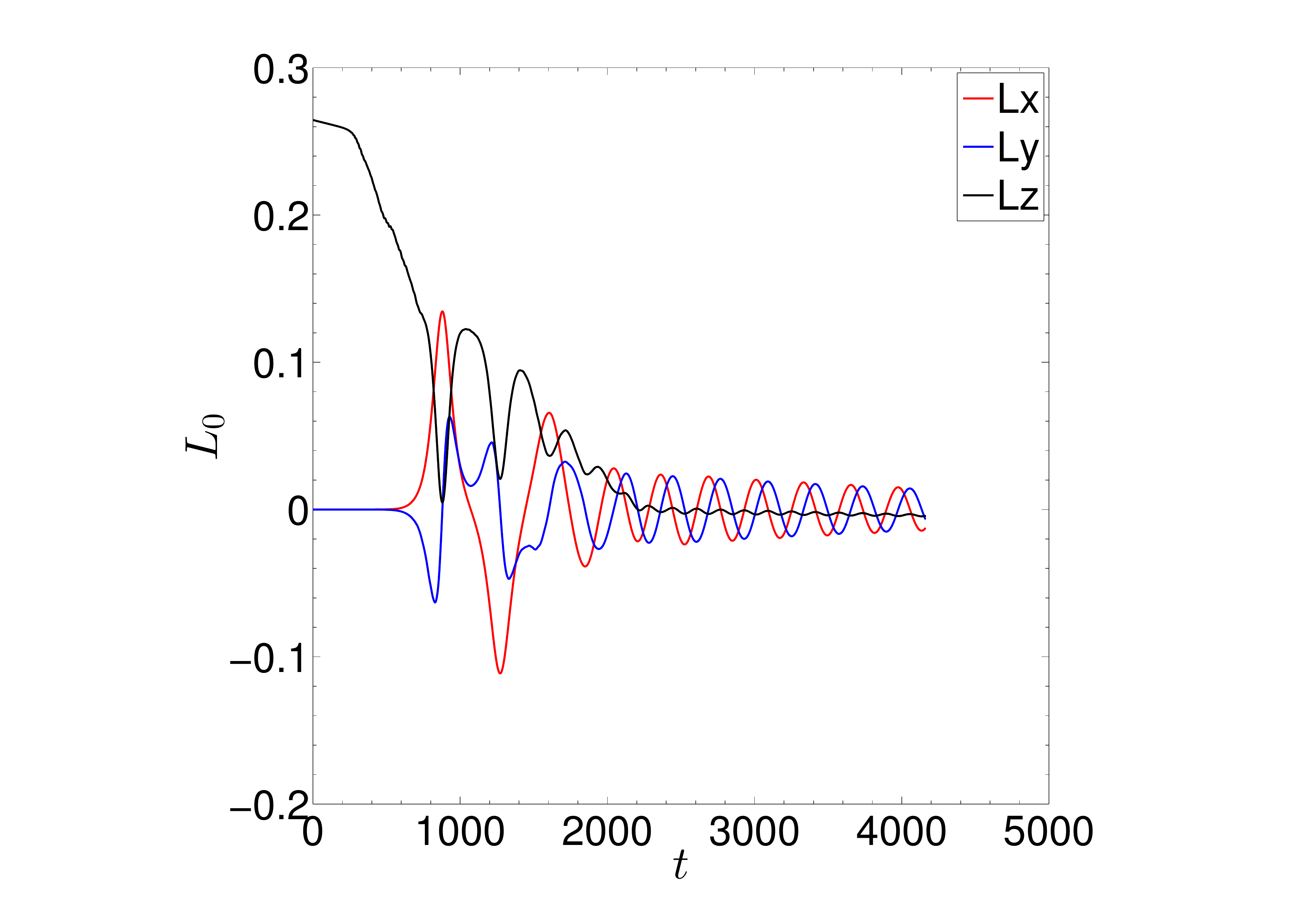} }
     \subfigure{\includegraphics[trim=6cm 0cm 7cm 1cm, clip=true,width=0.23\textwidth]{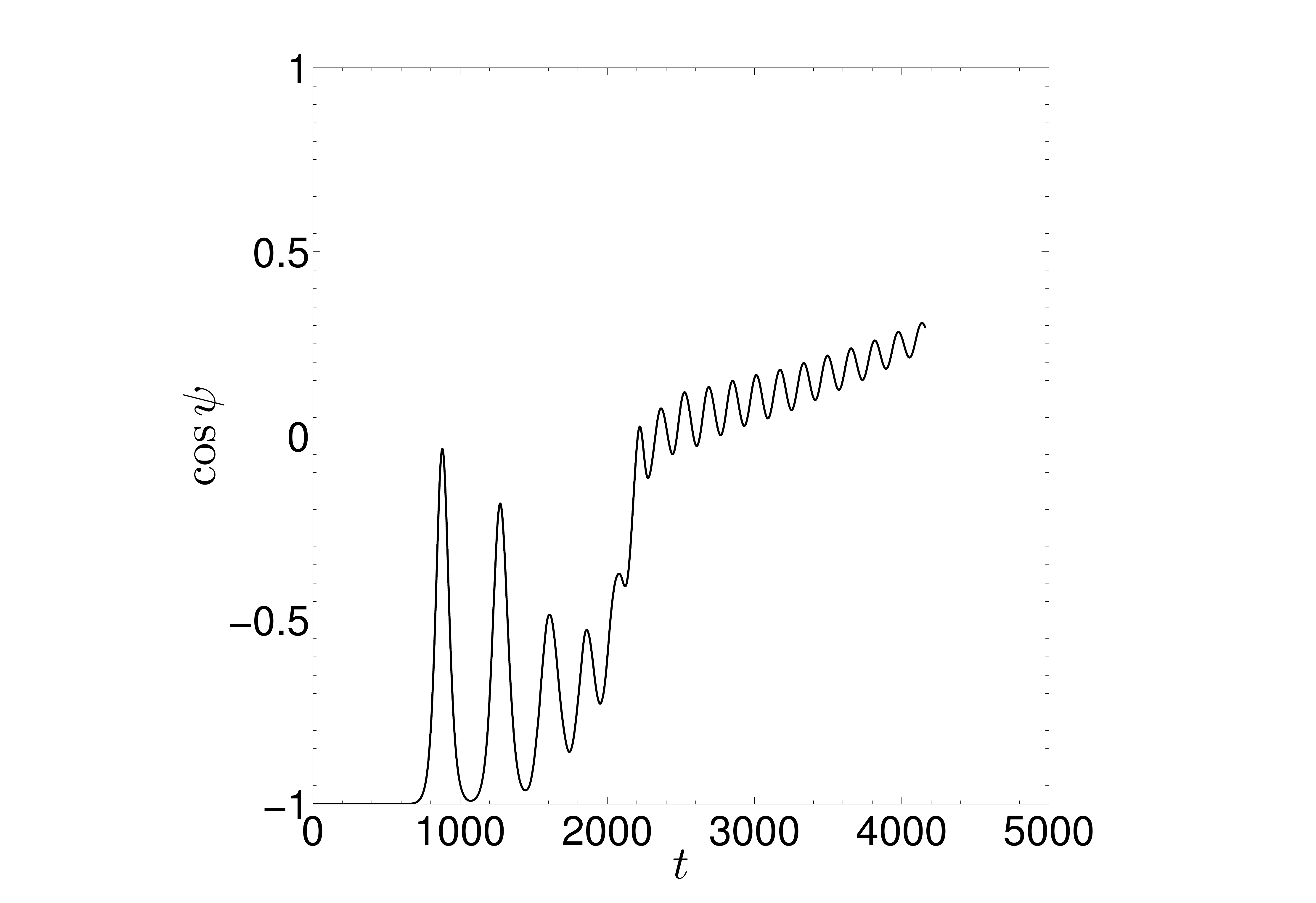}}
       \end{center}
  \caption{Evolution of various flow quantities with time for an initially anti-aligned simulation with a rigid boundary with $\Omega=0.2,n=-0.05,A=0.1$ and $\nu=10^{-4}$. Top: mean asynchronism of the flow $\langle\gamma\rangle$. Bottom left: Cartesian components of the angular momentum in the inertial frame. Bottom right: cosine of the spin-orbit angle $\psi$. This shows qualitatively similar evolution to those in \S~\ref{spinover}.}
 \label{15c}
\end{figure}

Previous laboratory experiments \citep{Lacaze2004,LeBars2007,LeBars2010} and global numerical simulations \citep{Cebron2010,Cebron2013} of the elliptical instability have adopted a rigid boundary. This is the only boundary condition that is possible in the laboratory. It it is also much simpler computationally than a free boundary, as well as being more appropriate for the cores of terrestrial planets. However, a free surface is the correct boundary condition at the surface of a giant planet or star. In this section, I briefly compare the results of global simulations with a free surface with an equivalent set of simulations that instead adopt a rigid boundary that is stress-free to represent the surface of the planet. The simulations with a rigid boundary are initialised in an identical manner to those with a free surface. 

I present a comparison of the spin evolution, as well as the $E_z$ and $E_\mathrm{dr}$ evolution, from two examples with both types of boundary condition in Figs.~\ref{6} and \ref{7}. In Fig.~\ref{6}, results are plotted from two simulations with $\Omega=0.2$, $n=0$, $A=0.15$ with $\nu=10^{-4}$, with the free surface simulation plotted as solid lines, and the rigid boundary simulation plotted as dashed lines. In Fig.~\ref{7}, both simulations have $\Omega=0.3$, $n=0.05$, $A=0.15$ with $\nu=10^{-4}$, similarly with the free surface simulation plotted as solid lines and the rigid boundary simulation plotted as dashed lines. I also plot the nonlinear evolution of a simulation with $\Omega=0.2,n=-0.05,A=0.1$ and $\nu=10^{-4}$ in Fig.~\ref{15c}, which can be qualitatively compared with those presented in \S~\ref{spinover}.

The eventual nonlinear evolution will differ to some extent between these cases because the geometry of the ellipsoid is fixed for all time in cases with a rigid boundary\footnote{In addition, the iterative solver for the pressure was only able to converge to $\sim 10^{-5}$ during each time-step in the rigid boundary simulations. I have been unable to resolve this discrepancy, which has led to some errors in the viscous dissipation as well as the flow in the polar regions. Nevertheless, the general agreement of these simulations with those that adopt a free surface is worth presenting in spite of this poor convergence.}, whereas it can evolve self-consistently as the internal flows evolve in the simulations with a free surface. Nevertheless, Figs.~\ref{6} and \ref{7} demonstrate that the nonlinear outcome of the elliptical instability in both cases is qualitatively similar, and it is also quantitatively similar in the main features of its evolution e.g. amplitude of zonal flows, rate of spin-synchronisation etc. This suggests that adopting a rigid \textit{but stress-free} ellipsoidal boundary to study the elliptical instability may not give misleading results. Future simulations with a rigid boundary could allow us to probe the outcome of the elliptical instability more deeply into the astrophysical regime of small viscosities, where simulations with a free surface would be prohibitively costly.

\section{Timescales for tidal evolution}
\label{timescales}
For reference, I collect together several expressions that have been used in \S~\ref{astrophysical} to estimate the tidal evolution timescales for short-period extrasolar planets (and close binary stars). I refer to \cite{BL2013} and assume Eq.~\ref{dmlt} is appropriate to describe the dissipation resulting from the elliptical instability, as suggested by the simulation results. The modified tidal quality factor due to elliptical instability-driven turbulence is \citep{BL2013}
\begin{eqnarray}
Q'=\frac{1}{\chi}\left(\frac{m_{\star}+m_p}{m_\star}\right)\frac{P^2P_\mathrm{tide}^2}{P_\mathrm{dyn}^4},
\end{eqnarray}
where $P_\mathrm{tide}$ is the appropriate tidal period for the problem. If we consider the circularisation problem for a spin-synchronised planet, the time to circularise its slightly eccentric orbit is
\begin{eqnarray}
\tau_e = \frac{4}{63}\frac{Q'}{2\pi}\frac{m_p}{m_\star}\left(\frac{m_p+m_\star}{m_p}\right)^{\frac{5}{3}}\frac{P^{\frac{13}{3}}}{P_\mathrm{dyn}^{\frac{10}{3}}},
\end{eqnarray}
where $P_\mathrm{tide}=P$. In a given timescale $\tau_e$ we would predict a planet to have a circular orbit if its orbital period is shorter than
\begin{eqnarray}
P=\left(\chi \tau_e \frac{126\pi}{4}\frac{m_\star^2 m_p^{\frac{2}{3}}}{(m_p+m_\star)^{\frac{8}{3}}}P_\mathrm{dyn}^{\frac{22}{3}}\right)^{\frac{3}{25}},
\end{eqnarray}
and we define $P\equiv P_e$ if $\tau_e=1$ Gyr.

If we consider the synchronisation (and spin-orbit alignment) problem, the time to synchronise (and align) the planetary spin is
\begin{eqnarray}
\tau_\Omega = \frac{4}{9}\frac{Q' r_g^2}{2\pi}\left(\frac{m_p+m_\star}{m_\star}\right)^{2}\frac{P^{4}}{P_\mathrm{rot}P_\mathrm{dyn}^{2}}.
\end{eqnarray}
If we take $P_{\mathrm{tide}}=\frac{P_\mathrm{rot}}{2}$ (i.e. if $\Omega\gg n$ then the magnitude of the tidal frequency is approximately $2\Omega$, which is independent of $P$), we would expect spin-orbit synchronisation for planets with orbital periods shorter than
\begin{eqnarray}
P=\left(\chi \tau_\Omega \frac{18\pi}{r_g^2}\left(\frac{m_\star}{m_p+m_\star}\right)^{3}\frac{P_\mathrm{dyn}^{6}}{P_\mathrm{rot}}\right)^{\frac{1}{6}},
\end{eqnarray}
and we define $P\equiv P_\Omega$ if $\tau_\Omega=1$ Gyr. On the other hand, if we were instead to assume $P_\mathrm{tide}=P$, then we would expect spin-orbit synchronisation for planets with orbital periods shorter than
\begin{eqnarray}
P=\left(\chi \tau_\Omega \frac{18\pi}{4r_g^2}\left(\frac{m_\star}{m_p+m_\star}\right)^{3}P_\mathrm{dyn}^{6}P_\mathrm{rot}\right)^{\frac{1}{8}},
\end{eqnarray}
which was used for the estimates reported in \cite{BL2014}.

\bibliography{tid}
\bibliographystyle{mn2e}
\label{lastpage}
\end{document}